\documentclass[useAMS,usenatbib]{mn2e}
\usepackage{graphicx,fleqn,fix2col}
\DeclareGraphicsExtensions{.eps,.ps,.eps.gz,.ps.gz,.eps.Z}
\title[Irradiation models for ULXs and fits to optical data]{Irradiation models for ULXs and fits to optical data}

\author[Chris Copperwheat, Mark Cropper, Roberto Soria, Kinwah Wu]{Christopher Copperwheat$^{1}$, Mark Cropper$^{1}$ Roberto Soria$^{1,2}$ and Kinwah Wu$^{1}$\\
$^{1}$ Mullard Space Science Laboratory, University College London,
Holmbury St. Mary, Dorking, Surrey, RH5 6NT, UK\\
$^{2}$ Harvard-Smithsonian Center for Astrophysics, 60 Garden Street, Cambridge, MA 02138, USA
}

\date{Received: }

\begin{document}

\newcommand{\dg} {^{\circ}}
\outer\def\gtae {$\buildrel {\lower3pt\hbox{$>$}} \over
{\lower2pt\hbox{$\sim$}} $}
\outer\def\ltae {$\buildrel {\lower3pt\hbox{$<$}} \over
{\lower2pt\hbox{$\sim$}} $}
\newcommand{\ergscm} {erg s$^{-1}$ cm$^{-2}$}
\newcommand{\ergss} {erg s$^{-1}$}
\newcommand{\ergsd} {erg s$^{-1}$ $d^{2}_{100}$}
\newcommand{\pcmsq} {cm$^{-2}$}
\newcommand{\ros} {{\it ROSAT}}
\newcommand{\xmm} {\mbox{{\it XMM-Newton}}}
\newcommand{\exo} {{\it EXOSAT}}
\newcommand{\sax} {{\it BeppoSAX}}
\newcommand{\chandra} {{\it Chandra}}
\newcommand{\hst} {{\it HST}}
\newcommand{\subaru} {{\it Subaru}}
\def\rchi{{${\chi}_{\nu}^{2}$}}
\newcommand{\Msun} {$M_{\odot}$}
\newcommand{\Mwd} {$M_{wd}$}
\newcommand{\Mbh} {$M_{\bullet}$}
\newcommand{\Lsun} {$L_{\odot}$}
\newcommand{\Rsun} {$R_{\odot}$}
\newcommand{\Zsun} {$Z_{\odot}$}
\def\Mdot{\hbox{$\dot M$}}
\def\mdot{\hbox{$\dot m$}}
\def\mincir{\raise -2.truept\hbox{\rlap{\hbox{$\sim$}}\raise5.truept
\hbox{$<$}\ }}
\def\magcir{\raise -4.truept\hbox{\rlap{\hbox{$\sim$}}\raise5.truept
\hbox{$>$}\ }}
\newcommand{\mnras} {MNRAS}
\newcommand{\aap} {A\&A}
\newcommand{\apj} {ApJ}
\newcommand{\apjl} {ApJL}
\newcommand{\apjs} {ApJS}
\maketitle

\begin{abstract} We have constructed a model which describes the optical
emission from ultra-luminous X-ray sources (ULXs), and have used it to
constrain the parameters of seven ULX systems. Our model assumes a binary
nature for ULXs, and accounts for optical emission from an X-ray irradiated
companion star and accretion disk. We apply our model to six different ULX
optical counterparts observed with \hst, and one observed with the ESO VLT, and
determine the mass, radius and age of the donor stars in these systems. In
addition, we obtained constraints for the black hole (BH) mass in some cases.
We use the mass accretion rate implied by the X-ray luminosity of these sources
as an additional constraint on the donor star, by assuming the mass transfer is
driven by the stellar nuclear evolution. We find that in general the donors are
older and less massive than previously thought, and are consistent with being
of spectral type B. We discuss how these results affect our understanding of
the evolution and history of ULXs. Where we can constrain the BH masses, we
find them to be consistent with stellar mass BHs or intermediate mass BHs of
order $\sim 100$\Msun. We make predictions for future observations of
optical/infrared ULX counterparts, calculating binary periods for different BH
masses in each of the seven sources. \end{abstract}


\maketitle

\section{INTRODUCTION} \label{sec:intro} Ultraluminous X-ray Sources (ULXs) are
point-like, non-nuclear X-ray sources which have inferred X-ray luminosities 
in excess of $10^{39}$\ergss, assuming that the emission is isotropic.  Some
sources have X-ray luminosities even greater than $10^{40}$\ergss,  which is
far in excess of the Eddington limit for a stellar mass black hole (BH) with
mass $3$--$20$\Msun. ULXs have been observed in many nearby galaxies
\citep{Fab03,Swartz04,Liu05}, implying that they are populous. To explain the
large inferred isotropic luminosities of these sources, it has been proposed
that the accreting object is a black hole of mass  $50 - 10000$\Msun -- an
intermediate mass black hole (IMBH) \citep{ColMus99,Maki00}. Such a mass range
fits between those of the two well-known populations:  the stellar mass BHs
commonly found in our galaxy and the supermassive BHs in active galactic
nuclei. The existence of IMBHs has important consequences for the theories of
BH formation and evolution. It is still a mystery how supermassive BHs were
formed. A continuous BH mass spectrum would suggest an evolutionary link
between the three classes,  thus allowing us to better understand BH formation
in the younger universe.   

Whether IMBHs exist or not is currently under intense debate. The emission from
ULXs may not be isotropic.  If the emission is collimated or relativistically
beamed  (\citealt{King01}; \citealt*{Kord02}; \citealt{Fabrika04}), the
accretion rates of the systems do not need to violate the Eddington limit.
However, there is evidence that some sources are truly ultraluminous
\citep{Fab04}. An example is the ULXs associated with extended, diffuse
H$\alpha$ nebulae, which are more naturally explained by models with isotropic
illumination of the interstellar medium by a central X-ray source 
\citep{Pakull02,Miller03}. Low-frequency quasi-periodic oscillations (QPOs) in
the X-ray emission from some sources (M82 X-1: \citealt{Stroh03}; Holmberg IX
X-1: \citealt*{Dewangan06}; NGC 5408 X-1: \citealt{Stroh07})
suggest non-beamed emission and hence a truly high luminosity for those
sources. Low-frequency breaks in the power density spectra of NGC4559 X-7
\citep{Cr04} and NGC5408 X-1 \citep{Soria04} are also consistent with accretors
more massive than galactic BHs. However, population studies based on the
luminosity function of ULXs suggest a more modest upper limit on the mass of
IMBHs than previously thought. The break in the luminosity function at $\sim
2.5 \times 10^{40}$\ergss \ (\citealt{Swartz04}; \citealt*{Gilfanov04})
suggests the bulk of the IMBH population lies below $\sim 150$\Msun, assuming
accretion at the Eddington limit. 

Another argument proposed for IMBH accretors is based on the detection of a
spectral component interpreted as emission from a cool accretion disk
\citep{Miller04}. However, this spectral argument has been disputed
(e.g., \citealt*{Stobbart06}; \citealt{Goncalves06}). Alternative
spectral models based on high-temperature, super-Eddington disks have been
proposed \citep{Stobbart06,Ebisawa03}. It has also been suggested that super-Eddington
accretion can occur in BH systems through a thin disk covered by a
Comptonizing corona \citep{Socrates05, Done05}.  An accretion disc dominated by
radiation pressure  may exhibit strong density clumping.  If the density
inhomogeneities are on length scales much smaller than the disc scale height, 
such clumpy accretion discs could permit the radiation luminosity to exceed
$L_{Edd}$ by factors of $\sim10-100$ \citep{Bege02}. 

Since the X-ray data alone can be interpreted in different ways, we have sought
an alternative channel to X-ray investigations by which the nature of ULXs can
be elucidated. The ULX may
contain a stellar mass BH with super-Eddington accretion or an IMBH. The
optical/infrared (optical/IR) properties of ULXs  will be strongly affected by
the intense X-ray radiation field heating the donor star and accretion disk.
The emission from the heated star may be the dominant optical component (as in
the case of high mass X-ray binaries). Alternatively, theoretical modelling has
shown that the heated accretion disc might dominate (\citealt*{Rapp05};
\citealt{Copp05}), analogous to low mass X-ray binaries. In \citet{Copp05} we
considered such a binary model for ULX  and derived a formulation  which
calculated the intensity and colour variations from the intrinsic and
re-radiated emisison as a function of several parameters such as X-ray spectral
hardness, black hole mass and inclination. The predictions of the calculations
can be used to identify the optical counterpart, to determine the intrinsic
properties of the component stars, and to determine the relative contributions
of the star and disk. We may also use the optical/IR emission  as a diagnostic
of the X-ray emission and the accretion processes in ULXs/IMBHs. 

In recent times, there have been a number of optical observations of ULXs  with
the Hubble Space Telescope (\hst)  and ground based 8-m class telescopes. 
These observations have resulted  in the isolation of optical counterparts
of some ULXs,  and have provided colours and magnitudes in the optical
wavebands  (see, e.g. \citealt*{Kaaret04}; \citealt*{Liu04};
\citealt{Kuntz05}).  In this paper, we apply our model to determine the
properties of the ULX counterparts observed by these telescopes. 

In Section \ref{sec:model} we briefly reprise the \citet{Copp05} model and
describe the recent refinements. In Section \ref{sec:results} we apply our
model to seven ULX sources. In Section \ref{sec:discussion} we discuss the
implication of our results on the current views regarding the nature of ULXs
and IMBHs. 


\section{MODEL}
\label{sec:model}

\subsection{Irradiation in ULX binaries} \label{sec:mod_irrad} The
\citet{Copp05} model assumes a binary system with a BH accreting matter from a
companion star. We assume the X-ray emission from the ULX is isotropic. Our
model considers the effects of radiative transport and radiative equilibrium in
the irradiated surfaces of both the star and an accretion disc, using the
prescription described in \citet{Wu01}. 

We assume the disc is geometrically thin, and for the star we include both
gravity and limb darkening effects, and the effects of radiation pressure.
We assume the companion star is filling its Roche lobe and the BH is accreting
matter via Roche lobe overflow. This provides a geometrical constraint on the
system that is key in our determination of the optical luminosity of both the
star and the disc. These assumptions are valid when the X-ray luminosity is in
excess of $10^{40}$\ergss, since the donor star cannot generate a sufficient
stellar wind in order to maintain the accretion rate required by this X-ray
luminosity. However, some of the sources we detail in this section have X-ray
luminosities of the order of $10^{39}$\ergss. At this point it is conceivable
that the accretion mechanism could be a wind from a massive supergiant. In this
case we cannot constrain the binary separation without temporal observations
that reveal the binary period, which means we cannot quantify the stellar
irradiation. However, a wind-fed BH would only intercept a few percent of the
stellar wind from a star, and so given the mass accretion rate of $\sim
10^{-7}$\Msun/yr necessary to provide the observed X-ray luminosity, a wind
loss of $\sim 10^{-5}$\Msun/yr would be required from the star, which is very
high. ULXs also tend to have lower column densities than the typical wind-fed
sources observed in the Galaxy/LMC \citep{Stobbart06}. We therefore consider
Roche lobe overflow as the accretion mechanism to be a reasonable assumption
for all of the sources down to $10^{39}$\ergss.

\citet{Copp05} showed that the results of the irradiative calculations could
depend sensitively on the assumed hardness of the X-rays, which is specified by
a parameter $\xi$, where $\xi=S_h/S_s$ and $S_h$ and $S_s$ are the hard and
soft components of the X-ray flux respectively. As in \citet{Copp05}, the
absorption coefficients of the soft and hard X-rays are $k_s\kappa$ and
$k_h\kappa$, where $\kappa$ is the absorption coefficient of the optical
radiation and $k_s > 1$ and $k_h < 1$ defines our soft/hard X-ray convention.
The band boundary is a parameter to be determined. In \citet{Copp05} we found
that if we set $k_s = 2.5$ and $k_h = 0.01$, then for an input spectrum
consisting of a blackbody and power-law component, this results in a soft/hard
band boundary of $1.5$keV.

Determining the hardness ratio of the incident X-rays on the accretion disk is
not straightforward and is complicated by the presence of absorption. Absorption
tends to harden the X-rays. If the absorbing region which produces this hardened
spectrum is intrinsic to the X-ray emitting region itself, the disk and star
will be irradiated by the same hard X-ray spectrum that we observe. If the
absorbing region is located between the binary system and the observer,
then the irradiating X-rays will be much softer than is observed.

By examining X-ray observations from ULX X-7 in NGC 4559 \citep{Cr04} we see
that for this source, reasonable physical values for the hardness ratio range
from $\xi \sim 0.1$ to $\sim 1$. In this paper, we set the hardness ratio to
$0.1$ since we expect a locally soft irradiating spectrum. In Section
\ref{sec:errors} we examine how varying this parameter affects our results.

In order to determine the mass accretion rates and irradiating flux for each
source, we use the characteristic average values of X-ray luminosity as given
by the authors listed in Table \ref{tab:photdat}. We note that this may be a
source of uncertainty in our results: since the X-ray observations are not
contemporary with the optical observations, the X-ray luminosity and spectrum
at the time of the optical observation may be different.

We assume the compact object in the binary system is a BH, but we do not assume
it to be an IMBH. For all sources we use BH masses of $10$ -- $1000$\Msun \ in
our model, a range that encompasses both a stellar mass and an intermediate
mass nature for the BH. We do not use BH masses of beyond $1000$\Msun \ since
at this point the optical emission is dominated by a large disc, and increasing
the BH mass further has little additional effect on our model results.
Similarly, results for a BH mass of $10$\Msun \ can be taken to apply to BHs
with masses less than this.

\subsection{Inclusion of stellar evolution models}   As well as detailing our
model in \citet{Copp05}, we applied it to the candidates for the optical
counterpart of ULX X-7 in NGC 4559. We took the masses, radii and
luminosities of various main sequence (MS) and supergiant stars from the tables
in \citet{Allen73}. We included these parameters in our model, and hence were
able to calculate the optical luminosity of the ULX if the companion star in
the binary system was described by those parameters. We plotted tracks on colour
magnitude diagrams, which describe the effect on the optical luminosity when
the BH mass was varied. We compared these tracks to the measured colours and V
band magnitudes of the candidates. This allowed us to constrain the properties
of possible candidates. 

In this paper we have further developed our method by using the Geneva stellar
evolution models of \citet{Lejeune01} to provide more realistic stellar
parameters, and also to provide constraints on the derived ages. We primarily
use the high mass-loss tracks which are recommended for use when dealing with
massive stars \citep{Maeder94}. These extend up to a stellar age of
$10^{7.5}$yr. We supplement these with the standard tracks when we wish to
consider less massive stars with an age greater than this. We input the set of
stellar parameters at each point along the evolutionary tracks into our model.
We use our model to produce colours and magnitudes appropriate for the 
irradiated star and disc. We do this for all relevant tracks, which enables
us to include stellar age as an additional parameter in our analysis. We repeat
this process as we vary the other important parameters, such as X-ray
luminosity, the inclination and orientation of the binary system, and BH mass.
We described the parameter space relatively completely in \citet{Copp05}. 

An additional parameter, introduced by our use of the Geneva tracks, is the
stellar metallicity. We use a sub-solar (Z$=0.2$\Zsun) metallicity throughout.
This is appropriate given that many ULXs are in low-metal environments such as
dwarf galaxies. Low-metal stars also lose less mass in stellar winds
\citep{Eldridge06}. Therefore, we speculate that they may end their lives with
bigger cores, which can more easily collapse directly into BHs
\citep{Heger03}. 

We note that we assume the Geneva models and metallicity to be correct, so we
have not taken into account any systematic error introduced by a difference
between the stellar evolution as described by those models, and the evolution of
a donor in a ULX binary system. A binary evolution code is not necessarily more
appropriate, since we do not know when the binary evolved to a semi-detached
state, and so we do not know to what extent the normal evolution of the star has
been disrupted by the mass transfer. We therefore use a single-star model in
this paper to describe the simplest case, where the donor star has not been
significantly perturbed by the mass transfer.

\subsection{Mass accretion rate as a constraint of the system parameters}

For some of the sources in our analysis  optical data alone is sufficient  to
determine the parameters of the donor star with good accuracy.   For
the others there is a large uncertainty  even in establishing the luminosity
class of the donor star,  because the data can be fitted by a wide range of
stellar parameters.   In this situation, we can use addition conditions to
constrain the parameter space.  

We consider a method that makes use of the information provided by the X-ray
data  and model stellar evolutionary calculations,  which is essentially
independent of the optical/IR photometric observations. The X-ray luminosity
$L_x$ of an accreting object is given by the relation 
\begin{equation}
   L_x = \eta {\dot M} c^2 \  ,  
\label{eqn:accrate}
\end{equation} 
where $\dot M$ is the mass accretion rate and  $\eta$ is the efficiency
parameter.  For accretion onto BH, we may take $\eta = 0.1$,  which  is
sufficient for the purpose of this study.   A ULX luminosity of $10^{40}$\ergss
\  therefore corresponds to  a mass loss rate of ${\dot M} \simeq 1.8 \times
10^{-6}$\Msun/yr for the donor star, if we assume all mass outflow from
the star will be accreted onto the BH.     

We suppose that mass transfer occurs  due to the donor star overfilling its
Roche-lobe,  and the transfer is driven by a gradual volume expansion of the
star as it evolves away from its main-sequence phase.   For a quasi-steady state,   the rate of mass loss from the donor star ${\dot M}_2$ is given by 
\begin{eqnarray} 
 {\dot M}_2 & = & \frac{M_2}{\zeta_{2s} - \zeta_{2r} } 
   \left[\frac{2}{\tau_{J}} - \frac{1}{\tau_{th}} -   \frac{1}{\tau_{nuc}} \right] \label{eqn:mdottimess} \ , 
\end{eqnarray} 
where $M_2$  is the mass of the donor star,  $\zeta_{2s}$ and $\zeta_{2r}$
are the adiabatic indices  of the mass donor star and its Roche lobe
respectively,  $\tau_{J}$ is the time scale of orbital angular momentum loss, 
$\tau_{th}$ is the thermal time scale of the donor star  and $\tau_{nuc}$ is
the nuclear evolutionary time scale  of the donor star
\citep*{Ritter88,Dantona89}. Stable mass transfer occurs when $\zeta_{2s} -
\zeta_{2r} > 0$. 

The orbital evolution of binary
undergoing mass transfer is governed by the following equation (see
\citealt{Wu97}): 
\begin{equation}
   {\dot a \over a} = 2{\dot J \over J} -
  2{\dot M_2 \over M_2}\Bigl[ 1 -  {\beta \over q} - {1 \over 2}\Bigl({{1-\beta}
\over {1+q}}\Bigr) -  {\alpha \over q}(1-\beta)(1+q)\Bigr] \label{eqn:massxfer} \ , 
\end{equation} 
where $a$ is the orbital separation, $J$ is the orbital angular momentum,  $q$
is the mass ratio $M_1/M_2$, $\beta$ is the fraction of mass loss from the
donor star accreted onto the BH, $\alpha$ is the specific angular momentum
carried away by mass loss from the system,  and ``dot'' denotes the time
derivative.   For conservative mass transfer, which is assumed in this study,  
we have $\dot J = \alpha =0$, and $\beta=1$.  It follows that  
\begin{equation}  
{\dot a \over a} = -2\Bigl({{q-1} \over {q}}\Bigr) {\dot M_2 \over M_2} .  
 \label{eqn:massxfer2} 
 \end{equation} 

The Roche-lobe radius $R_L$ of the donor star and the orbital separation $a$ is
well aproximated via 
\begin{equation} 
   {R_L \over a} = {{0.49 q^{-2/3}} \over {0.6 q^{-2/3} + \ln(1+q^{-1/3})}}
\label{eqn:rlapprox}
\end{equation}
   \citep{Eggleton83}.   
The Roche-lobe filling condition requires $R_L = R$,   where $R$ is the radius
of the donor star.    Suppose that the mass transfer is quasi-steady, {\it i.e.}
${\dot R}_L = {\dot R}$.     By combining these conditions with Equations
\ref{eqn:massxfer2} and \ref{eqn:rlapprox},  we obtain 
\begin{equation}  
  \frac{{\dot M}_2}{M_2} = \frac{\dot R}{R} 
      \left[  \frac{q}{2(1-q) + (1+q)\left[2/3 - g(q)\right]}  \right] \ , 
\label{eqn:dotM2} 
\end{equation} 
where the function $g(q)$ is given by 
\begin{equation}   
   g(q) = \frac{(2/5)q^{-2/3}+q^{-1/3}\left[ 3(1+q^{-1/3})\right]^{-1}}
      {(3/5)q^{-2/3}+\ln(1+q^{-1/3})}  \ .
\end{equation}  
For mass transfer driven by nuclear evolution,  
\begin{eqnarray} 
   {\dot M}_2 & \approx & - \gamma \frac{M_2}{\tau_{nuc}} \label{eqn:nuctimess} \ ,
\end{eqnarray} 
where $\gamma$ is a positive, slowly  varying parameter  depending on the
spectral type and mass of the donor star and the orbital parameters.     The
faster the donor star evolves, the higher the mass transfer rate will be,   and
how rapid mass transfer occurs is dictated by the evolutionary timescale. The rate of expansion of the stellar radius $\dot R$  and hence the mass
accretion rate $\dot M$ at particular evolutionary stage of the donor star can
be derived from stellar evolutionary tracks. 

Unless stated otherwise, we set the
the mass accretion rate for sources with X-ray luminosities  $L_x \sim
10^{40}$\ergss \  to be $\sim 10^{-6}$\Msun/yr and other sources with $L_x \sim
10^{39}$\ergss to be $\sim 10^{-7}$\Msun/yr.

\subsection{Mass transfer timescale considerations} \label{sec:ttmt} Some of the
models constructed to support stellar-mass BHs in ULXs propose they are
intermediate- or high-mass X-ray binaries undergoing a phase of mass transfer on
thermal timescales \citep{King01,King02}.  

Thermal-timescale mass transfer can occur when the ratio $q$ of donor mass to BH
mass exceeds a critical value $q_{crit} \sim 1$. The mass transfer is initiated
by orbital evolution of the binary or nuclear evolution of the star, and for
mass ratio of $q > q_{crit}$ the Roche lobe of the secondary will shrink in
response to the transfer. This will result in more of the donor spilling out of
the Roche equipotential and a phase of rapid mass transfer. In response to the
mass loss, the donor will decrease in size. In general, we find that the thermal
timescale is much shorter than the nuclear evolutionary timescale for early type
stars. On the other hand, relatively massive donors are required to achieve $q >
q_{crit}$. If a large fraction of ULXs are found to be undergoing
thermal-timescale mass transfer then the parameter space is much smaller than we
examine, since the runaway scenario can only occur with a mass ratio $q >
q_{crit}$.

In Equation \ref{eqn:nuctimess} we take the mass transfer to be proceeding on
the nuclear evolutionary timescale; thus we do not consider thermal-timescale transfer in
this paper explicitly. Our model breaks down when the thermal timescale is
comparable to the nuclear evolutionary timescale of the donor. In this case the star cannot
adjust to the mass loss fast enough to prevent runaway, unstable mass transfer.
Then the outer layers of the star cannot remain in radiative equilibrium and the
assumptions of our model are violated. 

Note from Equation \ref{eqn:mdottimess} that our formalism is versatile enough
to consider thermal timescale mass transfer (in this case the mass transfer
rate cannot be deduced directly from the  stellar evolutionary track).  Our
irradiation model is applicable irrespective of the mechanism and timescale of mass transfer.

\subsection{Inclination and orientation considerations} For all ULXs the
inclination with respect to the observer is currently unknown. At one extreme, the orbital
plane of a binary is perpendicular to the plane of the sky ($\cos(i)=0.0$). In
this case with a thin disk, all of the optical flux we observe will be from the
star. For any other inclination the optical flux will also contain a disc
component, the relative contribution of which will increase as $\cos(i)$ is
increased to $1.0$. The phase of the companion star at the time when the
observations were made is also unknown. If the star is in superior conjunction
with respect to the observer, the observation will be of the irradiated
hemisphere of the star. If the star and the observer are in inferior
conjunction, the observation will be of the hemisphere facing away from the BH
and therefore the flux from the star will contain little or no irradiated
component.

We noted in \citet{Copp05} that the geometric constraints of the binary system
determined whether reprocessed light from the disc or the star dominated
the optical/IR emission. The disc is truncated by tidal forces, and so when the
separation between the star and the BH is large, the disc is also large and
hence more likely to be the dominant optical component. A large separation is a
consequence of assuming a high BH mass, so a high BH mass implies disc
dominated optical/IR emission. We would expect that inclination would therefore
dominate the geometrical effects in high BH mass systems. A low BH mass
generally implies the emission is dominated by the donor star. We would
therefore expect the phase of the star to have a significant effect on the
results only in the cases where we assume a low BH mass.

We concentrate our analysis on the general case where there is an irradiated
component from both the star and the disc. We therefore start by assuming
$\cos(i)=0.5$.  As we demonstrated in \citet{Copp05} this orientation already
results in a strong contribution to the optical flux from the disc, and
increasing the inclination to $\cos(i)=1.0$ has little additional effect. By
considering the cases where the star is in superior and inferior conjunction we
examine the phases where the stellar contribution to the observed optical
emission is strongest and weakest for this inclination. We then additionally
consider superior conjunction and $\cos(i)=0.0$, describing the case when all of
the optical flux is from the irradiated hemisphere of the star. When we assume
inferior conjunction and take $\cos(i)=0.0$, we would see only the unirradiated
hemisphere of the star. In this case it would be appropriate to use a standard
unirradiated star. These extreme states allow us to explore relatively
completely the inclination and phase orientation parameter space.


\section{APPLICATION TO OBSERVATIONS} \label{sec:results} We now apply
our calculations to the observed optical counterparts of six ULXs observed with
\hst, and one observed with the ESO VLT telescope. The photometric values we
have used are in Table \ref{tab:photdat}. Where necessary, we have converted
the values to absolute magnitudes using the distances given in the table.

We have corrected the values for the NGC 4559 and the NGC 1313 ULXs using the
Galactic $E(B-V)$ values given in \citet{Soria05} and \citet{Mucciarelli05}
respectively, and $A_V / E(B-V) = 3.1$. For the M101 ULX we use the $E(B-V)$
from \citet{Kuntz05}, which includes Galactic reddening and reddening from the
disk of M101. In the case of the NGC 5204 ULX, we follow \citet*{Liu02} for the
reddening correction, using $n_H = 10^{21}$ cm$^{-2}$ and assume the Galactic
relation $n_H = 5.8 \times 10^{21}E(B-V)$ \citep*{Bohlin78}. The remaining
objects were already corrected for Galactic reddening. 

We determine the uncertainty in the values given in Table \ref{tab:photdat} by
taking the error given for the original photometric values, and combine this
with any additional uncertainty we have introduced by correcting for reddening.
We base our confidence intervals on these errors, and fix the X-ray luminosity
and hardness. We assume the distances given in Table \ref{tab:photdat} to be
correct, and so do not introduce any additional uncertainty where we have
converted from apparent to absolute magnitudes.

\begin{table*} \caption{The photometric data we have used for the ULX in this
work. Where necessary, we have converted the figures given in the references to
absolute magnitudes and have applied reddening corrections for galactic
absorption. } \label{tab:photdat} \begin{tabular}{l|l|l|l|l|l|l}
              	    &$M_{B}$        	&$M_{V}$            &$M_{R_{c}}$     	&$M_{I_{c}}$        &Distance (Mpc)\\
\hline
NGC 4559 X-7        &$-7.22 \pm 0.19$   &$-7.03 \pm 0.16$   &               	&$-6.98 \pm 0.16$   &$10$    	&\citep{Soria05}    \\ 
M81 X-6             &$-4.28 \pm 0.04$   &$-4.18 \pm 0.03$   &               	&$-4.20 \pm 0.07$   &$3.63$   	&\citep{Liu02}    \\
M101 ULX-1          &$-6.19 \pm 0.15$   &$-5.92 \pm 0.12$   &               	&$-5.81 \pm 0.16$   &$7.2$    	&\citep{Kuntz05}    \\
NGC 5408 ULX        &$-6.4 \pm 0.2$   	&$-6.4 \pm 0.2$     &               	&$-6.1 \pm 0.1$     &$4.8$    	&(\subaru \ and \hst \ archival data)\\
Holmberg II ULX     &$-6.03 \pm 0.19$   &$-5.78 \pm 0.11$   &               	&             	    &$3.05$   	&\citep{Kaaret04}   \\
NGC 1313 X-2 (C1)   &$-4.7 \pm 0.18$ 	&$-4.5 \pm 0.18$    &$-4.2 \pm 0.18$ 	&             	    &$3.7$      &\citep{Mucciarelli05}\\
\hline
\end{tabular}
\begin{tabular}{l|l|l|l|l|l|l}

\\
HSTMAG          &F$220$W     &F$435$W     &F$606$W     &F$814$W     &Distance (Mpc)\\
\hline
NGC 5204 ULX     &$-8.51 \pm 0.11$   &$-6.49 \pm 0.11$   &$-5.44 \pm 0.13$   &$-4.38 \pm 0.13$   &$4.3$     &\citep{Liu04}    \\
\end{tabular}
\end{table*}

For each source, we take each evolutionary track in turn, and comparing the available
photometric observations with our model calculations, calculate the variation
in the $\chi$-squared statistic along it. We then combine our tracks, and
determine the range of the important parameters to $68$\%, $90$\%, $95$\% and
$99$\% confidence levels. The results of our calculations are in Figures \ref
{fig:m81_0.5sc} to \ref{fig:5408colours} and Tables \ref{tab:zams} to
\ref{tab:ages}. We show all four confidence levels in the figures, but the
values we quote in the tables and the text are taken at the $90$\% confidence
level.


\subsection{ULX X-7 in NGC 4559}

\begin{figure*}
\centering

\begin{minipage}[c]{0.5\textwidth}
\includegraphics[angle=270,width=1.0\textwidth]{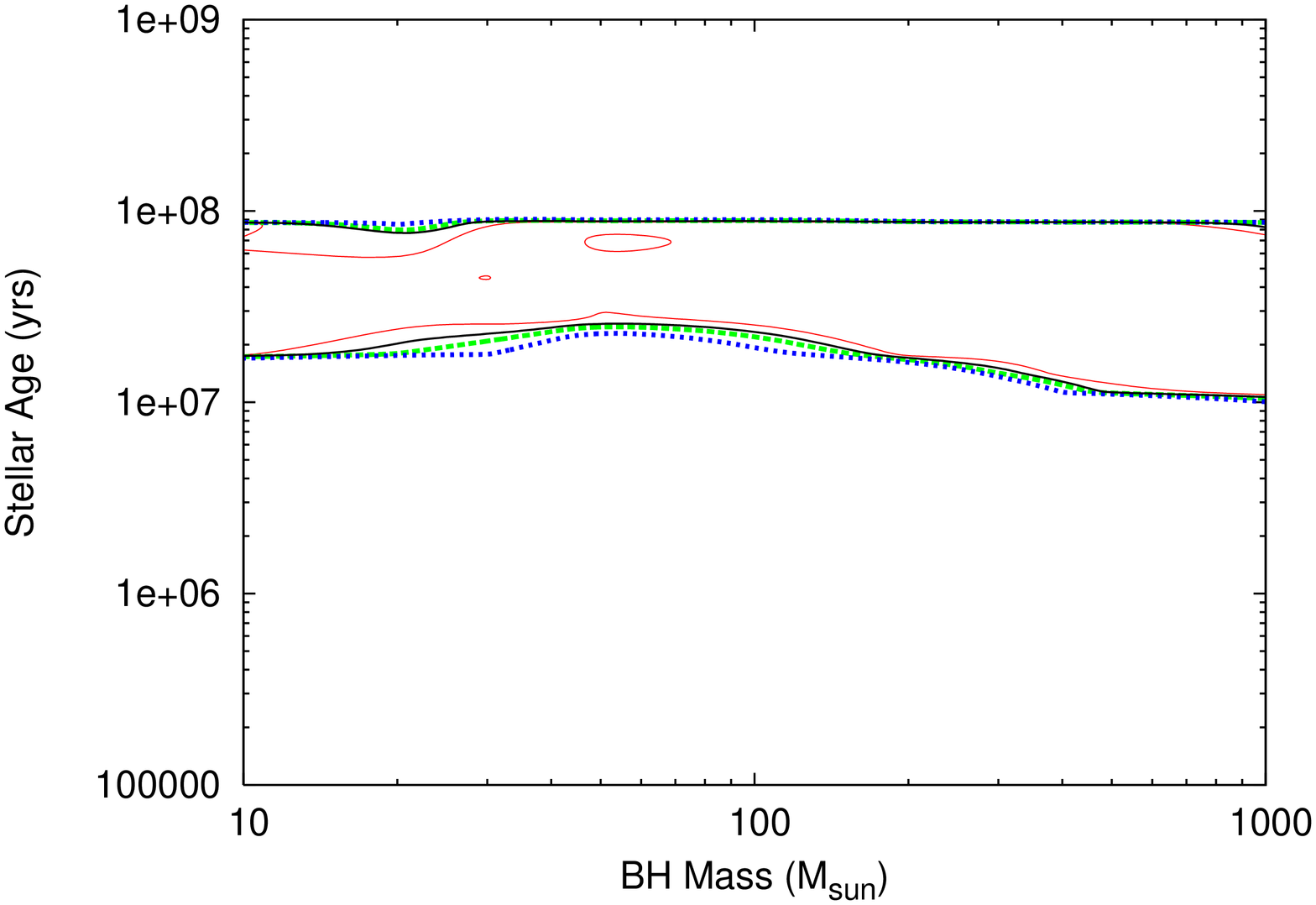}
\end{minipage}%
\begin{minipage}[c]{0.5\textwidth} \hfill
\includegraphics[angle=270,width=1.0\textwidth]{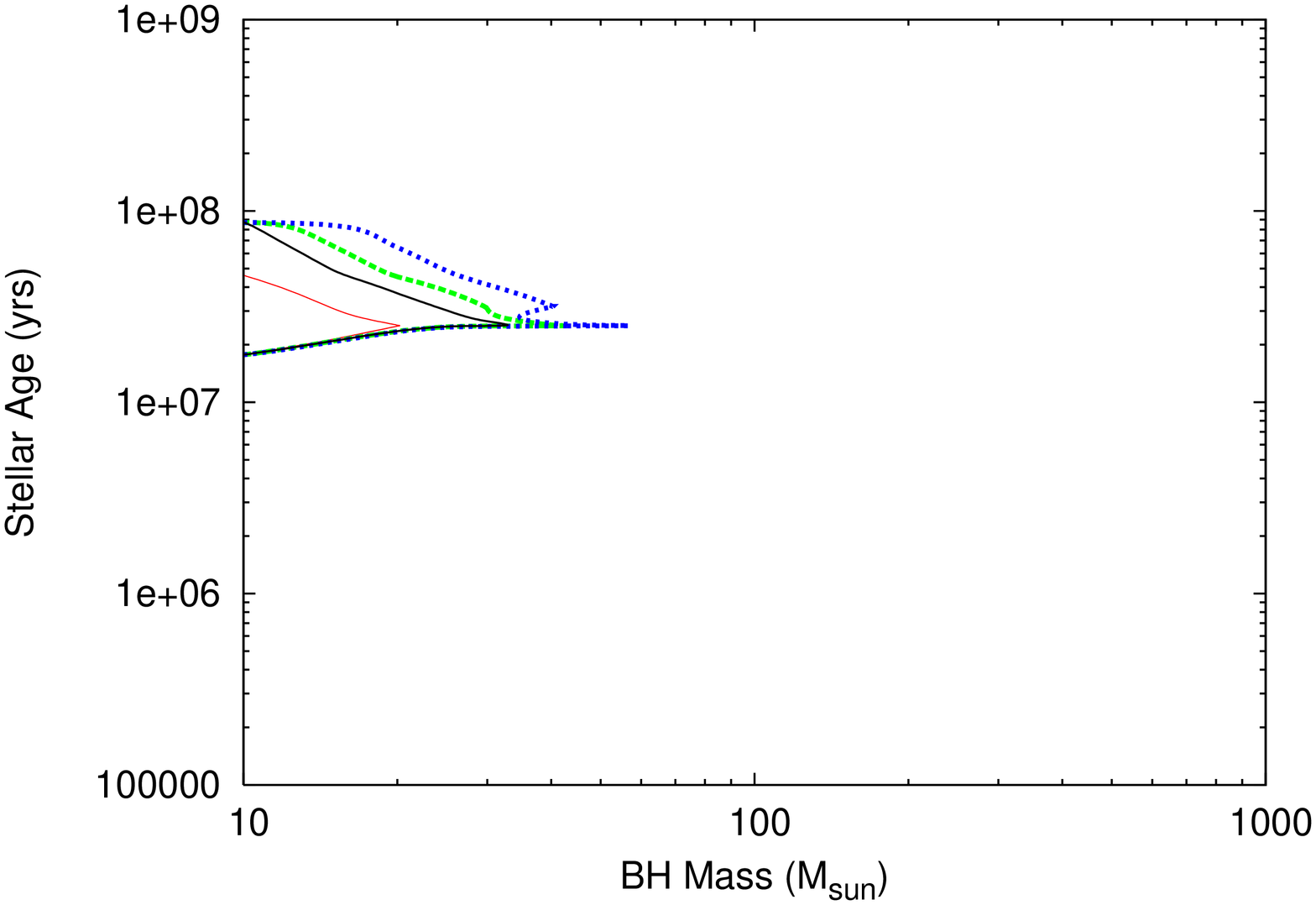}
\end{minipage} 

\begin{minipage}[c]{0.5\textwidth}
\includegraphics[angle=270,width=1.0\textwidth]{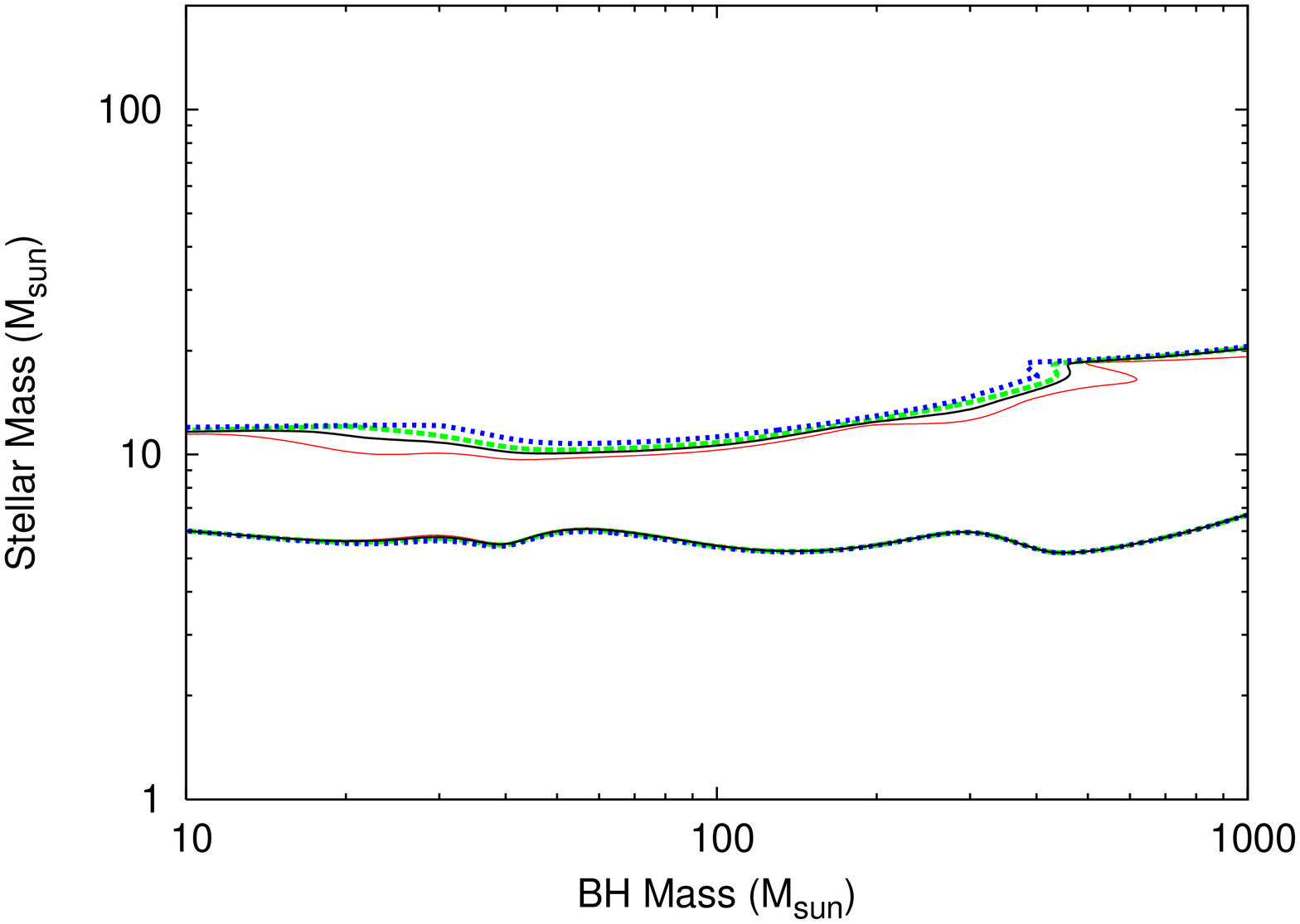}
\end{minipage}%
\begin{minipage}[c]{0.5\textwidth} \hfill
\includegraphics[angle=270,width=1.0\textwidth]{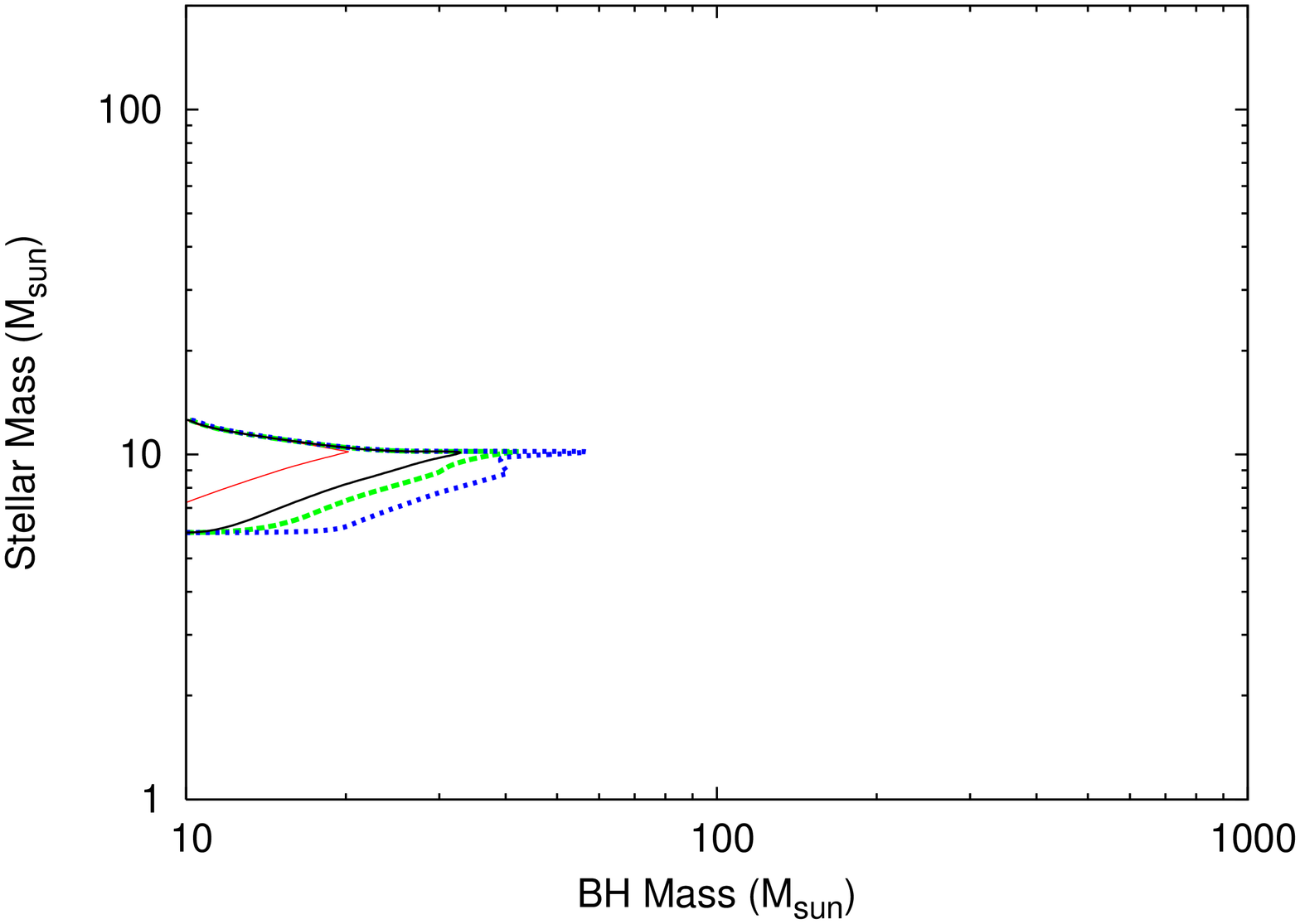}
\end{minipage} 

\begin{minipage}[c]{0.5\textwidth}
\includegraphics[angle=270,width=1.0\textwidth]{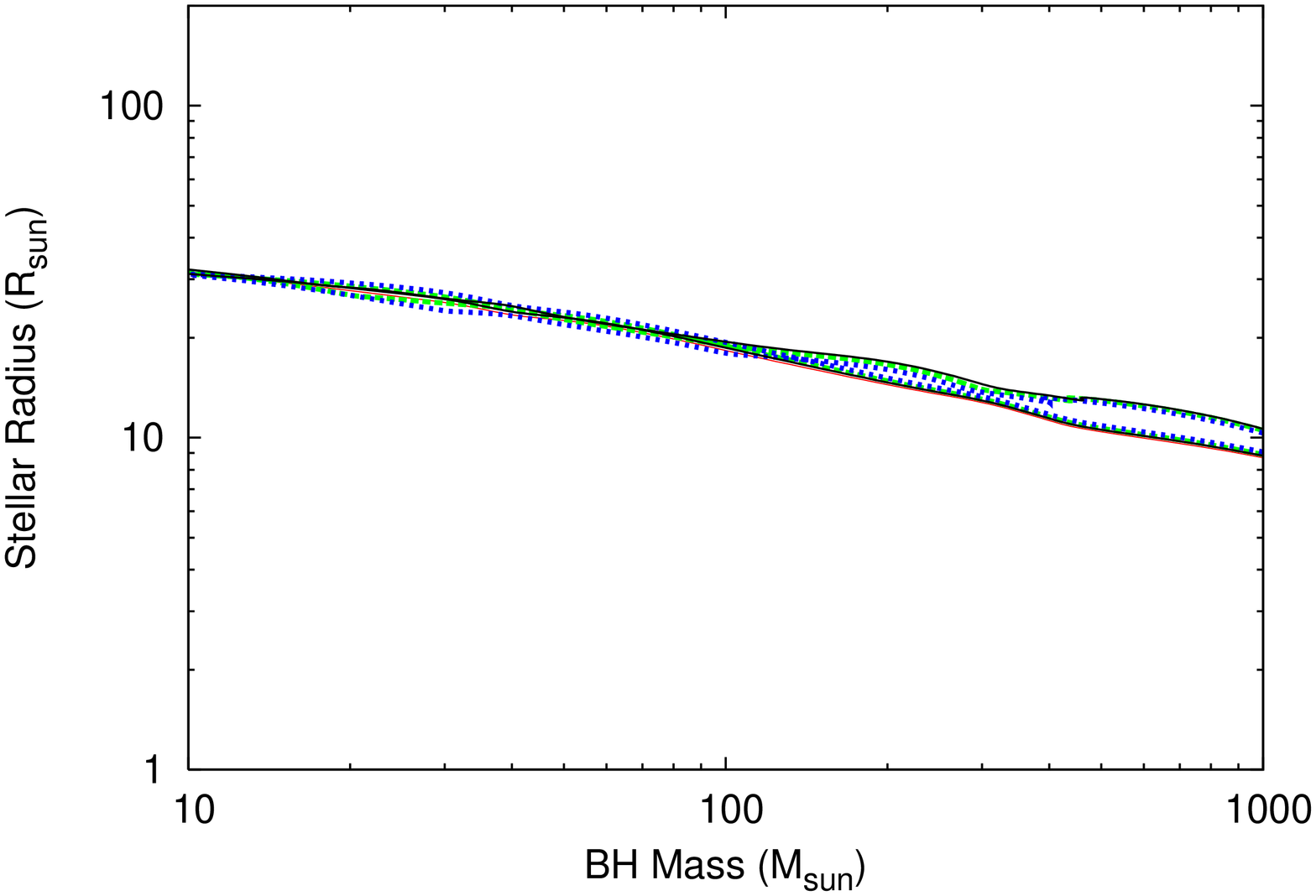}
\end{minipage}%
\begin{minipage}[c]{0.5\textwidth} \hfill
\includegraphics[angle=270,width=1.0\textwidth]{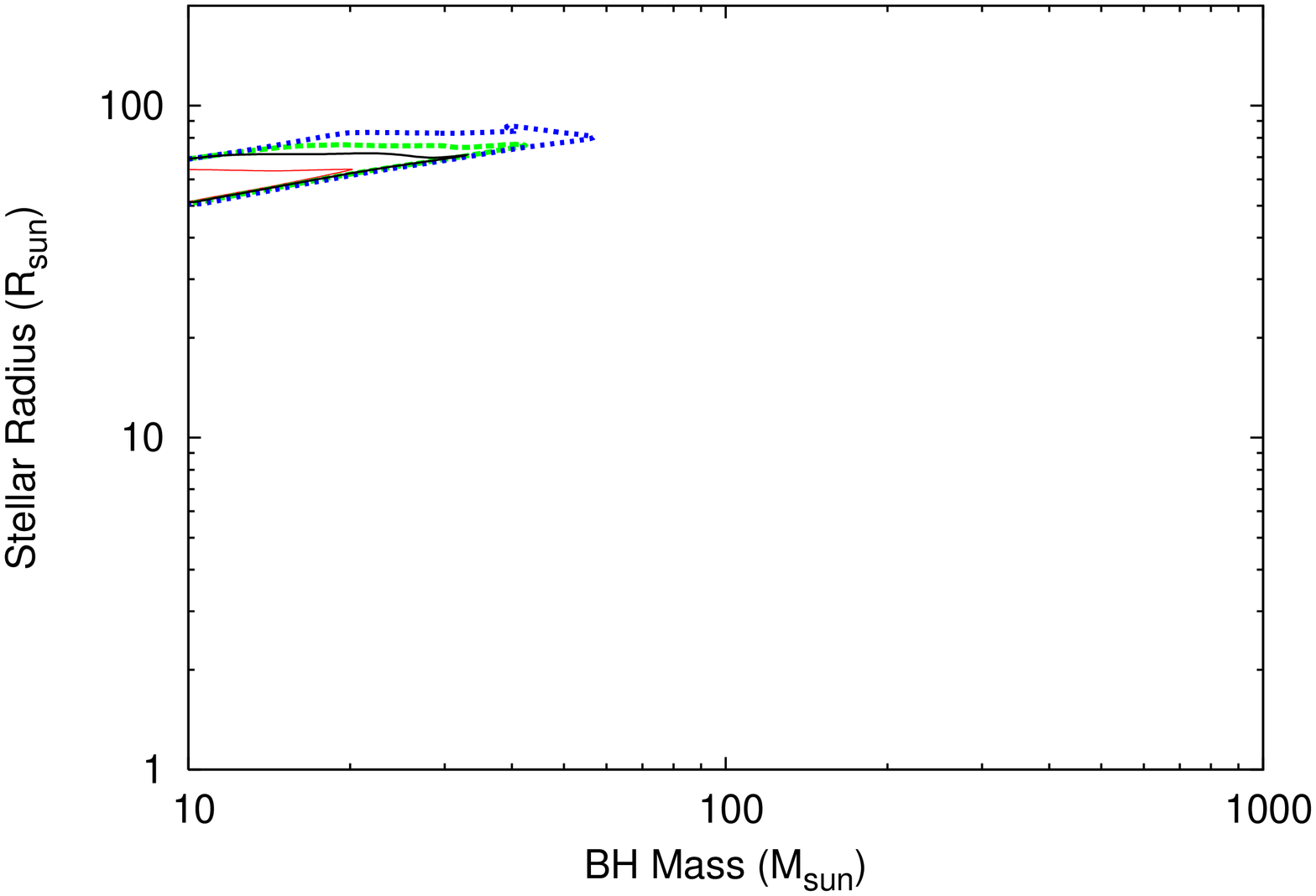}
\end{minipage}

\begin{minipage}[t]{0.49\textwidth}
\caption{Confidence contours for the binary parameters for the
source X-7 in NGC 4559, assuming candidate 1 is the optical counterpart.
We assume a binary inclination of $\cos(i)=0.5$, superior conjunction with
respect to the observer and the star, and a stellar metallicity of
Z$=0.2$\Zsun. We use an X-ray hardness ratio of
$\xi=0.1$. The red, black, green and blue lines denote the $68$\%, $90$\%,
$95$\% and $99$\% confidence intervals respectively.
}
\label{fig:4559_1_0.5}
\end{minipage}%
\begin{minipage}[t]{0.02\linewidth} \hfill
\end{minipage}%
\begin{minipage}[t]{0.49\textwidth} \hfill
\caption{As for Figure \ref{fig:4559_1_0.5}, but with a binary inclination of
$\cos(i)=0.0$.}
\label{fig:4559_1_0.0}
\end{minipage}

\end{figure*}

\citet{Soria05} used \hst \ WFPC2 observations to study the optical environment
of ULX X-7 in NGC 4559, a source with an average X-ray luminosity of
$10^{40}$\ergss \citep{Cr04}. They found eight possible candidates for the ULX
optical counterpart,  listing the $B$, $V$ and $I_{C}$ standard magnitudes for
each in table 2 of that paper. 

We applied our model to this source in \citet{Copp05}. We now do this again,
using the revised model as described in Section \ref{sec:model}. We apply the
extinction correction for absorption within our galaxy given in \citet{Soria05}
to the photometric observations. As in \citet{Copp05}, we find that candidates
2, 3 and 4 are consistent with our model only when we use a BH mass of $\simeq
1000$\Msun \ and an inclination of $\cos(i)=0.0$.

\citet{Soria05} suggested that candidate 1 was the most likely counterpart. The
findings of \citet{Copp05} are consistent with this and so for our further
analysis of this source we will assume candidate 1 is the optical counterpart.

In Figures \ref{fig:4559_1_0.5} and \ref{fig:4559_1_0.0} we plot the confidence
contours for the stellar age, mass and radius against the BH mass, for an
inclination of $\cos(i)=0.5$ in Figure \ref{fig:4559_1_0.5} and $\cos(i)=0.0$
in Figure \ref{fig:4559_1_0.0}. We assume the star is in superior conjunction
in both cases. 

We see first that there is an upper bound on the BH mass of $\sim 35$\Msun \ in
the $\cos(i)=0.0$ case. As for the donor star parameters, if we examine the
$\cos(i)=0.5$ case first we see that the age ranges from $10^{7}$ -- $10^8$yr,
the mass ranges from $5$ -- $20$\Msun \ and the radius is between $9$ and
$30$\Rsun, with the lower radii implying a higher BH mass. For the
$\cos(i)=0.0$ case the age and mass range is similar. The radius lies between
$51$ and $72$\Rsun \ when we assume this inclination.

These values are consistent with candidate 1 being of a similar mass and age to
the other candidates within the error circle \citep{Soria05}, with its increased luminosity due
to the effects of irradiative heating.

\subsection{ULX X-6 in M81}

\begin{figure*}
\centering

\begin{minipage}[c]{0.5\textwidth}
\includegraphics[angle=270,width=1.0\textwidth]{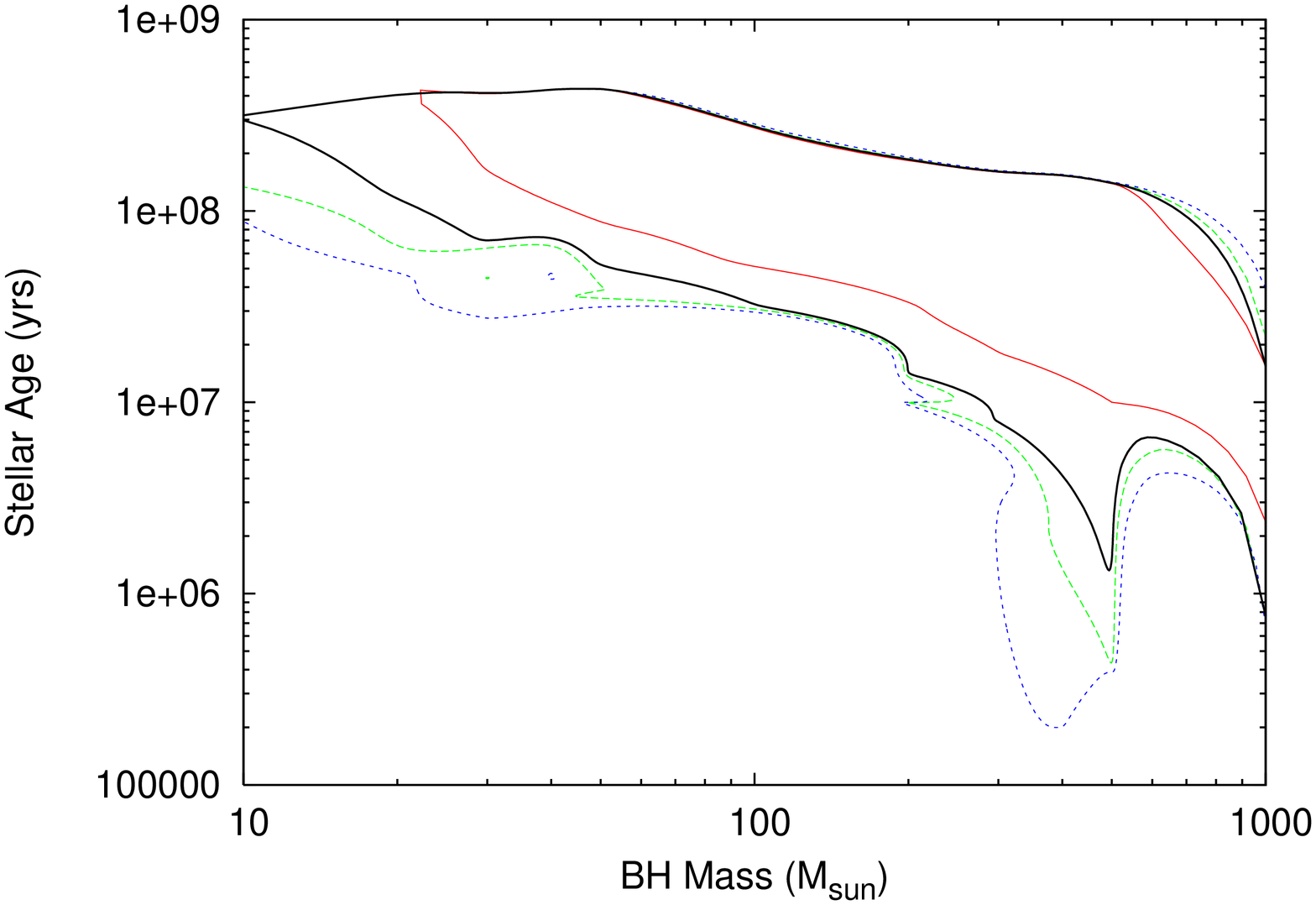}
\end{minipage}%
\begin{minipage}[c]{0.5\textwidth} \hfill
\includegraphics[angle=270,width=1.0\textwidth]{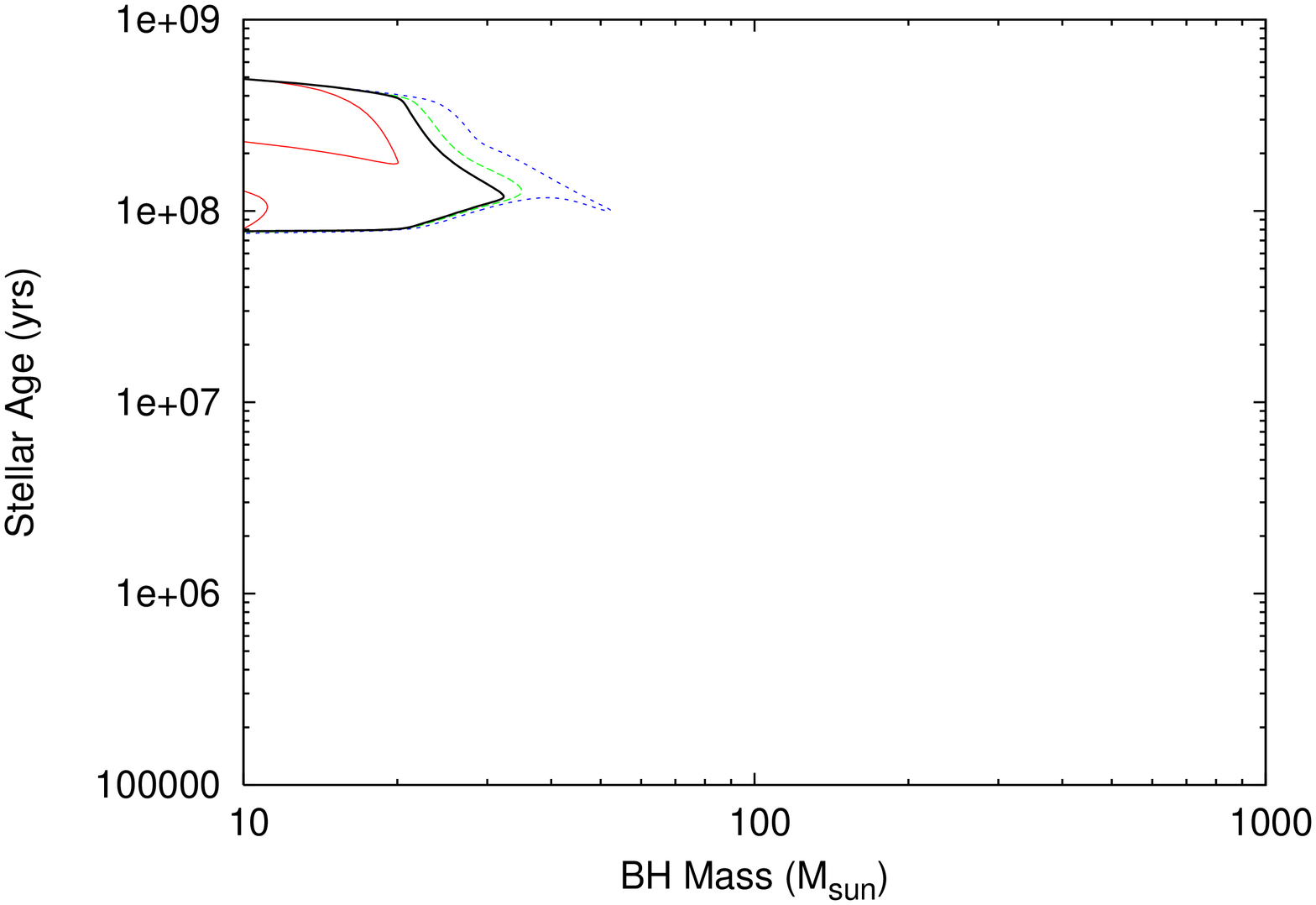}
\end{minipage} 

\begin{minipage}[c]{0.5\textwidth}
\includegraphics[angle=270,width=1.0\textwidth]{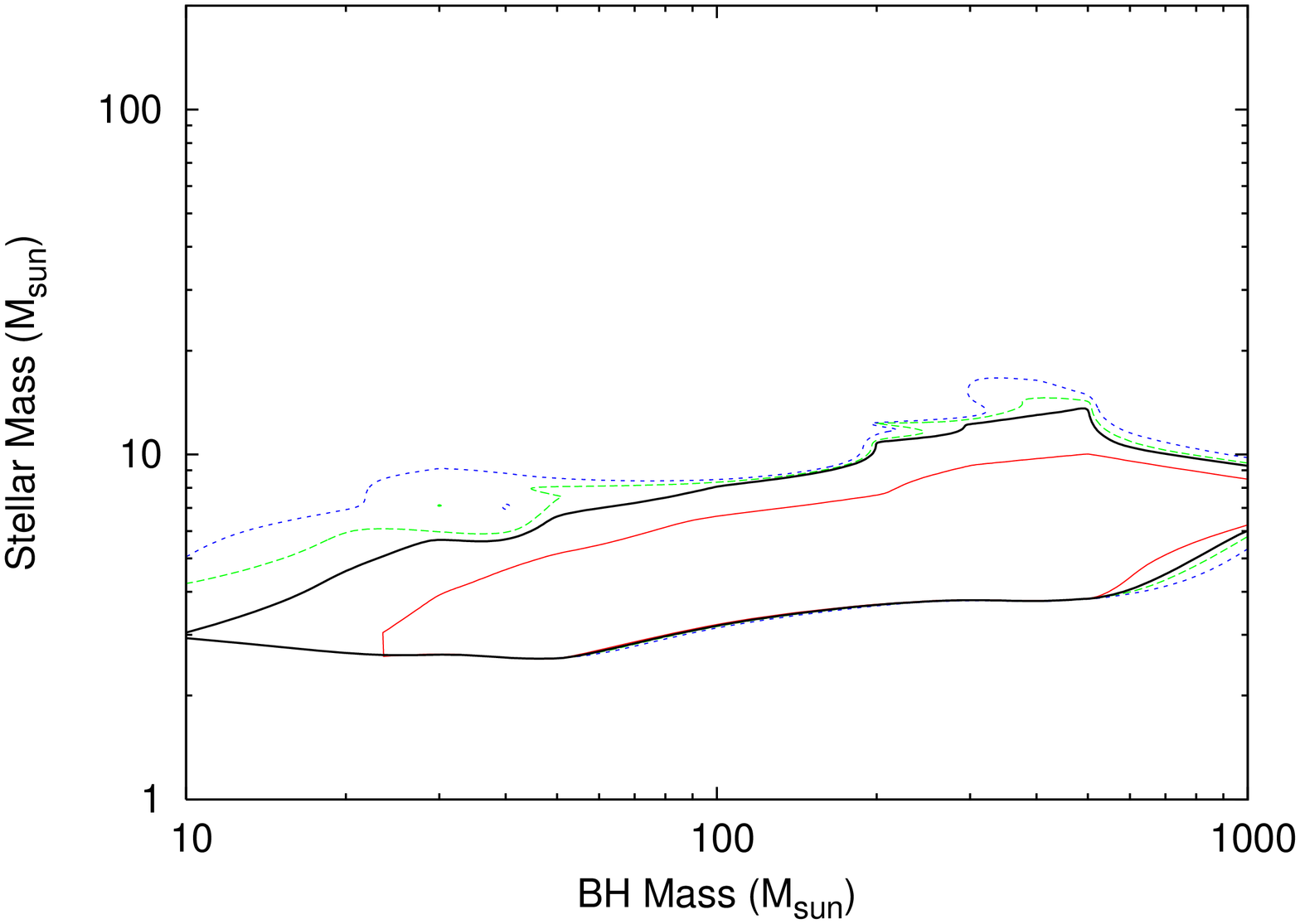}
\end{minipage}%
\begin{minipage}[c]{0.5\textwidth} \hfill
\includegraphics[angle=270,width=1.0\textwidth]{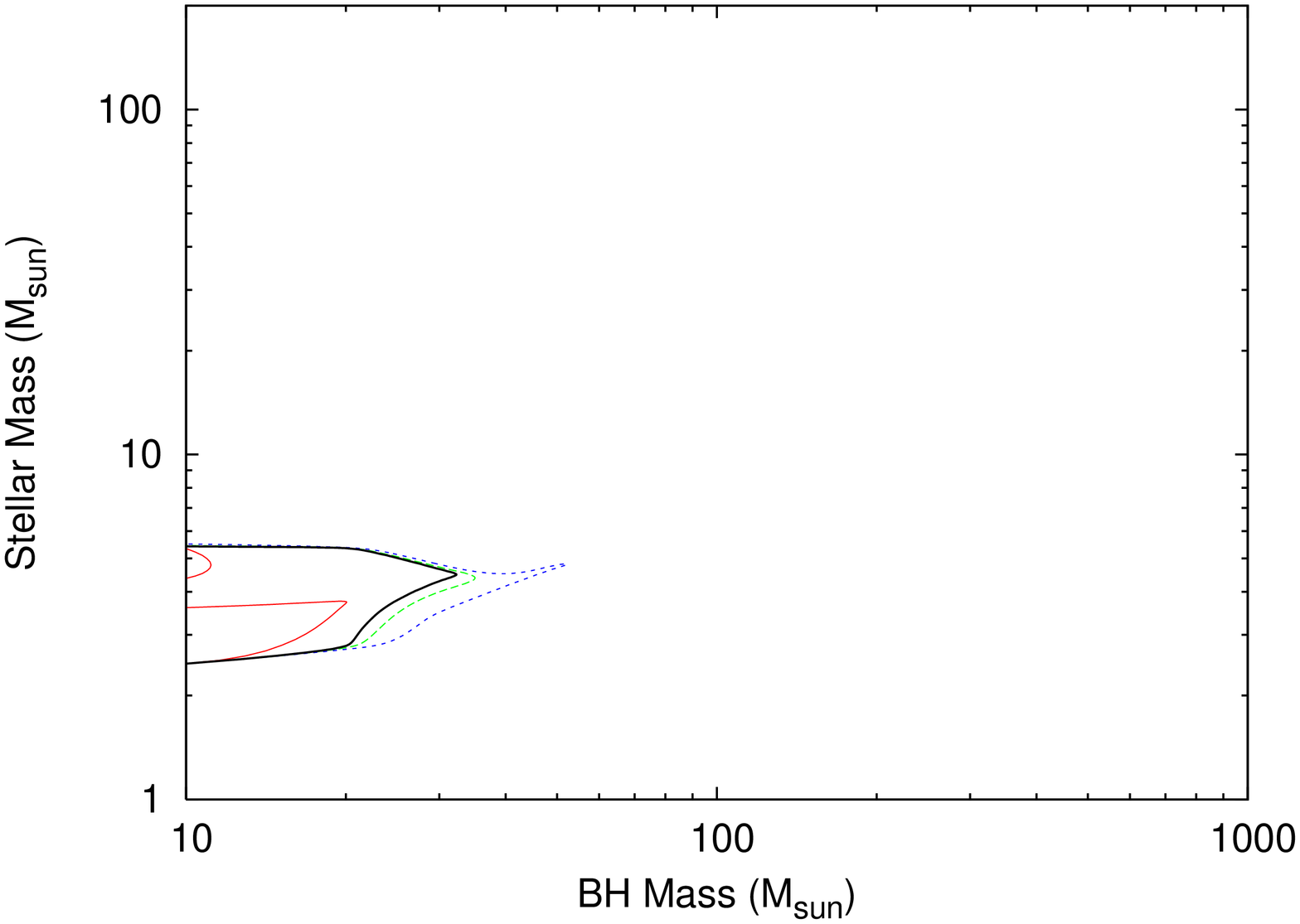}
\end{minipage} 

\begin{minipage}[c]{0.5\textwidth}
\includegraphics[angle=270,width=1.0\textwidth]{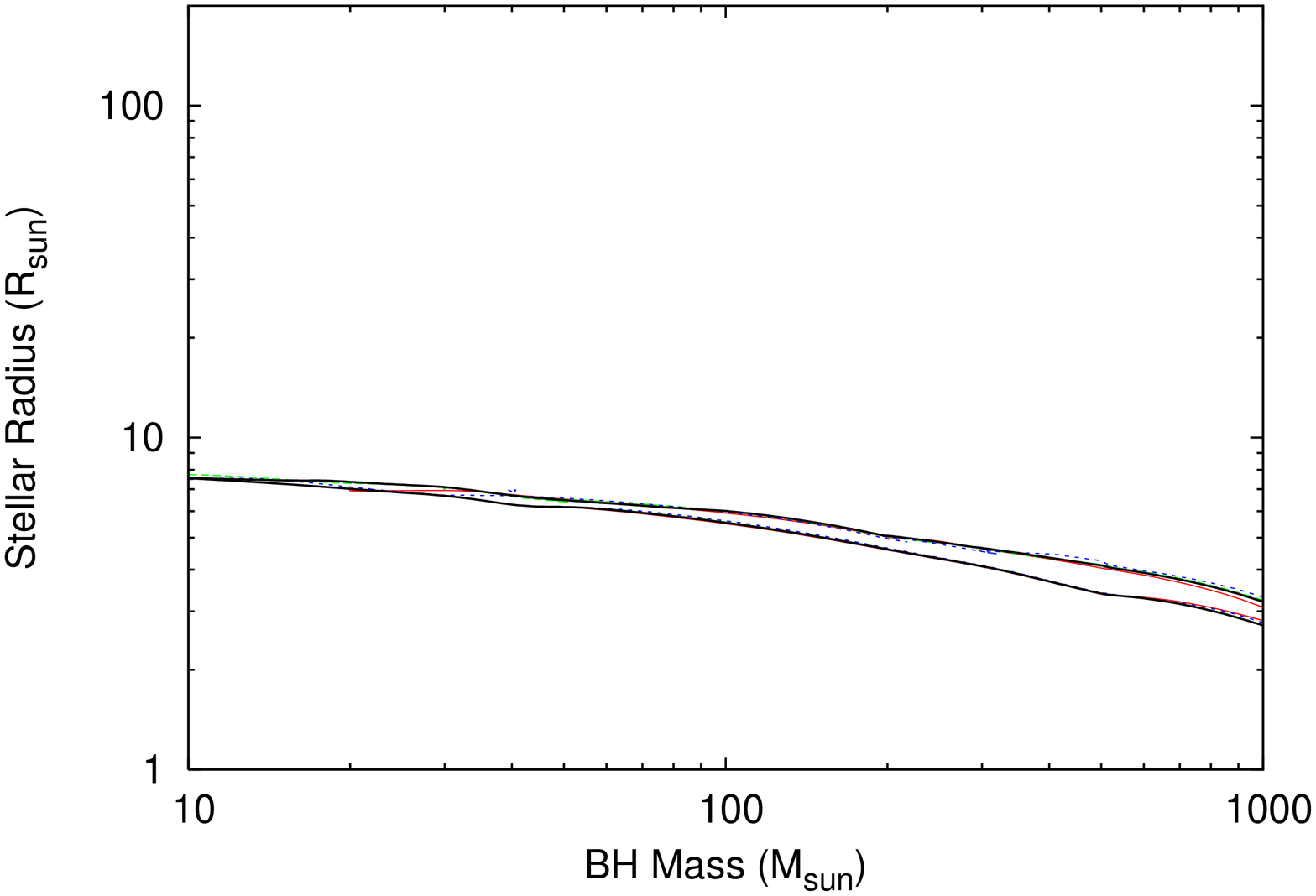}
\end{minipage}%
\begin{minipage}[c]{0.5\textwidth} \hfill
\includegraphics[angle=270,width=1.0\textwidth]{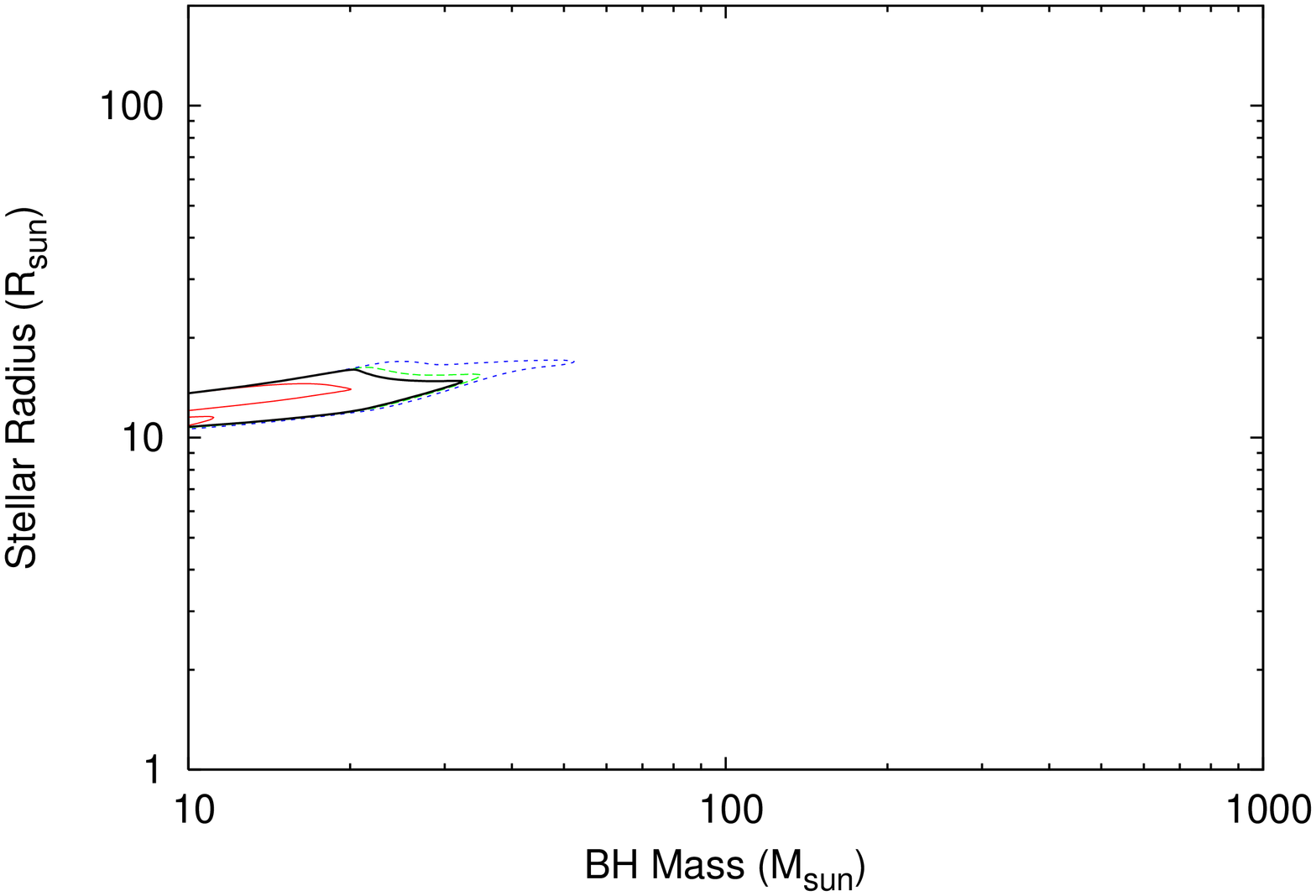}
\end{minipage}

\begin{minipage}[t]{0.49\textwidth}
\caption{Confidence contours for the binary parameters for the source X-6 in
M81. We assume a binary inclination of $\cos(i)=0.5$, superior conjunction with
respect to the observer and the star, and a stellar metallicity of
Z$=0.2$\Zsun. We use an X-ray hardness ratio of $\xi=0.01$. The red, black, green and
blue lines denote the $68$\%, $90$\%, $95$\% and $99$\% confidence intervals respectively.}
\label{fig:m81_0.5sc}
\end{minipage}%
\begin{minipage}[t]{0.02\linewidth} \hfill
\end{minipage}%
\begin{minipage}[t]{0.49\textwidth} \hfill
\caption{As for Figure \ref{fig:m81_0.5sc}, but with a binary inclination of
$\cos(i)=0.0$ and a hardness ratio of $\xi=0.1$.}
\label{fig:m81_0.5ic}
\end{minipage}

\end{figure*}

This source has an average X-ray luminosity of $2 \times 10^{39}$ \ergss
\citep{Roberts00}. \citet{Liu02} found an optical counterpart they considered
unique to this ULX (designated NGC 3031 X-11 in that paper), and reported $B$,
$V$ and $I$ magnitudes derived from HST ACS observations (Table
\ref{tab:photdat}).

We first examine the case where this source is at superior conjunction and
$\cos(i)=0.5$. We find that for this orientation our model is a poor fit to the
observation at the $90$\% confidence level: the irradiated disk/star are
together too luminous to match the observation for any combination of star and
BH. We therefore chose to adjust the hardness ratio of the irradiating X-ray
spectrum in this case in order to fit our model to the observation. When we
lower $\xi$ to $0.01$, we find a good fit to the data. We show the confidence
contours in Figure \ref{fig:m81_0.5sc}. We see that the stellar age ranges from
$10^6$ -- $10^{8.7}$yr, with a lower stellar age implying a higher BH mass. The
stellar mass ranges from $1.5$ -- $14$\Msun, and the stellar radius ranges from
$2.5$ -- $8$\Rsun.

When we assume the star to be in inferior conjunction, we again require a
reduced hardness ratio of $\xi = 0.01$ in order to obtain a good fit. We find
that most of the parameter space that we found fitted with our model for the
superior conjunction case is contained within that here. In addition, we find
that for low BH masses we can fit larger stars of radius $10$ -- $17$\Rsun \
with the observation. These parameters represent cases where the majority of
the optical emission is from the unirradiated hemisphere of the star. The
small, but not insignificant disc contribution, as well as the Roche lobe
shape of the star accounts for the difference between our fitted stellar
parameters and those of \citet{Liu02} in this inferior conjunction case. At
higher BH masses the emission is disk dominated and our results are not
dependent on the phase of the star.

We examine next the case when $\cos(i)=0.0$ (Figure \ref{fig:4559_1_0.0}). In
this orientation, we can fit our model with the observation for a hardness ratio
of $\xi = 0.1$. We note first of all that there is an upper bound on the BH mass
of $33$\Msun. The stellar age that fits with the observation ranges from
$10^{7.9}$ -- $10^{8.7}$yr, and the mass and radius range from $3$ -- $5.5$\Msun
\ and $10$ -- $15$\Rsun \ respectively. These values fit equally well, when we
set $\xi = 0.01$, since the stellar luminosity is much less sensitive to changes
in the X-ray hardness than the disk.

\citet{Liu02} found the field stars in the vicinity of this ULX range in age
from $1.0 \times 10^6$ -- $1.0 \times 10^8$yr. If we assume the donor in the ULX binary is of a
similar age, we see from Figure \ref{fig:4559_1_0.0} that the stellar
parameters are very tightly constrained in the $\cos(i)=0.0$ case. In the
$\cos(i)=0.5$ case (Figure \ref{fig:m81_0.5sc}), we see that there is a lower
limit on the BH mass of $20$\Msun \ if we constrain the stellar age in this
way.

\subsection{ULX in NGC 5204}

\hst \ WFPC2 and ACS observations of the optical counterpart to a ULX in NGC
5204 were described in \citet{Liu04}. This source has an X-ray luminosity of
$L_x \simeq 3 \times 10^{39}$\ergss. We took the photometric measurements from
\citet{Liu04} and corrected those magnitudes for interstellar absorption using
the $N_H$ column density given in \citet{Liu04}.

We find our model is a very poor fit to the observation when we orient the
system so as to include an irradiated disk and/or stellar component. This poor
fit is caused by our constraint on the mass accretion rate. When we remove this
constraint we find our model implies a companion star with a mass of $60$ to
$110$\Msun, a radius of $13$ -- $15$\Rsun \ and an age of $10^{6.3}$yr or less.
This is a young, massive star that is evolving rapidly in radius and is hence
transferring mass at a high rate, so when we set an upper limit on the mass
accretion rate as implied from the X-ray luminosity, this solution is precluded.
The best solution with this constraint applied is for a BH mass of $1000$\Msun
\ and a stellar age, mass and radius of $10^{5.3}$yr, $52$\Msun \ and $9$\Rsun
\ respectively. 

We note as an aside that, when we remove the constraint on the mass accretion
rate, there is an upper bound on the BH mass of $240$\Msun \ when we assume an
inclination of $\cos(i)=0.5$.

\subsection{M101 ULX-1} The source designated ULX-1 in M101 has a
peak X-ray luminosity of $\simeq 1.2\times 10^{39}$\ergss. \citet{Kuntz05}
reported a unique optical counterpart and reported B,V and I magnitudes
recorded with the \hst \ ACS instrument. From these data, they deduced the
companion star was a B type supergiant with a mass of $9$ to $12$\Msun. They
also observed no significant variation in the optical magnitudes over a sixty
day period.

When we use an inclination of $\cos(i)=0.5$ we find a stellar age of $10^8$ --
$10^{8.7}$yr, a mass of $2$ -- $7$\Msun \ and a radius of $6$ -- $30$\Rsun,
with lower radius values implying a more massive BH. 

As in the case of the ULX in NGC 5204, we find our model is a very poor fit to
the observation when we use an inclination of $\cos(i)=0.0$. Again, the poor
fit is caused by the upper bound on the mass accretion rate -- when we remove
this constraint we find a star of age $10^{7.0}$yr to $10^{7.3}$yr, mass $11$
-- $100$\Msun \ and radius  $12$ -- $33$\Rsun \ fits the observation. These are
much looser constraints on the stellar parameters than those reported by
\citet{Kuntz05}. By allowing emission from both an irradiated
star and disk component, the observation fits with a much wider range of binary
systems.

\subsection{ULX in NGC 5408}

\begin{figure*}
\centering

\begin{minipage}[c]{0.5\textwidth}
\includegraphics[angle=270,width=1.0\textwidth]{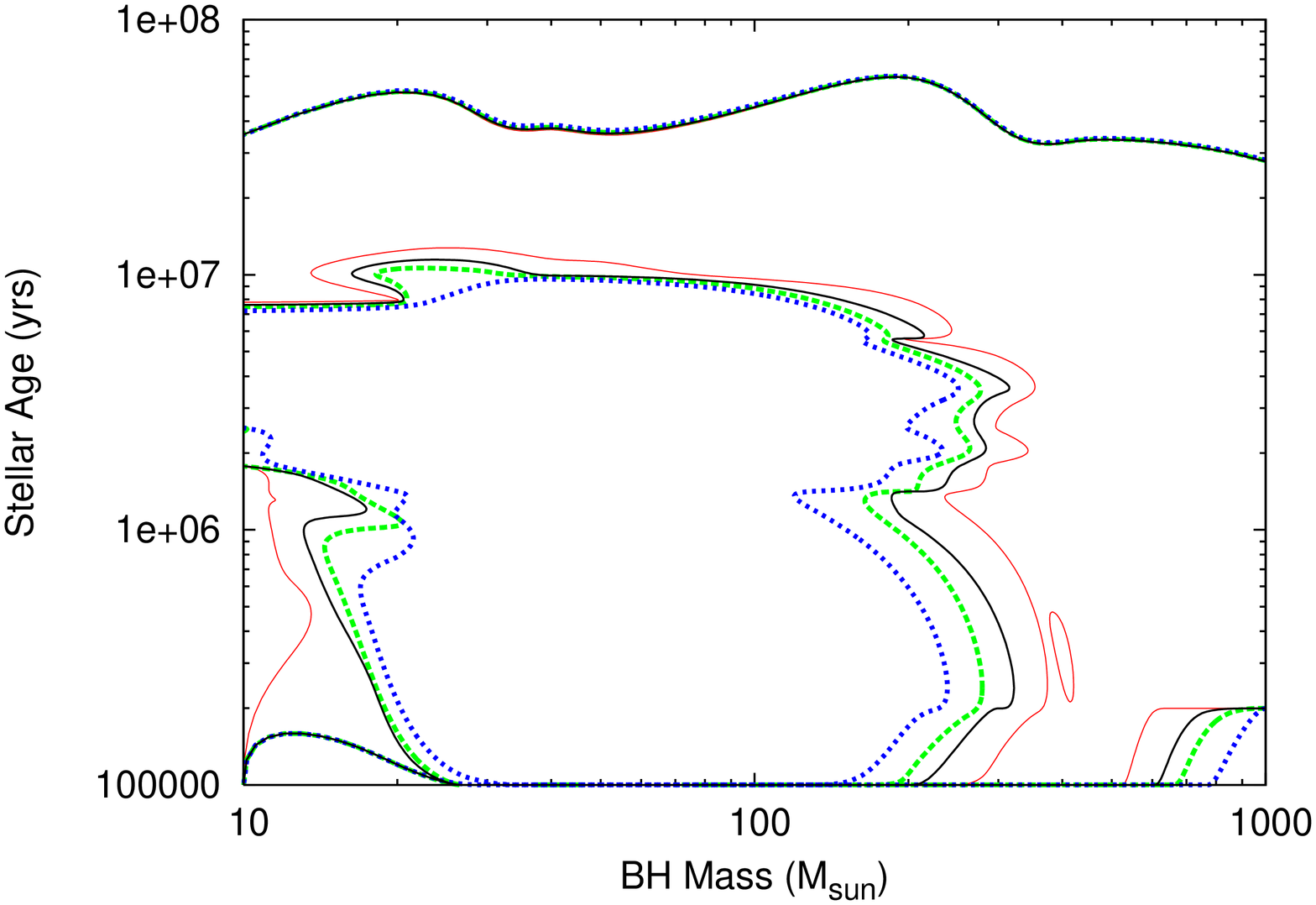}
\end{minipage}%
\begin{minipage}[c]{0.5\textwidth} \hfill
\includegraphics[angle=270,width=1.0\textwidth]{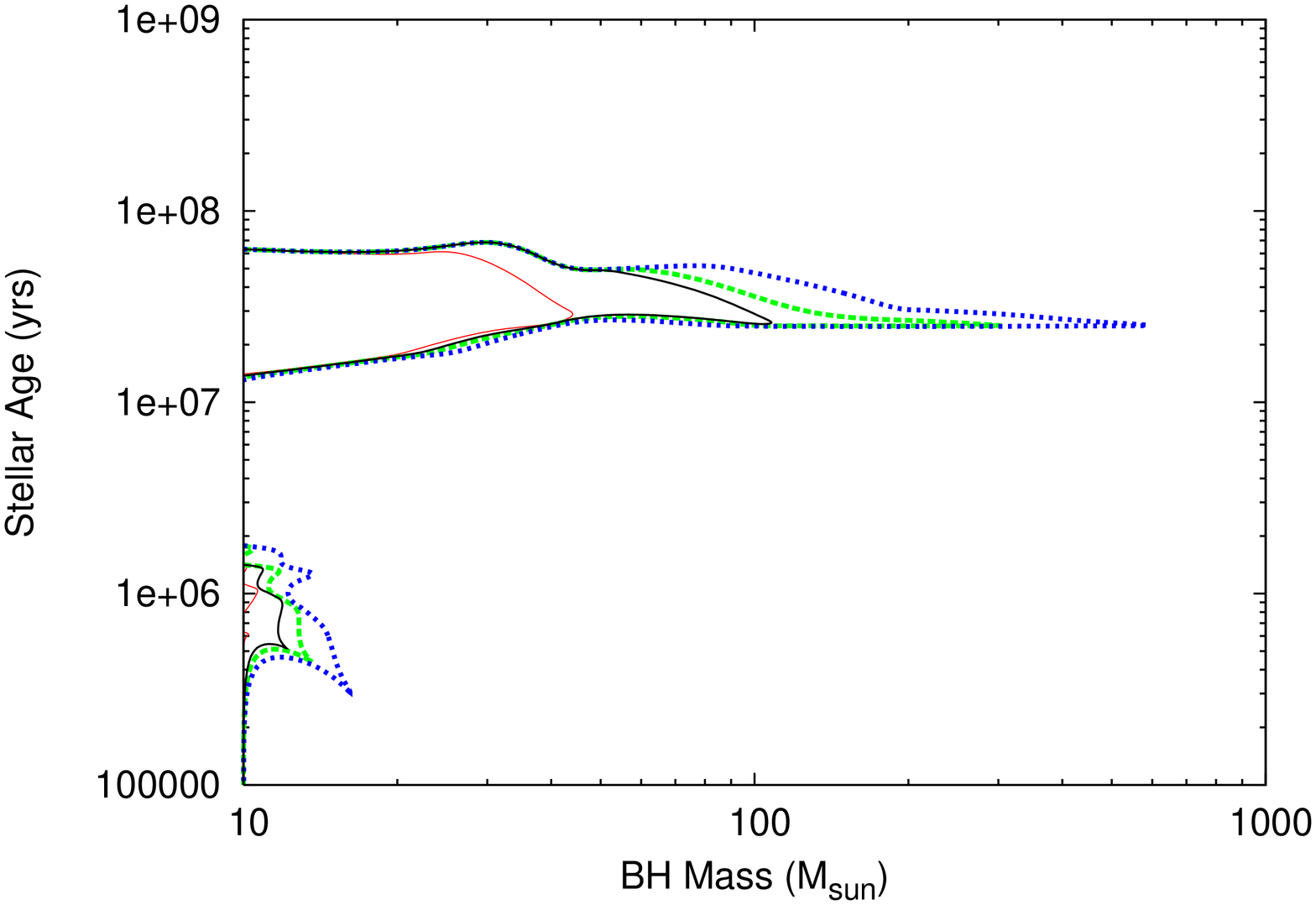}
\end{minipage} 

\begin{minipage}[c]{0.5\textwidth}
\includegraphics[angle=270,width=1.0\textwidth]{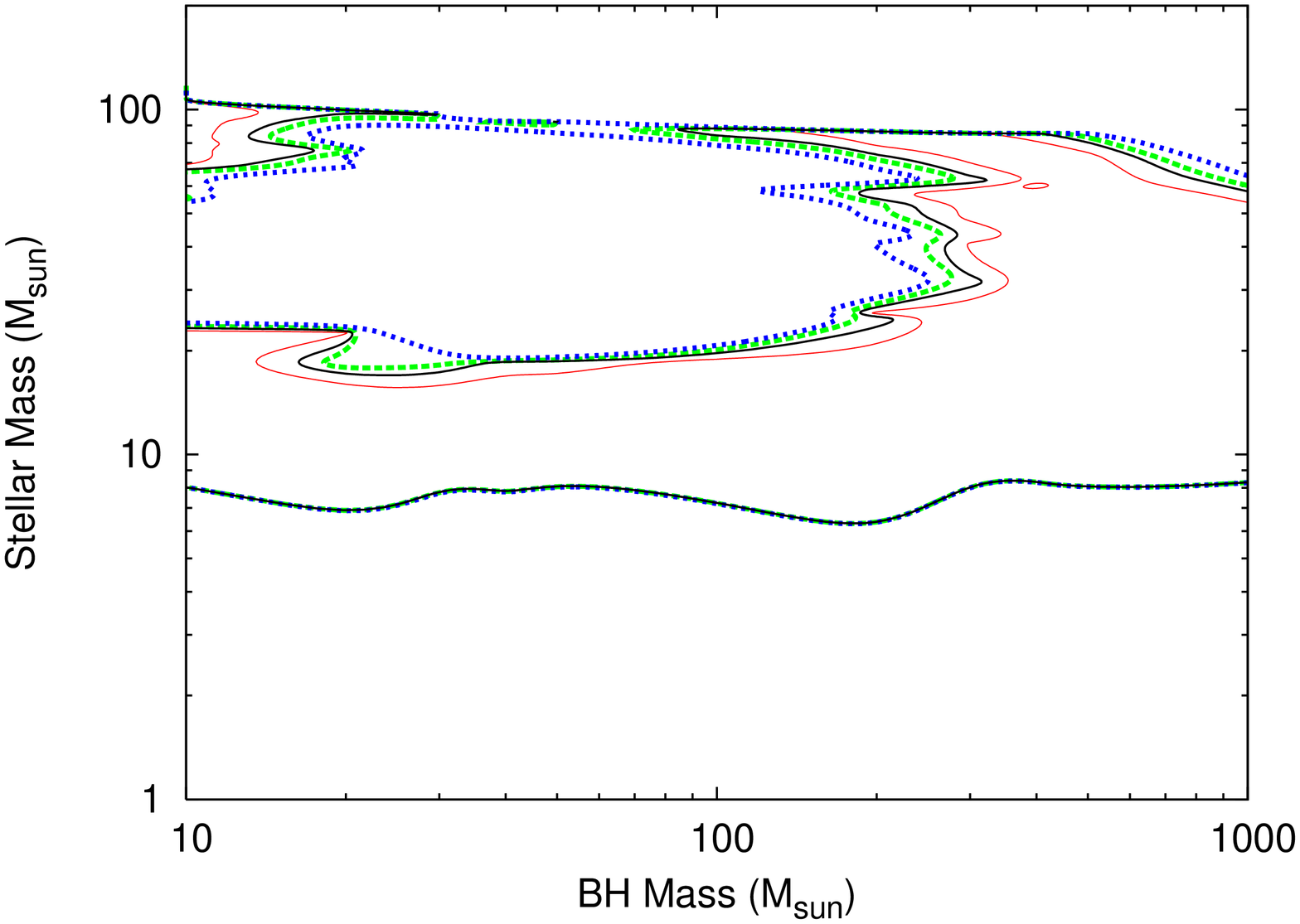}
\end{minipage}%
\begin{minipage}[c]{0.5\textwidth} \hfill
\includegraphics[angle=270,width=1.0\textwidth]{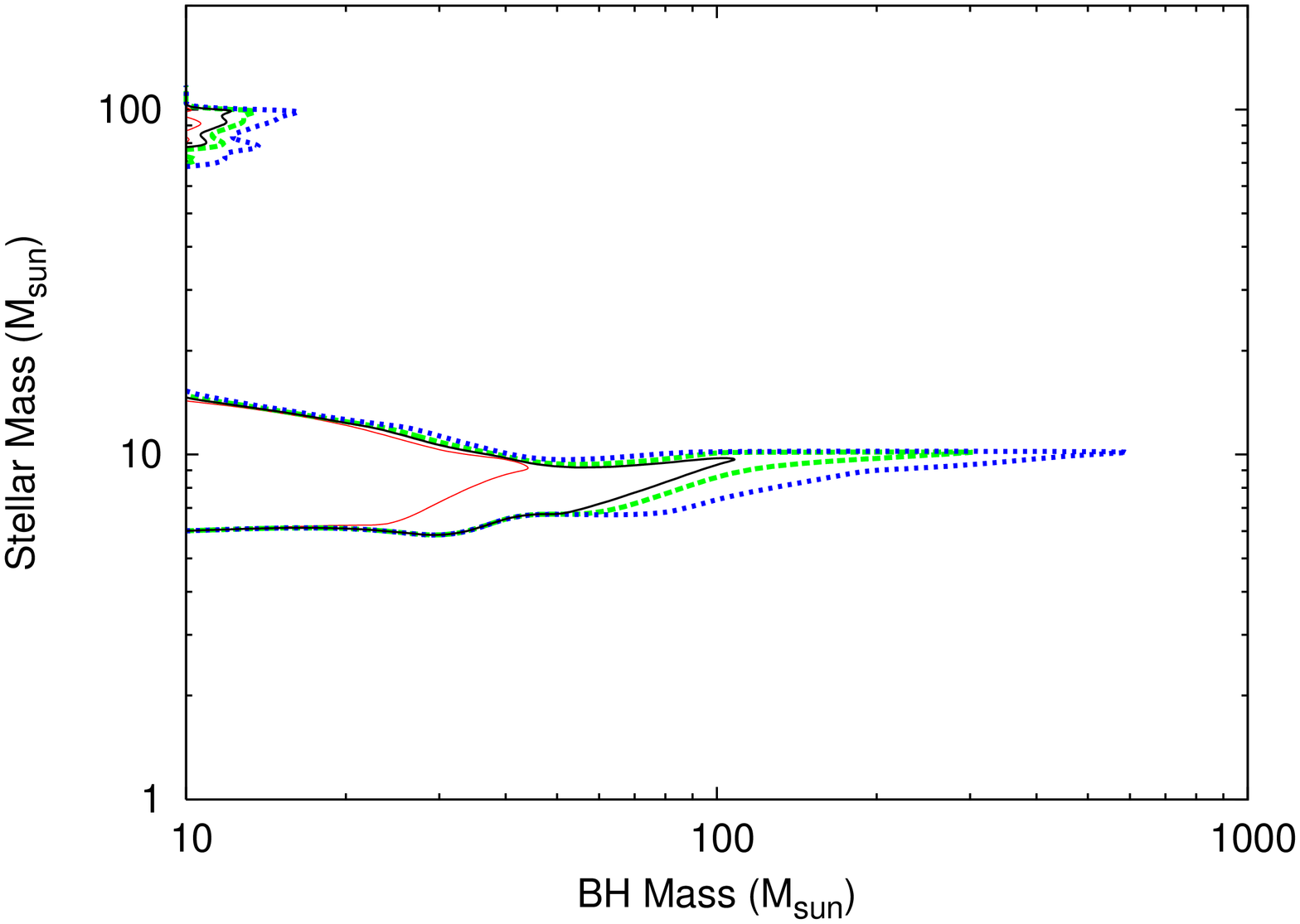}
\end{minipage} 

\begin{minipage}[c]{0.5\textwidth}
\includegraphics[angle=270,width=1.0\textwidth]{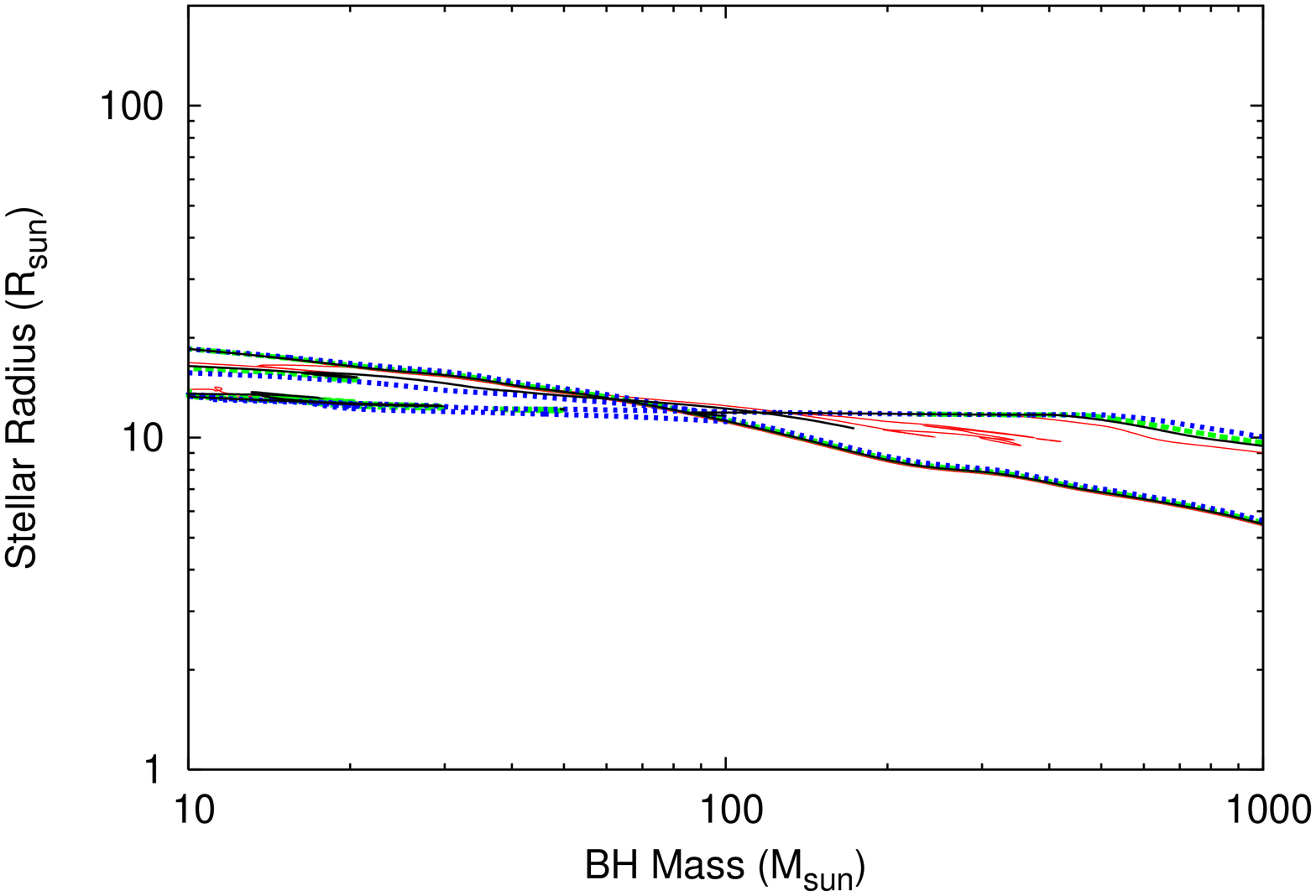}
\end{minipage}%
\begin{minipage}[c]{0.5\textwidth} \hfill
\includegraphics[angle=270,width=1.0\textwidth]{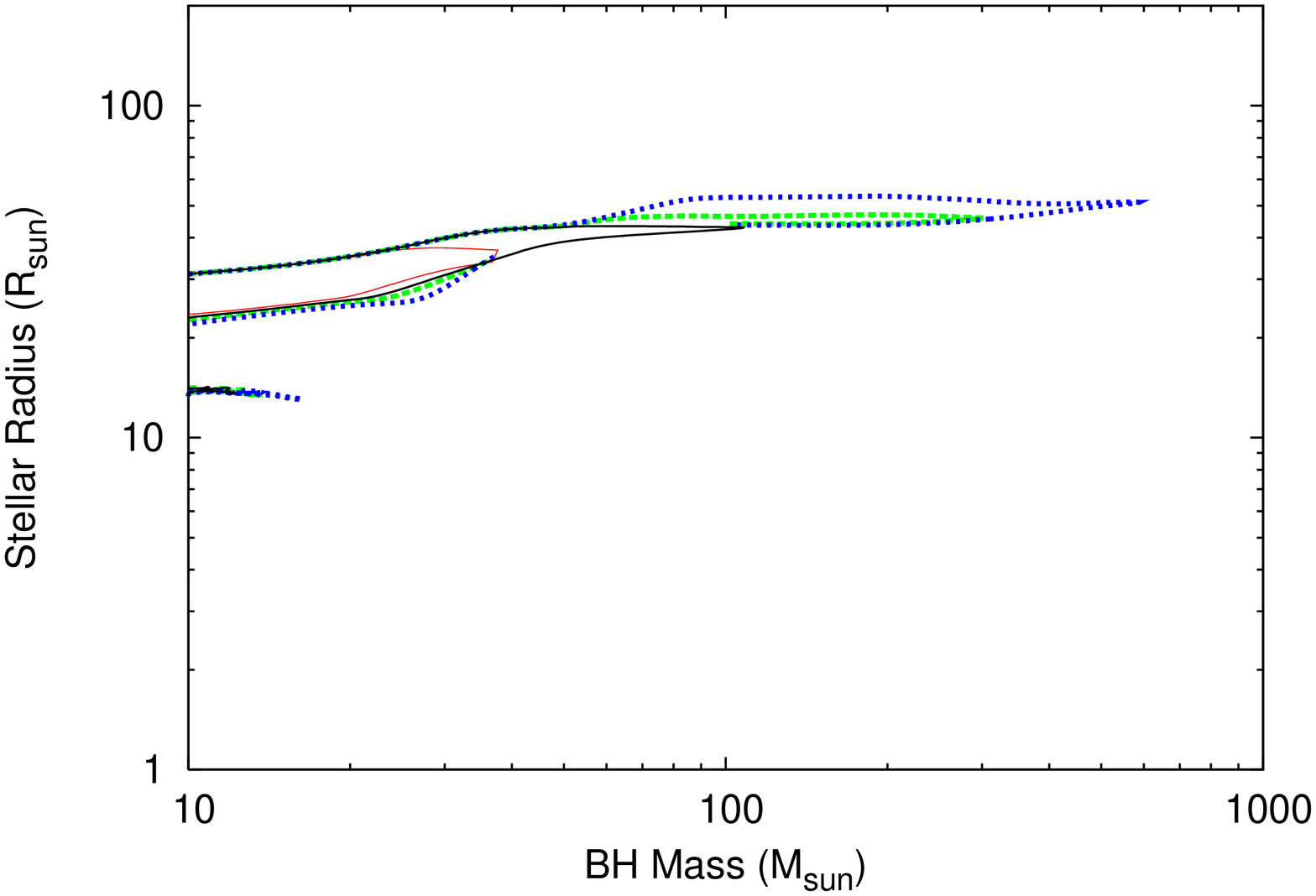}
\end{minipage}

\begin{minipage}[t]{0.49\textwidth}
\caption{Confidence contours for the binary parameters for the
ULX in NGC 5408.
We assume a binary inclination of $\cos(i)=0.5$, superior conjunction with
respect to the observer and the star, and a stellar metallicity of
Z$=0.2$\Zsun. We use an X-ray hardness ratio of
$\xi=0.1$. The red, black, green and blue lines denote the $68$\%, $90$\%,
$95$\% and $99$\% confidence intervals respectively.}
\label{fig:5408_0.5}
\end{minipage}%
\begin{minipage}[t]{0.02\linewidth} \hfill
\end{minipage}%
\begin{minipage}[t]{0.49\textwidth} \hfill
\caption{As for Figure \ref{fig:5408_0.5}, but with a binary inclination of
$\cos(i)=0.0$.}
\label{fig:5408_0.0}

\end{minipage}

\end{figure*}

NGC 5408 contains a ULX with an X-ray luminosity of $10^{40}$\ergss. This source
was initially thought to be consistent with a beamed microquasar
\citep{Kaaret03}. More recent studies have shown disk emission and QPOs, which
could be interpreted as evidence for an IMBH \citep{Soria04,Stroh07}. However, a
soft disk component may also be explained by alternative scenarios that do not
require an IMBH \citep{Stobbart06,Goncalves06}. Therefore, in this and other
sources characterized by an X-ray soft excess, optical constraints on the system
parameters can help break the degeneracy.

We use the archival HST/WFPC2 and Subaru observations to determine B, V and I
photometric magnitudes for the optical counterpart.There are in fact a number
of candidates for the optical counterpart within the \chandra \ error circle: we
assume here the counterpart is the source which appears most luminous in the V-band
\hst \ observation.

We examine the $\cos(i)=0.5$ case first (Figure \ref{fig:5408_0.5}). We see
that we can fit a donor star of mass $6$ -- $24$\Msun \ to the observational
data over the entire BH mass range. These stars have ages of order $10^7$yr and
radii of $23$ -- $44$\Rsun. We see also that when we assume a BH mass of
greater than $100$\Msun, we can fit more massive ($< 100$\Msun), younger and
more compact stars to the observation. A very massive ($> 67$\Msun) donor is
also possible when we use a BH mass of less than $30$\Msun.

In the $\cos(i)=0.0$ case, we see from Figure \ref{fig:5408_0.0}
that an upper bound of $110$\Msun \ exists. The donor star has an age of $10^7$
-- $10^{7.8}$yr, a mass of $6$ -- $15$\Msun \ and a radius of $23$ --
$43$\Rsun. A very massive ($>80$\Msun) donor is also possible when we use a low
BH mass.

\subsection{ULX X-2 in NGC 1313}

\begin{figure*}
\centering

\begin{minipage}[c]{0.5\textwidth}
\includegraphics[angle=270,width=1.0\textwidth]{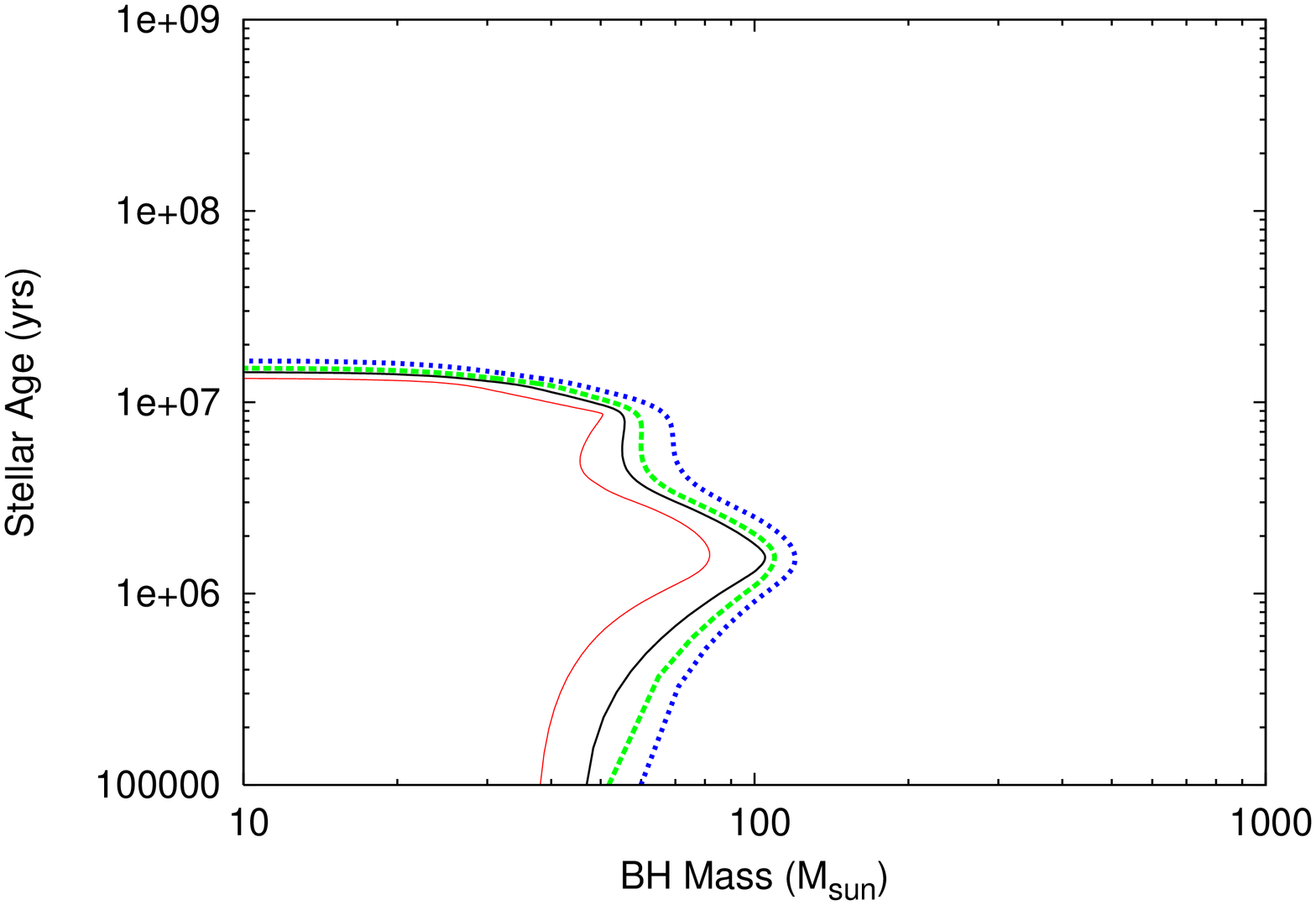}
\end{minipage}%
\begin{minipage}[c]{0.5\textwidth} \hfill
\includegraphics[angle=270,width=1.0\textwidth]{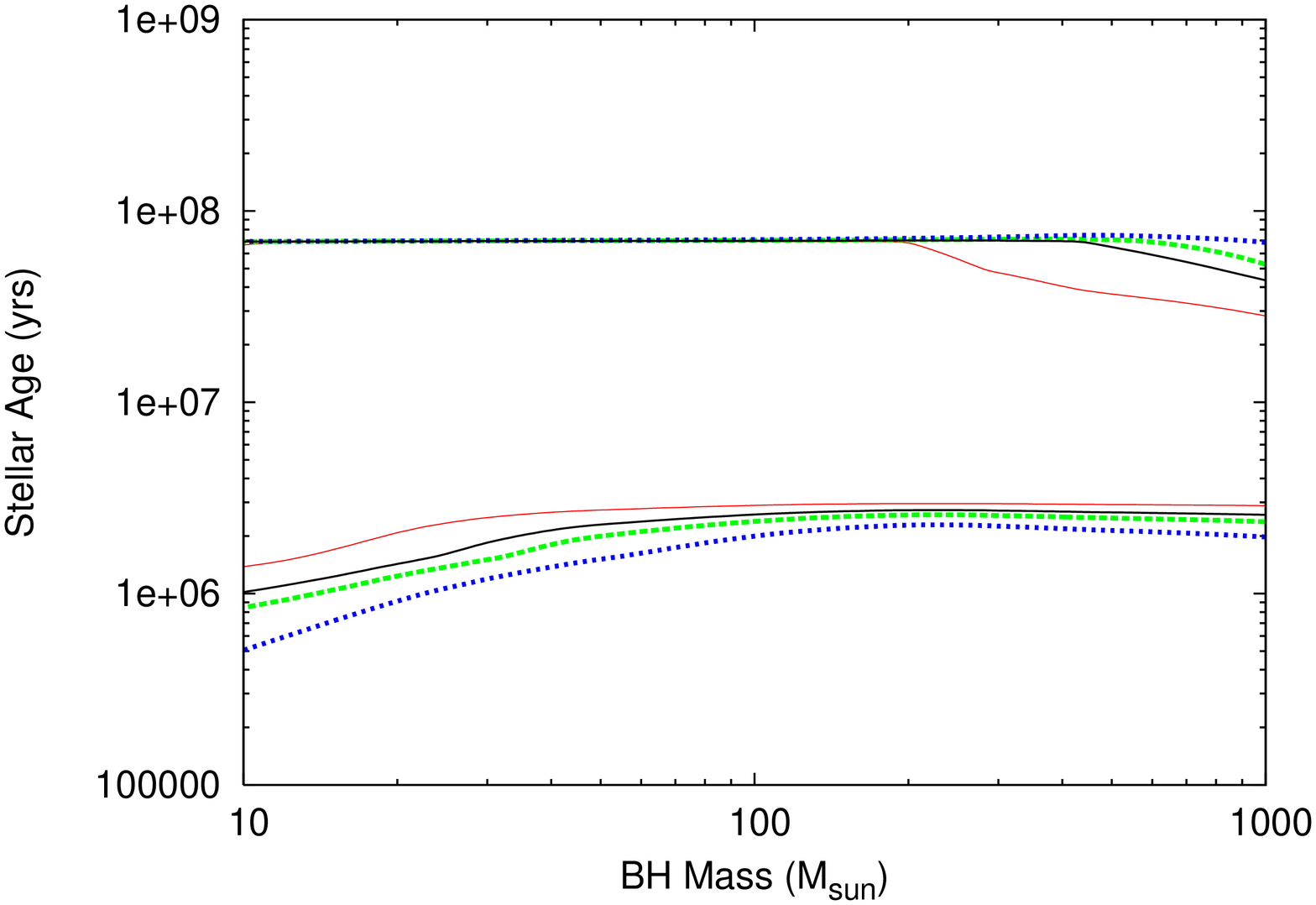}
\end{minipage} 

\begin{minipage}[c]{0.5\textwidth}
\includegraphics[angle=270,width=1.0\textwidth]{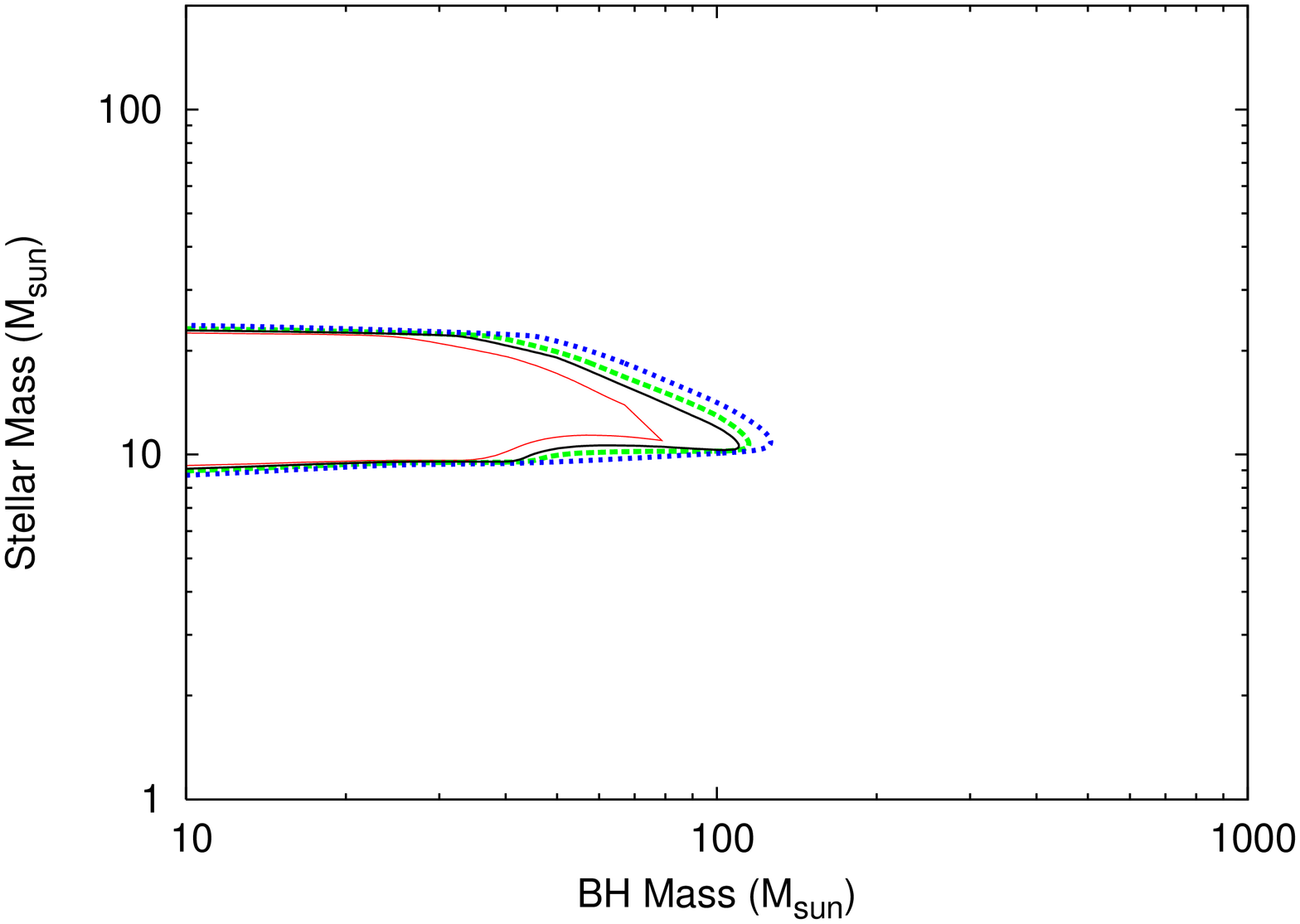}
\end{minipage}%
\begin{minipage}[c]{0.5\textwidth} \hfill
\includegraphics[angle=270,width=1.0\textwidth]{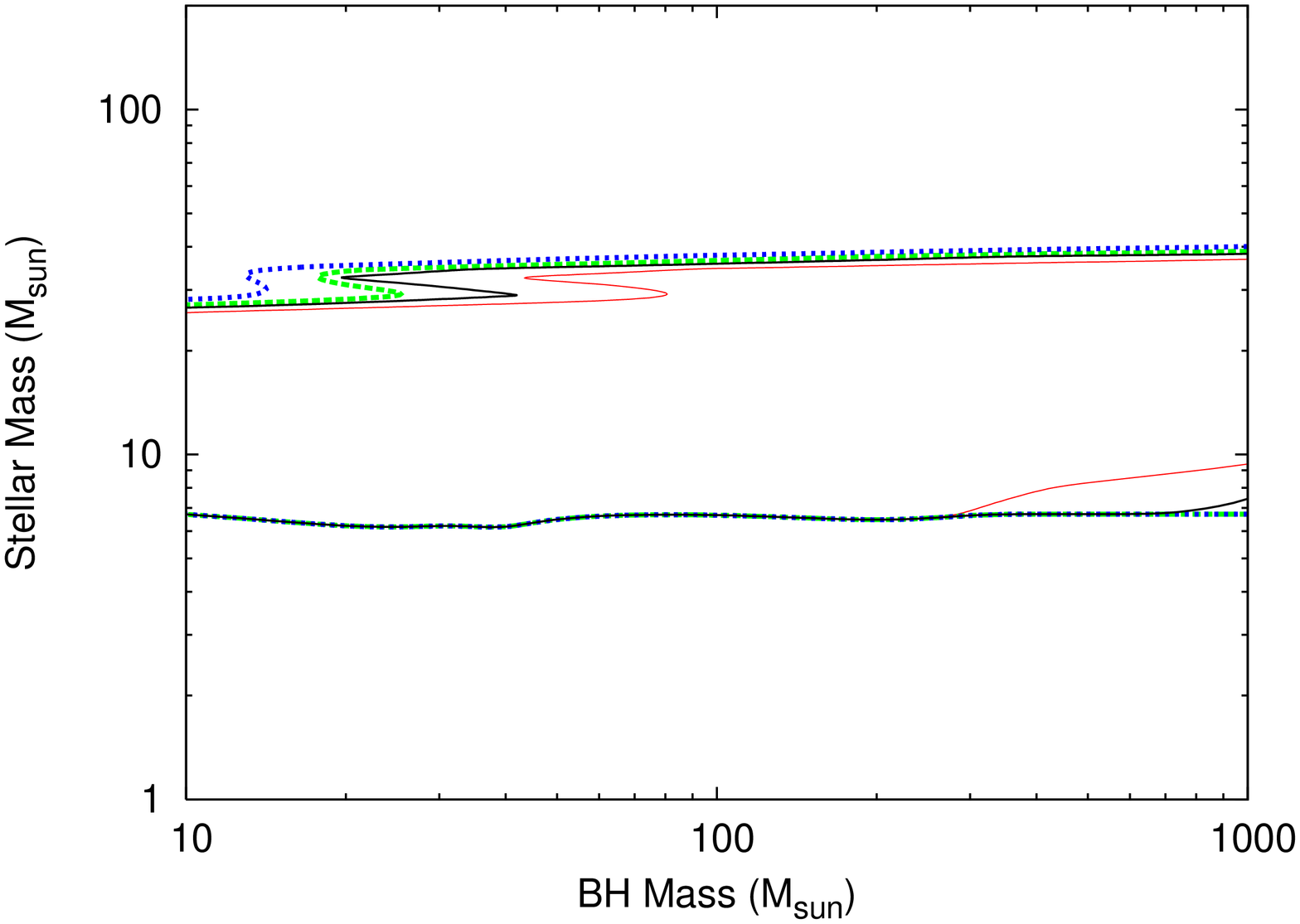}
\end{minipage} 

\begin{minipage}[c]{0.5\textwidth}
\includegraphics[angle=270,width=1.0\textwidth]{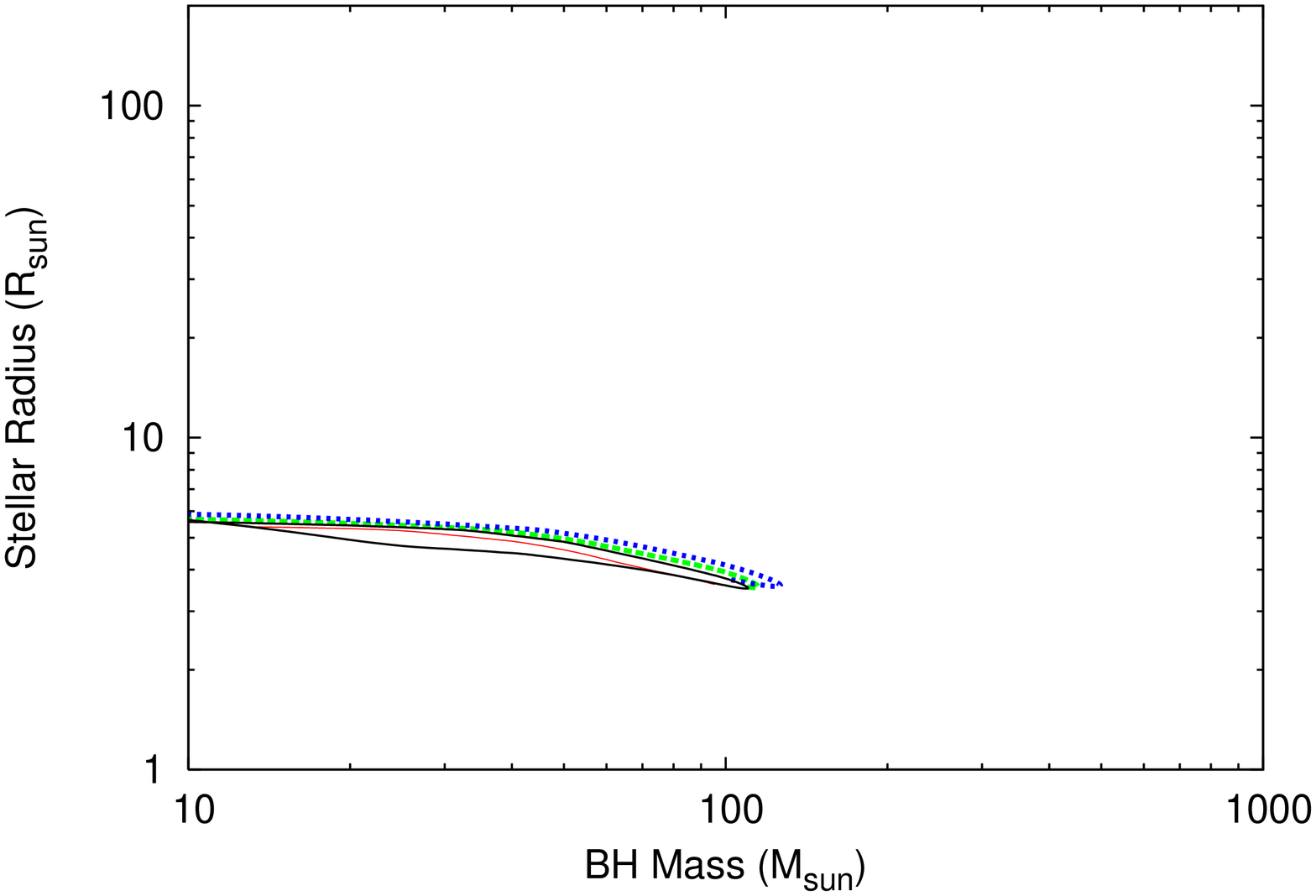}
\end{minipage}%
\begin{minipage}[c]{0.5\textwidth} \hfill
\includegraphics[angle=270,width=1.0\textwidth]{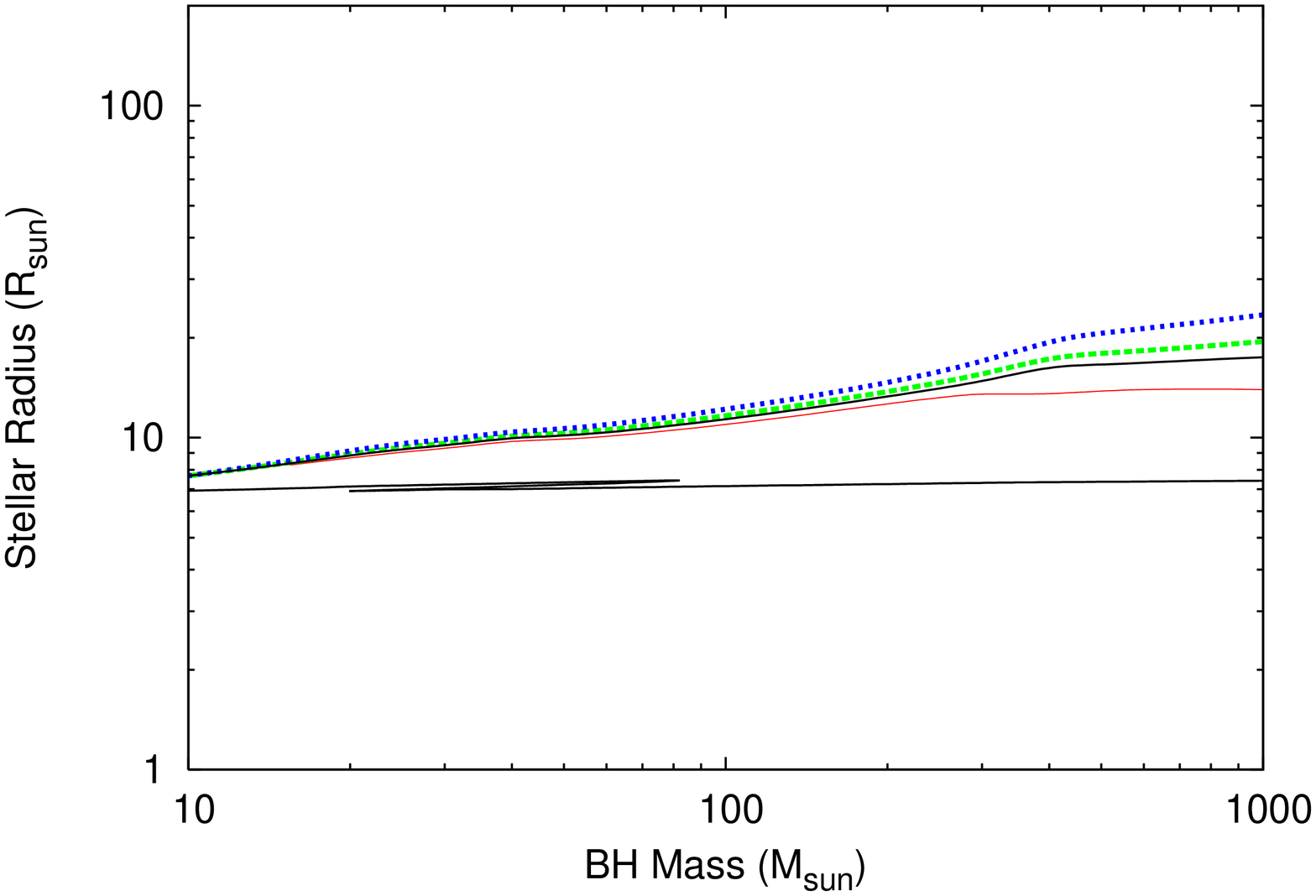}
\end{minipage}

\begin{minipage}[t]{0.49\textwidth}
\caption{Confidence contours for the binary parameters for ULX X-2 in NGC 1313, assuming candidate C1 is the optical counterpart.
We assume a binary inclination of $\cos(i)=0.5$, superior conjunction with
respect to the observer and the star, and a stellar metallicity of
Z$=0.2$\Zsun. We use an X-ray hardness ratio of
$\xi=0.1$. The red, black, green and blue lines denote the $68$\%, $90$\%,
$95$\% and $99$\% confidence intervals respectively.
}
\label{fig:1313_0.5}
\end{minipage}%
\begin{minipage}[t]{0.02\linewidth} \hfill
\end{minipage}%
\begin{minipage}[t]{0.49\textwidth} \hfill
\caption{As for Figure \ref{fig:1313_0.5}, but with a binary inclination of
$\cos(i)=0.0$.}
\label{fig:1313_0.0}
\end{minipage}

\end{figure*}

\citet{Mucciarelli05} analysed archive ESO VLT photometric data of this $L_x =
10^{40}$\ergss \ ULX. They found two possible candidates for the optical
counterpart of this ULX within the \chandra \ error circle. They gave B, V and
R magnitudes for the candidate designated C1 in that paper, and V and R
magnitudes for the candidate C2. This second candidate was not detected in the
B-band.

We applied our model to the photometric data of both candidates. We find the
non-detection of C2 in the B-band makes it inconsistent with being an
irradiated donor star and/or disc. We therefore take candidate C1 to be the
counterpart. This determination has been confirmed by a recent, more accurate
astrometric study with the \hst \ ACS instrument \citep{Liu07}. We see in
Figure \ref{fig:1313_0.5} that an upper bound of $100$\Msun \ exists for this
source when we set the inclination to be $\cos(i)=0.5$. If we take Figures
\ref{fig:1313_0.5} and \ref{fig:1313_0.0} together we see that the donor star
mass ranges from $6$ to $38$\Msun, with the upper bound decreasing to $23$\Msun
\ in the $\cos(i)=0.5$ case. For $\cos(i)=0.5$ the radius is $3$ -- $6$\Rsun.
Setting $\cos(i)=0.0$ results in the radius increasing to $7$ -- $18$\Rsun.

\subsection{ULX in Holmberg II}

\begin{figure*}
\centering

\begin{minipage}[c]{0.5\textwidth}
\includegraphics[angle=270,width=1.0\textwidth]{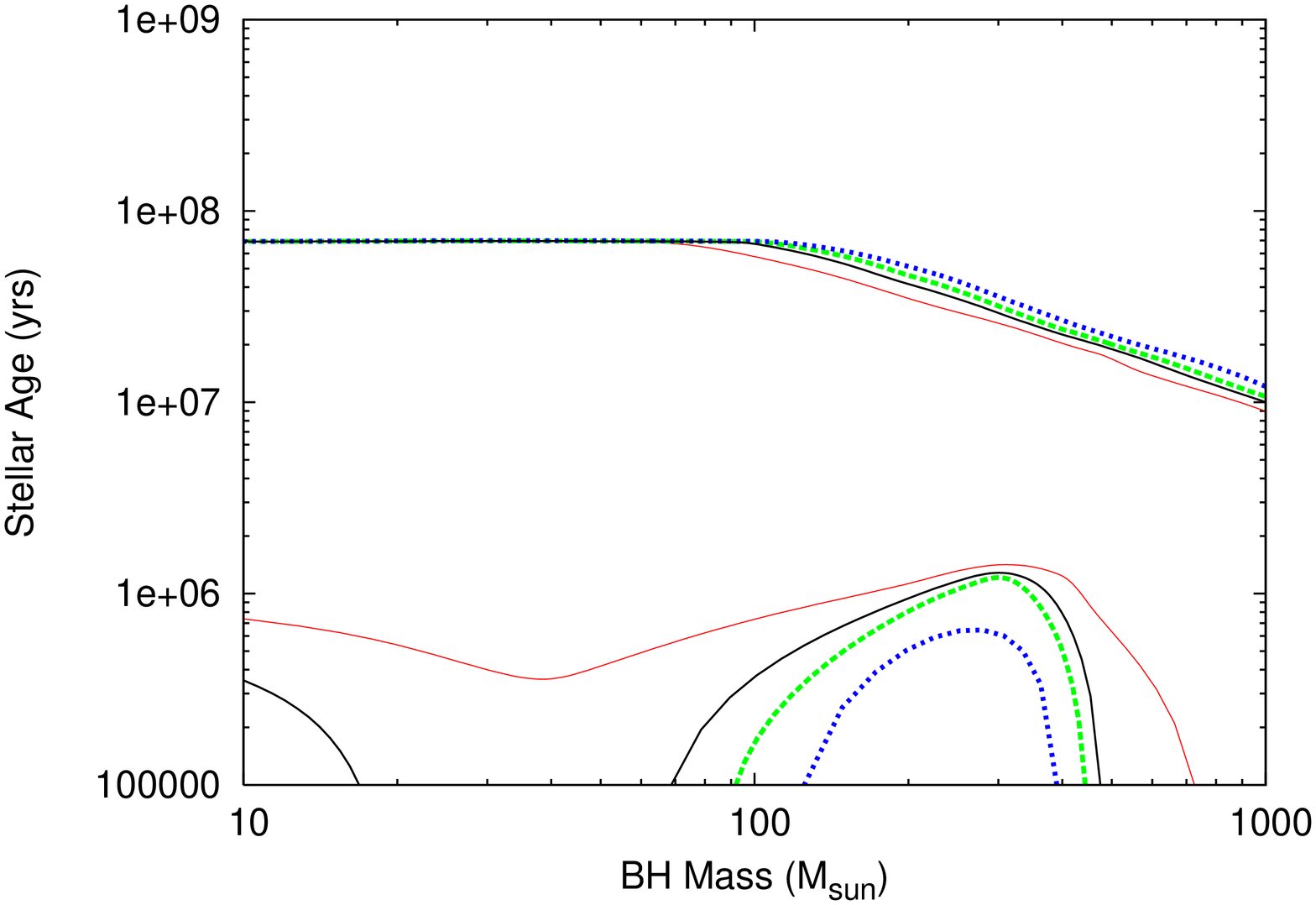}
\end{minipage}%
\begin{minipage}[c]{0.5\textwidth} \hfill
\includegraphics[angle=270,width=1.0\textwidth]{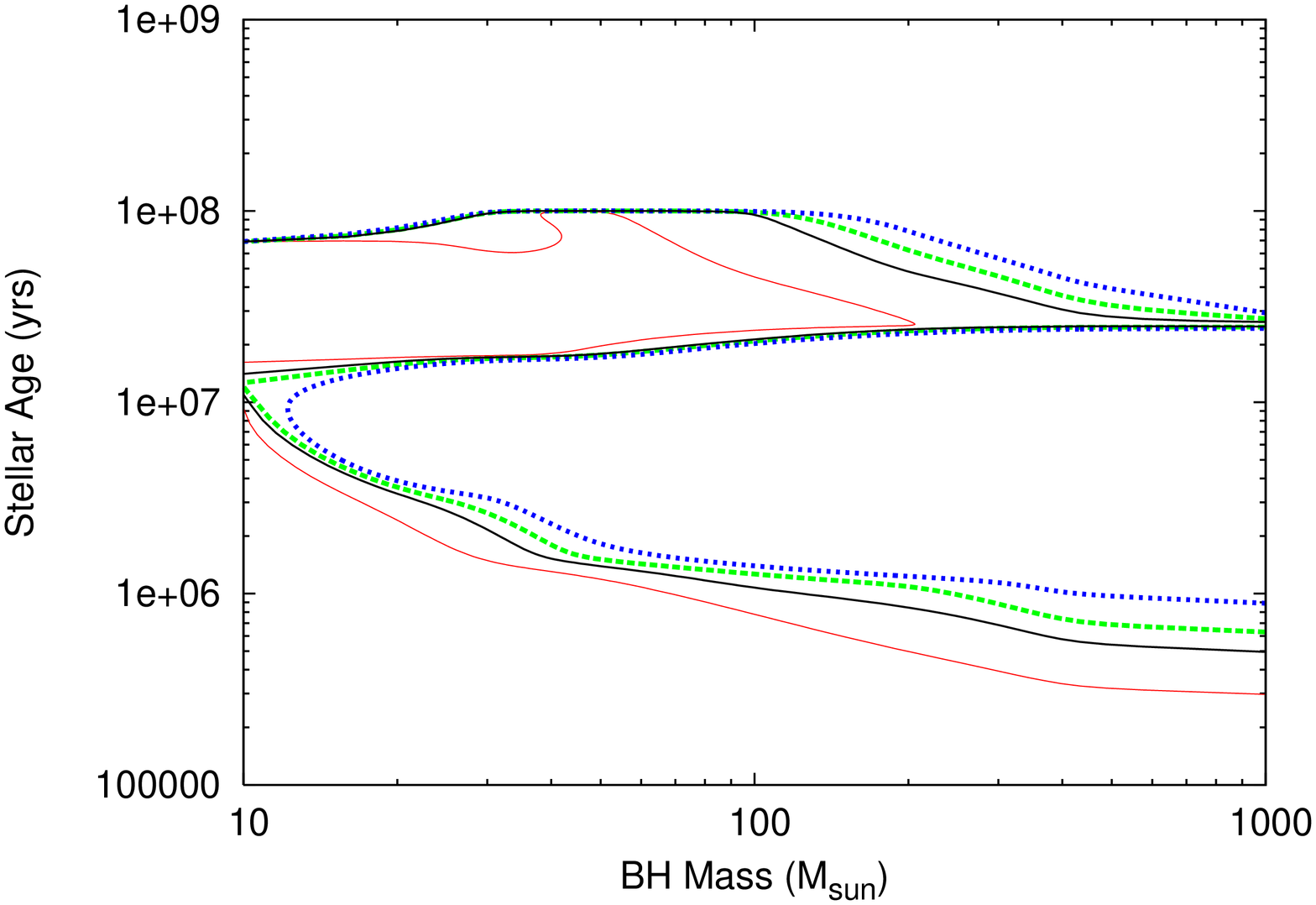}
\end{minipage} 

\begin{minipage}[c]{0.5\textwidth}
\includegraphics[angle=270,width=1.0\textwidth]{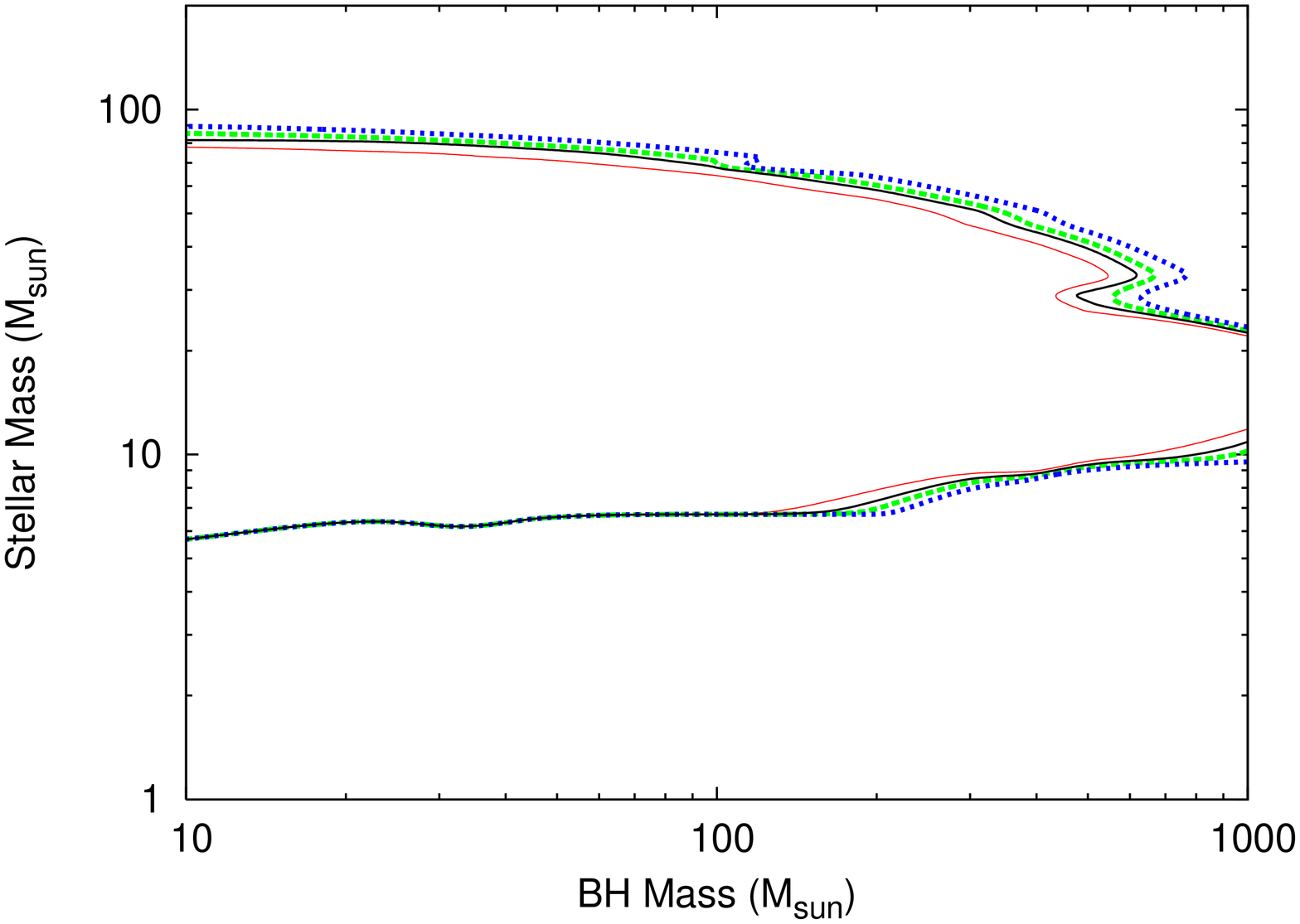}
\end{minipage}%
\begin{minipage}[c]{0.5\textwidth} \hfill
\includegraphics[angle=270,width=1.0\textwidth]{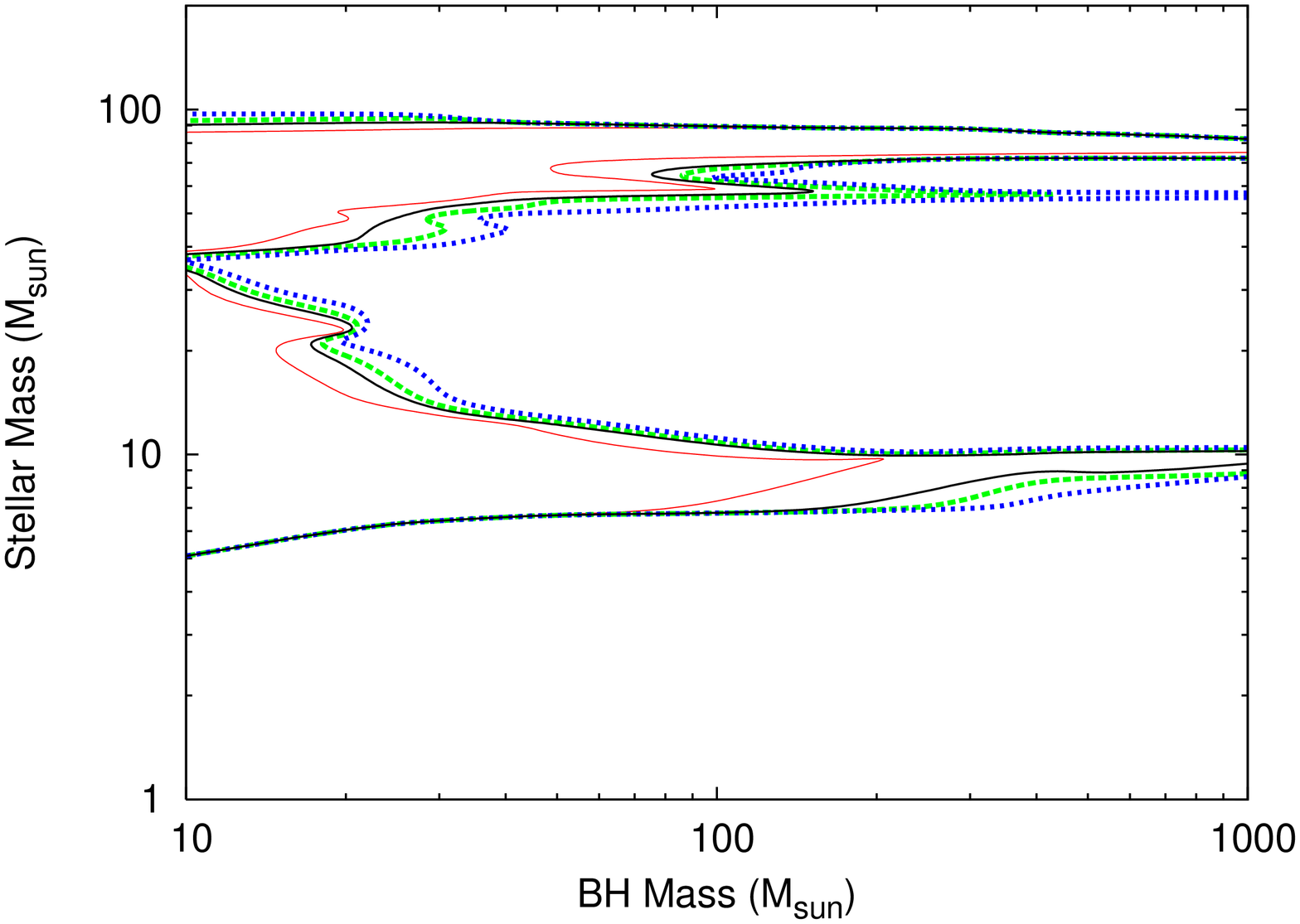}
\end{minipage} 

\begin{minipage}[c]{0.5\textwidth}
\includegraphics[angle=270,width=1.0\textwidth]{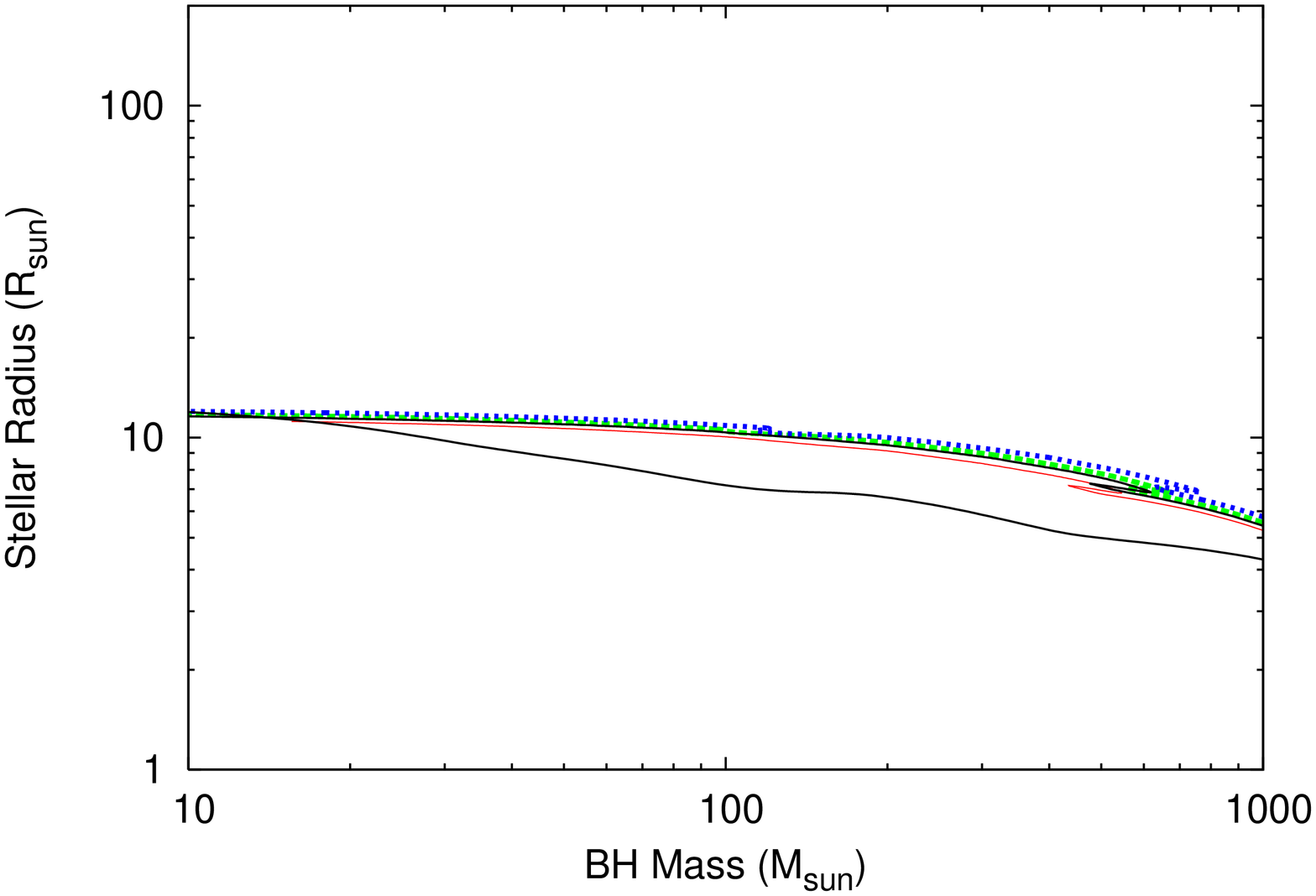}
\end{minipage}%
\begin{minipage}[c]{0.5\textwidth} \hfill
\includegraphics[angle=270,width=1.0\textwidth]{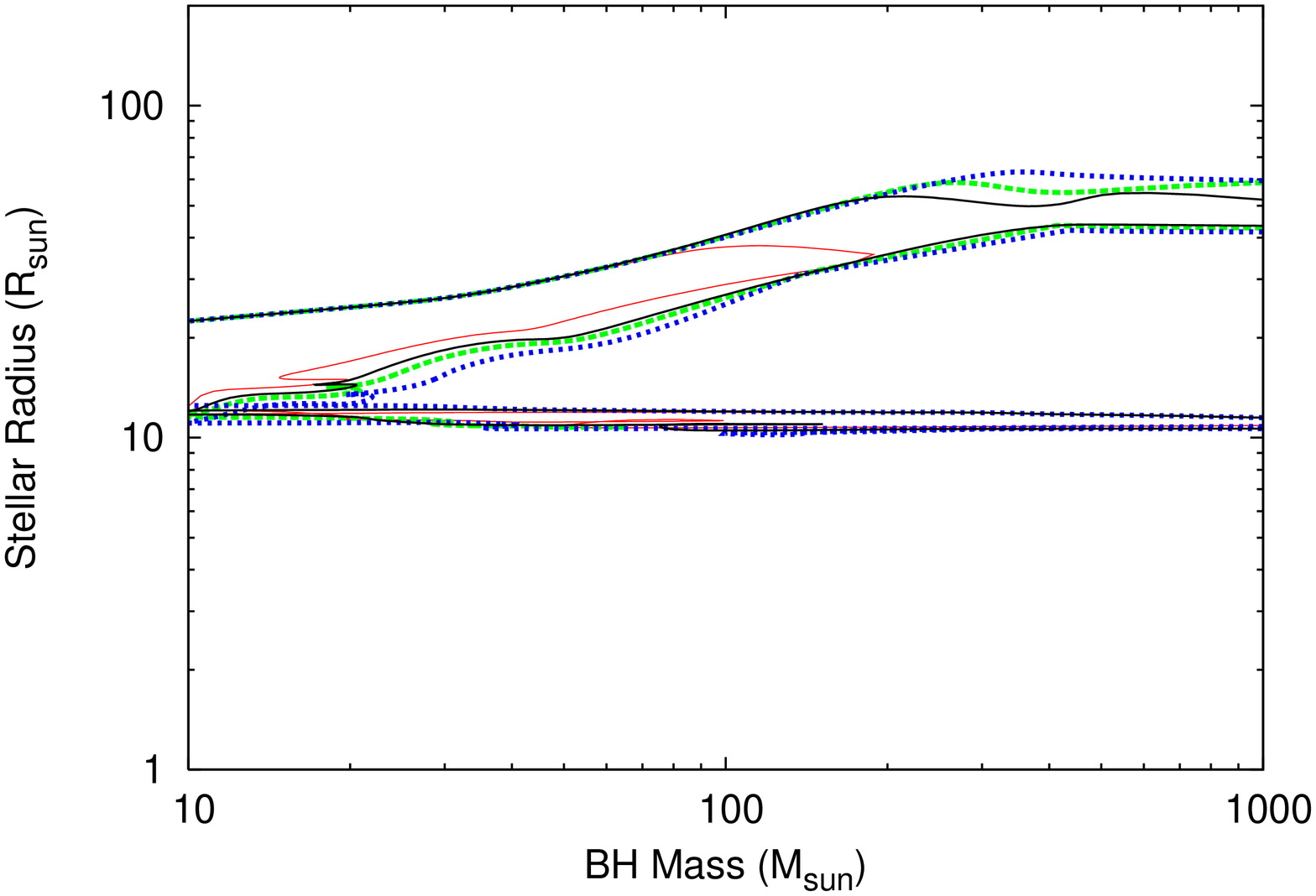}
\end{minipage}

\begin{minipage}[t]{0.49\textwidth}
\caption{Confidence contours for the binary parameters for the
ULX in Holmberg II.
We assume a binary inclination of $\cos(i)=0.5$, superior conjunction with
respect to the observer and the star, and a stellar metallicity of
Z$=0.2$\Zsun. We use an X-ray hardness ratio of
$\xi=0.1$. The red, black, green and blue lines denote the $68$\%, $90$\%,
$95$\% and $99$\% confidence intervals respectively.}
\label{fig:hoii_0.5}
\end{minipage}%
\begin{minipage}[t]{0.02\linewidth} \hfill
\end{minipage}%
\begin{minipage}[t]{0.49\textwidth} \hfill
\caption{As for Figure \ref{fig:hoii_0.5}, but with a binary inclination of
$\cos(i)=0.0$.}
\label{fig:hoii_0.0}
\end{minipage}

\end{figure*}

This ULX  has an X-ray luminosity measured at up to $10^{40}$\ergss \
\citep{Kaaret04}, although it is highly variable. It is associated with a
diffuse, photoionised nebula, whose energetics suggests that the X-ray emission
from the accreting source is truly luminous and not strongly beamed.
\citet{Kaaret04} gives both the V magnitude of the counterpart and its (B-V)
colour.

We find the available optical data does not allow us to constrain the BH mass
for any orientation. As regards the parameters of the donor star, we see in Figure
\ref{fig:hoii_0.5} that if we set $\cos(i)=0.5$ the stellar age and mass are
quite poorly constrained, with the mass ranging from $6$ to $82$\Msun \ and the
age ranging from $10^5$ to $10^{7.85}$yr. The stellar radius is better defined,
and lies between $4$ and $12$\Rsun.

For the $\cos(i)=0.0$ case (Figure \ref{fig:hoii_0.0}), the picture is more
complicated. Our results suggest two possibilities for the donor star
parameters. It can either be a star of mass $37$ -- $92$\Msun \ and radius $10$
-- $12$\Rsun, or a much older object with mass $5$ -- $34$\Msun \ and radius
$12$ -- $55$\Rsun. These two possibilites are more tightly constrained as the BH
mass increases.


\section{Discussion}
\label{sec:discussion}

\subsection{Classification of the donor stars}

In Table \ref{tab:zams} we list the Zero Age Main Sequence (ZAMS) stellar
masses and temperatures we calculate for the donor stars in the seven systems
we have investigated. These values do not account for any loss due to to the
mass accretion, and so should be increased by an amount depending on when the
mass transfer began. In Table \ref{tab:spectype} we list the most likely
currently observed spectral type we obtained for each source . We find the
donor stars are in general consistent with main sequence or evolved
giant/supergiant stars of type B, except in the NGC 5204 ULX. Not suprisingly,
we find the donor stars to be larger, less massive and older than inferred when
irradiation is not taken into account. We can also rule out donors of type A or
later in most cases. We now comment on individual sources, giving the currently
observed spectral type in each case. 

For ULX X-7 in NGC 4559, our determination of the mass and radius suggest the
star is a supergiant, of spectral type B (if we assume an inclination such that
there is a disk component to the emission) or spectral type A (if we assume the
plane of the disk is perpendicular to the sky -- $\cos(i)=0.0$). The donor has
an age and mass consistent with the stars in its immediate neighbourhood
\citep{Soria05}.

\citet{Liu02} found the photometric observations of the counterpart to M81 X-6
fitted with an MS O-star. They noted that although the photometric data can be
fit by considering an O9 MS star the colour of the data is redder than would be
expected. They corrected for this by assuming intrinsic extinction by the dusty
environment of the ULX, which changed their determination of the spectral type
to an O8 star. We argue instead that the red excess could be explained in terms
of a disc component adding to the optical emission (as in LMC X-3,
\citealt{vParadijs87}). The stellar parameters we calculate instead identify
the donor as an MS or giant evolved B-star.

For the ULX in NGC 5408, our best fit model is a giant, evolved B star, but we
cannot rule out the possibility of  very massive ($> 45$\Msun) O-type donor
star. When we assume the inclination is such that the plane of the disk is
perpendicular to the plane of the sky, the radius implied by our model
increases, and matches that of a B-type supergiant.

For ULX X-2 in NGC 1313, our calculations suggest the candidate designated C1
in \citet{Mucciarelli05} is most likely to be the optical counterpart. The low
stellar radius we find from the model fit suggest the donor is a main sequence
star. The most likely spectral type is B, although a late O-type is also a
possibility, especially when we set $\cos(i)=0.0$. \citet{Mucciarelli05}
suggest that the donor is an MS star of type O9 -- B0, which is consistent with
our findings.

For the Holmberg II ULX, when we set $\cos(i)=0.5$ we find a tight constraint on
the stellar radius but a large mass range. These results are consistent with
the donor being either an MS O-star or a giant B-star. We see two distinct
possibilities in the $\cos(i)=0.0$ case as well, but here the solutions
corresponding to a lower stellar mass give a higher stellar radius, so the
donor can be classifed as either an MS O-star or a B supergiant.
\citet{Kaaret04} suggested the donor was of type O4V or B3Ib, which is
consistent with our findings. 

For the ULX in NGC 5204, \citet{Liu04} reported the multi-band photometry to be
consistent with stars of type O5 V, O7 III or B0 Ib. They also reported
\hst/STIS far-ultraviolet spectral observations, and on the basis of those they
suggest the star is most likely to be of type B0 Ib, although they note the
spectrum does contain some peculiarities for a star of this type. The STIS data
are also consistent with the presence of an X-ray illuminated accretion disc.
Our model does not provide good fits, because we have applied a  constraint in
which the mass accretion rate is fixed. When we relax this constraint we find
the observation is consistent with a very massive O-type MS star, and a mass
accretion rate of an order of magnitude or more greater than we expect from the
observed X-ray luminosity.

The conclusion that the donor is an MS O-star disagrees with \citet{Liu04},
(and this would be the only source where we have found the donor to be a more
massive and compact star than originally thought). \citet{Liu04} found the
photometric observations to be consistent with an MS or giant O-type donor
star, but concluded that the star is a B-type supergiant is based on the
additional \hst \ STIS MAMA/FUV data. Specifically, the equivalent width of the
Si$_{III}$ $\lambda1299$ line is suggestive of a star cooler than $25000$K. An
MS O-star would be hotter than this, even on its unirradiated hemisphere. If we
assume our interpretation of the optical data is correct, we must therefore
suggest that this line originates somewhere other than on the surface of the
star, or that the surface layers are modified by the Roche lobe and material
loss by accretion. 

For the ULX in M101, we find the donor to be a B-type star if the optical
emission has an observable disk and a star component. If the disk is
perpendicular to the plane of the sky, we have the same mass accretion
constrain problem as for the NGC 5204 ULX. If we relax this constraint, the
observation is consistent with the donor being a late O or B supergiant, or
possibly an MS O star.

\subsection{Comments on mass accretion rate constraints} In the NGC 5204 ULX we
have found that the mass accretion rate as inferred from the nuclear evolution
timescale (via the evolutionary tracks) is larger than we would infer from the
X-ray luminosity. We can explain this in one of three ways. Firstly, we can
assume that our interpretation of the optical data is correct and the donor is
an MS O-star. This implies that the accretion efficiency $\eta \sim 0.01$, an
order of magnitude less than for standard disk accretion. Mass outflows or
advective inflows are well-known possible reasons for sub-standard accretion
efficiency.

The second possibility is that the BH is mostly fed by stellar winds rather than
Roche lobe overflow. This would invalidate both the geometry used in our model
and our calculation of the mass accretion rate, since both of them are dependent
on Roche lobe overflow as the accretion mechanism. Given that this source is a
lower-luminosity ULX ($3 \times 10^{39}$\ergss) when compared to the others in
our sample, a supergiant donor feeding the ULX via a wind is a reasonable
possibility. However, this is inconsistent with the findings of \citet{Liu04},
as they report the FUV spectrum shows evidence of Roche lobe overflow.

A third possibility is that these objects are confused at some level in current observations
at optical wavelengths, with more than one star contributing to an unresolved
counterpart.

For the ULX in M101, we have the same problem as for the NGC 5204 ULX if the
inclination of the system is set to $\cos(i)=0.0$. We can explain this by
ruling out an inclination of $\cos(i)=0.0$ in this case. Alternatively, this
can be exlained by either of the three possibilities detailed above. Given that
this source has the weakest X-ray luminosity ($1 \times 10^{39}$\ergss) of
those in our sample, the possibility that this system is wind fed is relatively
more likely.

\begin{table*}

\caption{Calculated stellar ZAMS parameters for the donor stars. These values
do not account for any loss due to to the mass accretion, and so should be
increased by an amount depending on when the mass transfer began (* applies only when we do not
apply a constraint on the mass accretion rate. ** applies only for a hardness
ratio of $\xi=0.01$.)}

\label{tab:zams}
\begin{tabular}{l|l|l|l|l|l|l}
& \multicolumn{2}{|c|}{BH Mass = 10\Msun} & \multicolumn{2}{|c|}{BH Mass = 100\Msun} &
\multicolumn{2}{|c|}{BH Mass = 1000\Msun} \\
    	    	    	    	    	& Mass (\Msun)  	& Log Temp (K)      & Mass (\Msun)  	& Log Temp (K)      & Mass (\Msun)  	& Log Temp (K)\\
\hline
NGC 4559 X-7&&&&&&\\
$\cos(i)=0.5$, superior conjunction 	&$5.08$ -- $12.19$	&$4.30$ -- $4.48$   &$5.08$ -- $9.80$	&$4.30$ -- $4.44$   &$5.67$ -- $20.73$	&$4.32$ -- $4.57$\\
$\cos(i)=0.5$, inferior conjunction 	&$5.68$ -- $12.68$	&$4.32$ -- $4.49$   &$5.68$ -- $9.11$	&$4.32$ -- $4.42$   &$5.67$ -- $20.73$	&$4.32$ -- $4.57$\\
$\cos(i)=0.0$, superior conjunction 	&$5.97$ -- $12.77$	&$4.33$ -- $4.49$   &--		    	&--		    &--		    	&--\\\\
M81 X-6&&&&&&\\
$\cos(i)=0.5$, superior conjunction **	&--	    	    	&--		    &$3.38$ -- $7.10$	&$4.20$ -- $4.37$   &$6.36$ -- $9.10$	&$4.35$ -- $4.42$\\
$\cos(i)=0.5$, inferior conjunction **	&--	    	    	&--		    &$2.56$ -- $8.03$	&$4.12$ -- $4.39$   &$6.56$ -- $9.10$	&$4.35$ -- $4.42$\\
$\cos(i)=0.0$, superior conjunction 	&$2.48$ -- $5.37$   	&$4.11$ -- $4.31$   &--		    	&--		    &--		    	&--\\\\
NGC 5204 ULX&&&&&&\\
$\cos(i)=0.5$, superior conjunction *	&$68.4$ -- $109.3$  	&$4.71$ -- $4.74$   &$72.0$ -- $95.3$	&$4.71$ -- $4.73$   &--		    	&--\\
$\cos(i)=0.5$, inferior conjunction *	&$69.4$ -- $110.3$  	&$4.71$ -- $4.74$   &$72.0$ -- $95.3$	&$4.71$ -- $4.73$   &--		    	&--\\
$\cos(i)=0.0$, superior conjunction *	&$65.1$ -- $116.3$  	&$4.70$ -- $4.75$   &$69.6$ -- $117.3$	&$4.71$ -- $4.75$   &$73.4$ -- $118.3$	&$4.71$ -- $4.75$\\\\
M101 ULX-1&&&&&&\\
$\cos(i)=0.5$, superior conjunction 	&$2.46$ -- $4.13$	&$4.11$ -- $4.24$   &$2.48$ -- $4.81$	&$4.11$ -- $4.28$   &$2.93$ -- $6.35$	&$4.16$ -- $4.35$\\
    	    	    	    	    	&$81.3$ -- $84.3$	&$4.72$     	    &              	&		    &		    	&\\
$\cos(i)=0.5$, inferior conjunction 	&$3.24$ -- $4.03$   	&$4.18$ -- $4.24$   &$2.94$ -- $4.58$	&$4.16$ -- $4.27$   &$2.93$ -- $6.35$	&$4.16$ -- $4.35$\\
    	    	    	    	    	&$81.3$ -- $84.3$   	&$4.72$		    &		    	&		    &$45.0$ -- $52.4$	&$4.67$ -- $4.69$\\
$\cos(i)=0.0$, superior conjunction *	&$15.7$ -- $101.3$  	&$4.52$ -- $4.74$   &$19.0$ -- $100.4$	&$4.56$ -- $4.74$   &$15.7$ -- $104.3$	&$4.52$ -- $4.74$\\\\
NGC 5408 ULX&&&&&&\\
$\cos(i)=0.5$, superior conjunction 	&$5.08$ -- $23.49$	&$4.30$ -- $4.59$   &$5.68$ -- $18.98$	&$4.32$ -- $4.56$   &$8.32$ -- $57.4$	&$4.40$ -- $4.69$\\
    	    	    	    	    	&$68.4$ -- $117.3$	&$4.71$ -- $4.75$   &$84.3$ -- $88.3$	&$4.72$ -- $4.73$   &		    	&\\
$\cos(i)=0.5$, inferior conjunction 	&$5.37$ -- $14.3$	&$4.31$ -- $4.51$   &$5.67$ -- $19.0$	&$4.32$ -- $4.56$   &$8.32$ -- $57.4$	&$4.40$ -- $4.69$\\
    	    	    	    	    	&$70.4$ -- $117.3$	&$4.71$ -- $4.75$   &$86.3$ -- $88.3$	&$4.72$ -- $4.73$   &		    	&\\
$\cos(i)=0.0$, superior conjunction 	&$5.08$ -- $14.35$  	&$4.30$ -- $4.51$   &$9.80$  	    	&$4.44$		    &--		    	&--\\
    	    	    	    	    	&$79.2$ -- $117.3$	&$4.72$ -- $4.75$   &  		    	&		    &  		    	&\\\\
NGC 1313 X-2&&&&&&\\
$\cos(i)=0.5$, superior conjunction 	&$9.29$ -- $23.1$	&$4.42$ -- $4.59$   &$10.3$ -- $12.1$	&$4.45$ -- $4.48$   &--	    	    	&--\\
$\cos(i)=0.5$, inferior conjunction 	&$6.71$ -- $27.11$	&$4.36$ -- $4.61$   &$11.0$ -- $15.3$	&$4.46$ -- $4.52$   &--	    	    	&--\\
$\cos(i)=0.0$, superior conjunction 	&$5.68$ -- $26.11$	&$4.32$ -- $4.61$   &$5.37$ -- $35.5$	&$4.32$ -- $4.65$   &$7.56$ - $37.5$	&$4.38$ -- $4.65$\\\\
Holmberg II ULX&&&&&&\\
$\cos(i)=0.5$, superior conjunction 	&$5.68$ -- $81.4$	&$4.32$ -- $4.72$   &$6.71$ -- $68.4$	&$4.36$ -- $4.71$   &$10.82$ -- $22.1$	&$4.45$ -- $4.58$\\
$\cos(i)=0.5$, inferior conjunction 	&$5.68$ -- $88.3$	&$4.32$	-- $4.73$   &$5.67$ -- $68.4$	&$4.32$	-- $4.71$   &$10.82$ -- $22.1$	&$4.45$ -- $4.58$\\
$\cos(i)=0.0$, superior conjunction 	&$5.08$ -- $90.3$	&$4.30$	-- $4.73$   &$4.81$ -- $9.11$	&$4.28$	-- $4.42$   &$10.28$	    	&$4.45$\\
    	    	    	    	    	&	    	    	&   	    	    &$57.2$ -- $88.3$	&$4.70$	-- $4.73$   &$72.3$ -- $84.3$	&$4.71$ -- $4.72$\\\\
\end{tabular}
\end{table*}

\begin{table*} \caption{Spectral type of the donor stars. We compare the
classification given by previous authors (as listed in table \ref{tab:photdat})
with our determination of the current and ZAMS spectral types. We give the most
likely type in each case - we elaborate on this in the text. (* applies only
when we do not apply a constraint on the mass accretion rate.). References:
$^1$\citet{Soria05}, $^2$\citet{Liu02}, $^3$\citet{Liu04}, $^4$\citet{Kuntz05},
$^5$\citet {Mucciarelli05}, $^6$\citet {Kaaret04}.}\label{tab:spectype} 
\begin{tabular}{l|l|l|l|l}
    	    	    & Previous              & \multicolumn{2}{|c|}{Our determination}  \\
              	    & spectral type 	    & Current             	    & ZAMS \\      
\hline
NGC 4559 X-7        &O / B supergiant$^1$   &Late B -- A giant/supergiant     	    &B5 -- B0\\
M81 X-6             &O8V / O9V$^2$	    &B MS/giant 	    	    	    &B9 -- B2\\
NGC 5204 ULX        &B0Ib$^3$               &O MS *     	    	    	    &O5 or earlier *\\
M101 ULX-1          &B supergiant$^4$       &A/B MS/giant     	    	   	    &A0 -- O4\\
NGC 5408 ULX        &--                     &B giant/supergiant	(or O MS)      	    &B4 or earlier\\
NGC 1313 X-2        &B0 -- O9 MS$^5$        &B MS	    	    	    	    &B6 -- O6\\
Holmberg II ULX     &O4V / B3Ib$^6$         &B giant/supergiant	(or O MS)      	    &B5 or earlier\\
\end{tabular}
\end{table*}

\subsection{Constraining the BH mass}
A key to understanding the nature of ULXs is the determination of the BH mass.
In five out of seven systems, we can constrain the mass of
the BH based on the optical observations, for certain inclinations. We
list these five cases in Table \ref{tab:bhmass}.

If we assume the ULX X-7 in NGC 4559 has an inclination of $\cos(i)=0.0$, we
find an upper limit on the BH mass of $\simeq 35$\Msun. Analysis of the X-ray
data has suggested a lower limit on the BH mass of $50$\Msun \citep{Cr04}.
This inconsistency can be accounted for by inclining the binary system in our
model so that the optical emission includes a disk component. This results in
the upper limit on the BH mass increasing. We see in Figure
\ref{fig:4559_1_0.5} that for an inclination of $\cos(i)=0.5$ the upper limit
has disappeared.

When we assume an inclination of $\cos(i)=0.5$ for the ULX in NGC 1313, we find
an upper limit on the BH mass of $100$\Msun. This upper limit increases as the
disk component is reduced, and vanishes when we set $\cos(i)=0.0$. This BH range
is consistent with a BH mass of $\sim 100$\Msun \ previously inferred from the
X-ray data \citep{Zamp04}.

If we use an inclination of $\cos(i)=0.0$ for the ULX in NGC 5408, we find a
maximum BH mass of $110$\Msun. When we incline the system so as to include a
disk component, the upper limit disappears. In both cases this is consistent
with the X-ray data, which implies a BH mass of $\sim 100$\Msun, assuming
accretion at the Eddington limit \citep{Soria04}.

We note also that the ULX in NGC 5204 has a maximum BH mass of $\simeq
240$\Msun \ when we assume an inclination of $\cos(i)=0.5$ and a very low
accretion efficiency. The constraint disappears when we set $\cos(i)=0.0$. The
X-ray data supports the presence of a cool thermal disk component
\citep{Roberts05} which may be produced by an IMBH or by a stellar-mass disk
cooled by other processes. Unfortunately, optical mass constraints are not
strong enough to discriminate between the stellar-mass and IMBH scenarios.

The fifth source for which we have a BH mass constraint is ULX X-6 in M81. We
see that when we set $\cos(i)=0.0$ there is an upper limit on the BH mass of
$33$\Msun, but if we set $\cos(i)=0.5$ we find a lower limit on the BH mass of
$20$\Msun. The existence of this lower limit is dependent on the irradiating
X-ray spectrum being softer than we have elsewhere assumed and the age of the
donor being comparable to the of the field stars. We note also that this BH has
been suggested to have a mass of $18$\Msun \ based
on analysis of X-ray data \citep{Liu02}, but the authors found it to be model
dependent.

\begin{table}
\caption{BH mass constraints, and the inclination for which they apply (* applies only when we do not apply a constraint
on the mass accretion rate. ** applies only for a hardness ratio of $\xi=0.01$.)}
\label{tab:bhmass}
\begin{tabular}{l|l}
\hline
NGC 4559 X-7 &$< 35$\Msun for $\cos(i)=0.0$ \\
M81 X-6      &$> 20$\Msun for $\cos(i)=0.5$ ** \\
    	     &$< 33$\Msun for $\cos(i)=0.0$ \\
NGC 5204 ULX &$< 240$\Msun for $\cos(i)=0.5$ * \\
NGC 5408 ULX &$< 110$\Msun for $\cos(i)=0.0$ \\
NGC 1313 X-1 &$< 100$\Msun for $\cos(i)=0.5$ \\

\end{tabular}
\end{table}

\subsection{Predictions for further opt/IR observations}

We now look at what these new identifications for the donor stars mean for
future optical observations of these sources. We show our predictions for the
binary periods of the seven systems in Table \ref{tab:ages}. The ULX in M81 is
a good prospect for determination of the BH mass, so we base this
discussion around this source. \citet{Liu02} suggested the donor star was of
type O8V/O9V. A typical O8V star has a mass of $23$\Msun \ and a radius of
$8.5$\Rsun. Assuming a Roche-lobe filling donor, the binary period for such a
source will be $\simeq 39$ hours if we assume a $10$\Msun \ BH and $\simeq 44$
hours if we assume a $1000$\Msun BH. The amplitude of the optical lightcurve will
be dependent on the BH mass, binary inclination and other parameters, but in
figure 3 of \citet{Copp05} we see that the optical amplitude of the lightcurve
of a typical O-star irradiated by an X-ray source of $L_x = 10^{40}$\ergss \ is
of the order of a tenth of a magnitude. 

As we see in Table \ref{tab:ages}, our predictions for the binary period of
the M81 ULX change significantly with both inclination and BH mass. A measured
period of $\sim 100$ -- $300$ hours would imply an inclination of close to
$\cos(i)=0.0$ and a low ($\sim 10$\Msun) BH mass. A shorter period would imply
both a more face-on system and a more massive BH, with a period of around $16$
hours suggesting a BH mass of $1000$\Msun. We note also that the donor star in
the $\cos(i)=0.0$ / $10$\Msun case is a larger, more evolved object. We would
expect the variation in optical luminosity to be observable in this case, since
the variation is not diluted by the accretion disk, and the optical luminosity
of evolved stars is more signifcantly modified by X-ray irradiation than for MS
stars \citep{Copp05}. A small or unobservable optical variation would imply an
inclined system, an MS donor and a more massive BH.

\begin{figure*} \centering 
\begin{minipage}[c]{0.5\textwidth}
\includegraphics[angle=270,width=1.0\textwidth]{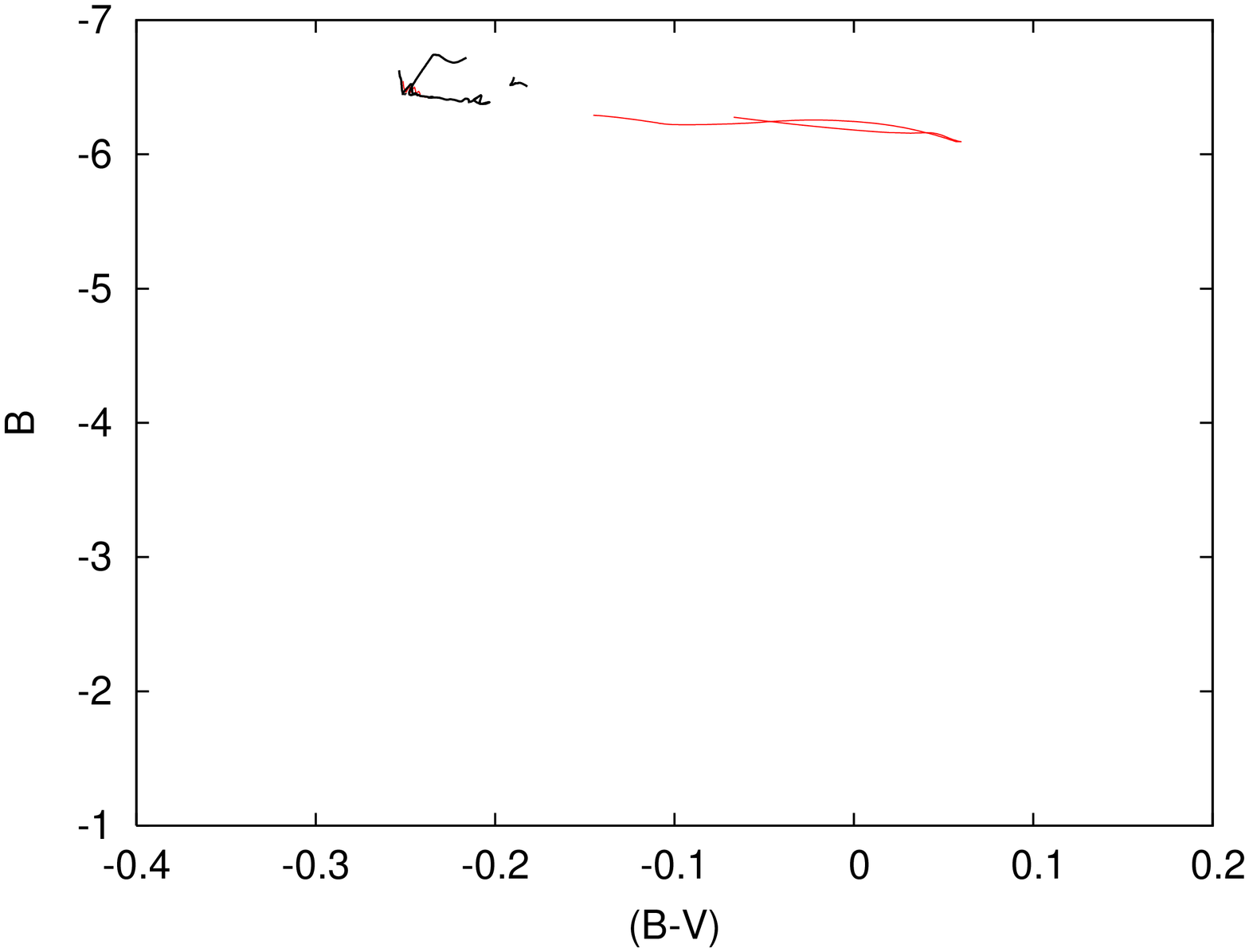}
\end{minipage}%
\begin{minipage}[c]{0.5\textwidth} \hfill
\includegraphics[angle=270,width=1.0\textwidth]{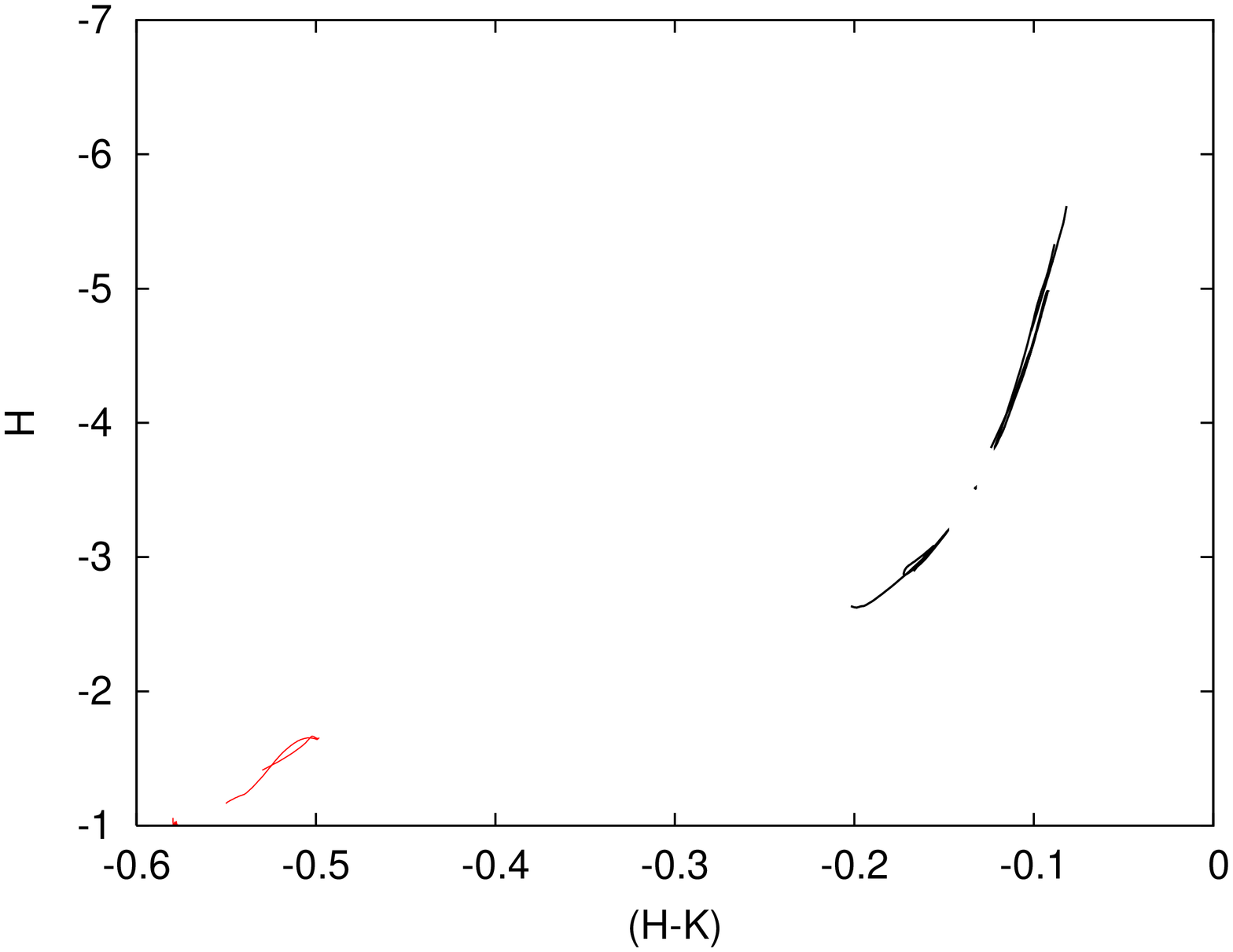}
\end{minipage} 
\caption{Colour-magnitude diagrams showing the $90$\% confidence contours for
the optical counterpart of the ULX in NGC 5408, for inclinations of $\cos(i)=0.5$
(black) and $\cos(i)=0.0$ (red) with the star in superior conjunction.} \label{fig:5408colours}
\end{figure*}

In \citet{Copp05} we noted that infrared observations might be an important
diagnostic in determining the nature of ULXs. We illustrate this point in Figure
\ref{fig:5408colours}. We show here colour-magnitude diagrams for $B$ against
$(B-V)$ and $H$ against $(H-K)$ for the optical counterpart to the ULX in NGC
5408. In both cases, we plot the sections of the evolutionary tracks which fit
the optical data to the $90$\% confidence level. We plot tracks for inclinations
of $\cos(i)=0.5$ and $\cos(i)=0.0$ with the star in superior conjunction. In
the $B$ versus $(B-V)$ plot, the sets of tracks for the two inclinations are
necessarily similar in colour and magnitude. However, we see that at infrared
wavelengths there is a clear distinction between the $\cos(i)=0.5$ tracks and
the $\cos(i)=0.0$ tracks in both colour and magnitude. It is clear therefore
that the combination of IR and optical observations will allow us to constrain
parameters in ULX systems with much higher precision.

\subsection{The evolution and history of ULXs}

\begin{table} 
\caption{Predictions of the orbital periods in these systems, in hours, based on our
determinations of the donor star parameters (* applies only when we do not
apply a constraint on the mass accretion rate. ** applies only for a hardness ratio of $\xi=0.01$.)}
\label{tab:ages} 
\begin{tabular}{l|l|l|l|l} 

BH mass     	    &$10$\Msun      	&$100$\Msun         &$1000$\Msun    \\ 
\hline
\multicolumn{4}{c}{$\cos(i)=0.5$, superior conjunction}\\\\ 
NGC 4559 X-7        &$418$ -- $732$ 	&$260$ -- $276$	    &$67.0$ -- $78.8$	\\
M81 X-6 **          &--	    	    	&$47.9$ -- $59.5$   &$15.9$ -- $16.2$	\\
NGC 5204 ULX *      &$27.4$ -- $42.0$	&$37.8$ -- $44.5$   &--	    	    	\\
M101 ULX-1          &$767$ -- $873$ 	&$271$ -- $295$	    &$77.6$ -- $82.9$	\\
    	    	    &$26.6$ -- $26.9$	&   	    	    &	    	    	\\
NGC 5408 ULX        &$114$ -- $277$ 	&$94.9$ -- $112$    &$33.4$ -- $36.6$	\\
    	    	    &$27.5$ -- $39.3$	&$37.1$ -- $37.4$   &	    	    	\\
Holmberg II ULX     &$27.2$ -- $152$	&$35.4$ -- $61.5$   &$21.1$ -- $22.1$	\\
NGC 1313 X-2 C1     &$22.6$ -- $36.4$	&$21.8$ -- $27.4$   &--	    	    	\\

\hline
\multicolumn{4}{c}{$\cos(i)=0.0$, superior conjunction}\\\\
NGC 4559 X-7        &$864$ -- $2060$	&-- 	    	    &--	    	    	\\
M81 X-6             &$136$ -- $287$	&--   	    	    &--	    	    	\\   
NGC 5204 ULX *      &$27.5$ -- $46.7$	&$38.8$ -- $56.4$   &$41.8$ -- $57.7$	\\
M101 ULX-1 *        &$27.2$ -- $387$	&$38.2$ -- $394$    &$40.6$ -- $376$	\\
NGC 5408 ULX        &$247$ -- $732$	&$760$   	    &--	    	    	\\
    	    	    &$27.5$ -- $41.0$ 	&   	    	    &	    	    	\\
Holmberg II ULX     &$27.0$ -- $417$	&$542$ -- $1500$    &$783$	    	\\
   	    	    &	    	    	&$35.3$ -- $37.3$   &$36.2$ -- $38.5$  	\\
NGC 1313 X-2 C1     &$26.0$ -- $96.2$  	&$28.4$ -- $157.5$  &$28.3$ -- $190.4$	\\
\end{tabular}
\end{table}

If IMBHs do indeed exist, it is of great importance to clarify how the ULX/IMBH
and the star formation process in their vicinity are related. The open question
to resolve is whether the donor star is coeval to the BH progenitor or captured
by the BH some time after formation. If the star and the BH formed together,
determining the age of the donor star also determines the age of the BH. If the
star was captured by the BH, then the statistics of the spectral type and mass
distribution of the donor stars can be used to set constraints on the capture
rate  and hence provide estimates to the IMBH populations.  

In our study, we found that the donor stars are mainly of spectral type B, and
are significantly older than previously determined. For example, \citet{Liu02}
inferred the donor star in M81 X-6 ULX to have an age of less than
$10^{6.7}$yr. We find the minimum stellar age to be an order of magnitude
greater, if the BH mass is assumed to be $<\sim 100$\Msun. In a
number of cases, the photometric data alone allows us to infer the donor is of
spectral type B. In other cases, a range of (more massive) possibilities exist,
but by applying constraints on the mass accretion rate we find a B-type star to
be the most likely donor. In the case where the optical data points to a
massive, O-type donor, the implied mass accretion rate is inconsistent with
that which would be expected from the X-ray observation, given our assumed
accretion efficiency of $\eta = 0.1$. We suggest therefore, that donor stars of
a narrow spectral and mass range are necessary to produce a very luminous,
Roche lobe fed ULX, and our finding of large fraction of B-type stars in the
ULX sample may be significant. 

If the compact objects in these systems are indeed IMBHs, and if we assume the
capture scenario, then the fact that a B-type donor is sufficient to fuel a ULX
allows a lower spatial density for IMBHs for the observed population of ULXs
than if the donors were found to be of type O, since B-type stars are more
common and so the chances of forming a ULX binary are higher. Various authors
have modelled the tidal capture of a donor star by an IMBH (see {\it e.g}
\citealt*{Hopman04}; \citealt{Blecha06}). The capture rate appears to be $\sim$
stellar number density, but is only weakly dependent on stellar mass. This
would imply that more B stars than O stars should be captured. However, a
competing effect is the requirement that the star comes close enough to be
captured but far enough not to be tidally destroyed. This requirement may
favour O stars, since they  may more easily survive tidal squeezing. We note
also that tidal capture of isolated stars is only one process through which an
IMBH might aquire a companion; another process is by capturing stars in binary
systems, which may have a different frequency of occurence and period
distribution for B or O stars.

It is also interesting to note that the two lowest luminosity ULXs in our
sample, those in M81 and M101, are also those where an old, less massive
donor of age $\sim 10^8$Myr is a possibility. This may be related to the
existence of a population of low-luminosity sources ($<\sim
2\times10^{39}$\ergss) also found in old elliptical galaxies, probably
identified as low-mass XRBs. Conversely, the ULXs more luminous than
$2\times10^{39}$\ergss \ are almost always found in star-forming galaxies, and
we find the donor stars to be of age $\sim 10^7$Myr or less.

\subsection{Other systematic effects}
\label{sec:errors}

We have made a number of assumptions in this work. We reexamine here some of
the effects that will have an influence on our conclusions.

\subsubsection{System geometry} We have assumed a thin disk in our model, but
as we noted in Section \ref{sec:intro}, some authors have suggested more
complicated disk models, with a thin disk covered by a Comptonised corona
\citep{Socrates05}. These models have been developed to support the possibility
of super-Eddington accretion in ULX systems. The corona emits hard X-rays which
are reflected by the ionized surface of the inner disk. As in our disk model,
the bulk of the optical emission comes from the outer regions of the disk, due
to these regions having a much larger surface area than the inner parts of the
disk. To a large extent therefore the optical/IR emission is similar. However,
the corona model will result in a harder X-ray spectrum incident on the outer
disk. We discuss the effect of changing the X-ray hardness in more detail in
Section \ref{sec:hardness}. Additionally, since the corona extends away from
the disk surface the angles of incidence of the X-rays on the outer disk
regions are changed somewhat, but the extent of the corona is not large enough
for this to have a significant effect.

\subsubsection{Radiation pressure} In the results we have presented here, we
have not included the effects of radiation pressure. The effects of radiation
pressure on the shape of a Roche lobe filling star are unclear: some authors
suggest that the shape will be unaffected \citep{Howarth97}, whereas others
predict a significant effect in very X-ray luminous binaries
\citep{Phillips02}. We showed in \citet{Copp05} that under the
\citet{Phillips02} formulation, radiation pressure has an increasing effect on
the shape and luminosity of the donor as the BH mass is decreased. This is
because as we decrease the BH mass the binary separation also decreases, and
the X-ray flux incident on the stellar surface increases. We found also that
the effect decreases as we use a donor of later spectral type, for the same
reasons. We find therefore that for most of our fitted solutions, where the BH
mass lies between $100$ and $1000$\Msun \ and the donor is found to be a star
of type B or later, the effect of including radiation pressure on our results
is small. We begin to see an appreciable deviation from the results we have
reported when we use a BH mass of $\sim 100$\Msun \ and a main sequence, O-type
donor. Even in this case, the fitted stellar mass only changes by $1$ or
$2$\Msun, which is a small percentage of the total stellar mass and not enough
to alter our classification of the star. We would expect a more significant
deviation for low ($\sim 10$\Msun) BH masses, but the \citet{Phillips02}
formulation we use for radiation pressure becomes inappropriate at this point,
since it does not allow for any circulatory currents in the stellar surface
driven by the irradiation. This means it tends to represent an extreme case of
maximum stellar distortion, and in this extreme case the radiation pressure is
very large, to the point of stripping the outer layers away from the donor so
that it looks quite unlike an ordinary star. A radiation pressure formulation
which included circulatory currents would predict a lessened effect on the
stellar shape. Shielding by the disk will mitigate this effect further. 

Given that the actual effect of radiation pressure on the stellar shape in ULX
systems is not fully understood, we consider our omission of this component in
this paper to be appropriate. We do however note this introduces an additional
source of uncertainty, particularly for low BH masses. 

\subsubsection{X-ray hardness}
\label{sec:hardness}

\begin{figure*} \centering 
\begin{minipage}[c]{0.5\textwidth}
\includegraphics[angle=270,width=1.0\textwidth]{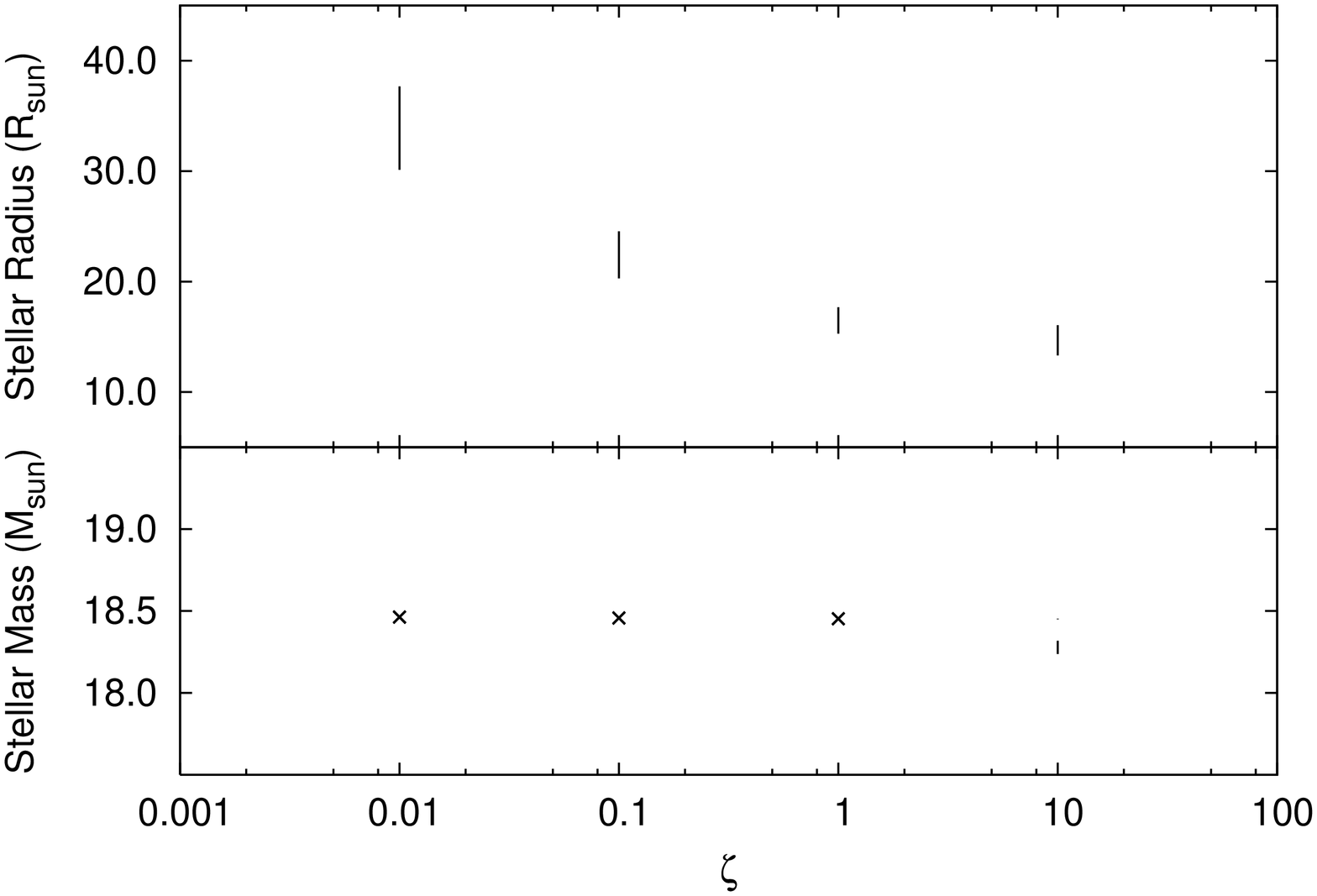}
\end{minipage}%
\begin{minipage}[c]{0.5\textwidth} \hfill
\includegraphics[angle=270,width=1.0\textwidth]{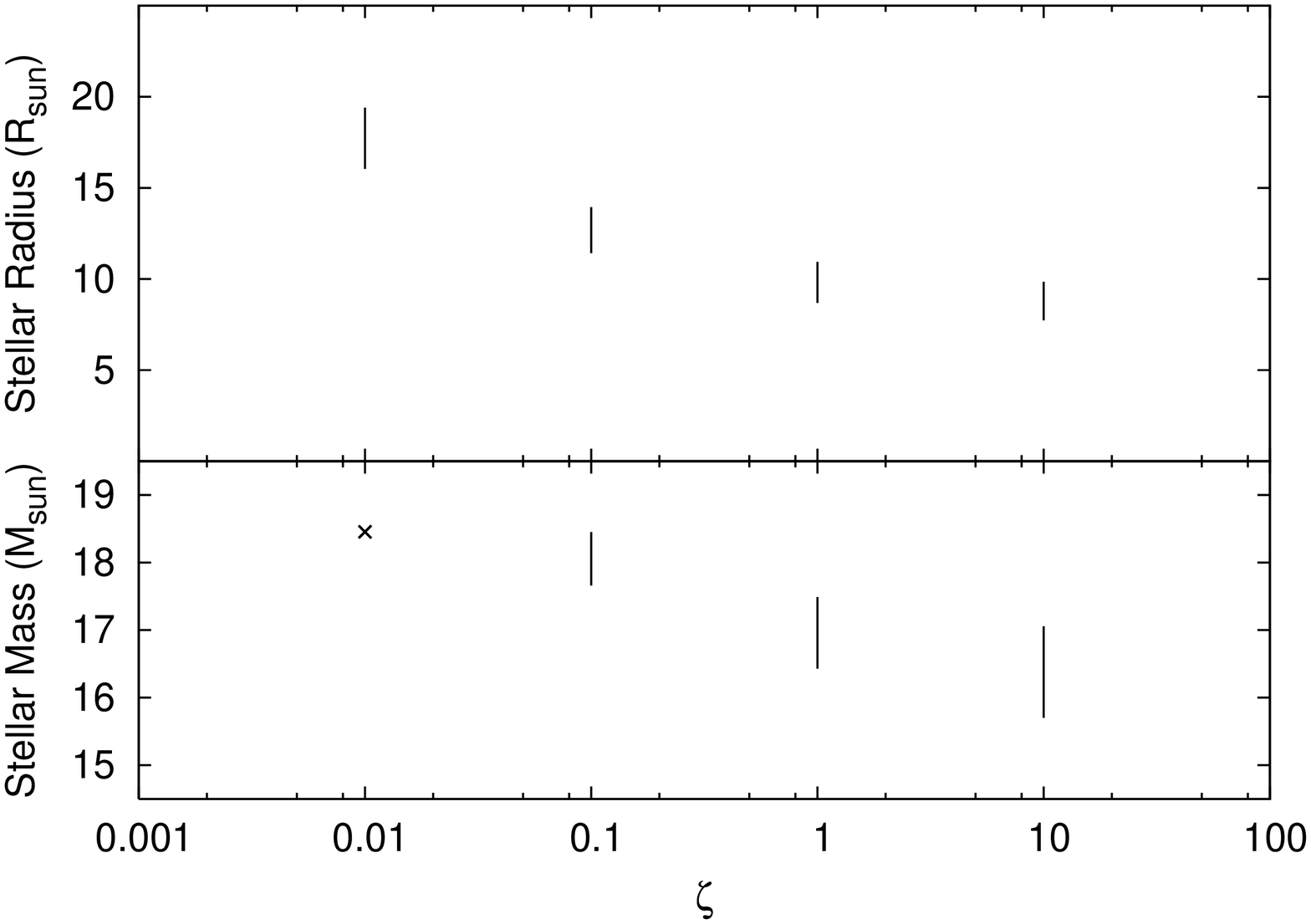}
\end{minipage} 
\caption{The change in stellar mass/radius with hardness ratio $\xi$ for ULX
X-7 in NGC 4559 (left) and the ULX in NGC 5408 (right). We use an inclination
of $\cos(i)=0.5$, a BH mass of $100$\Msun \ and a stellar age of 10Myr. The
lines show the range of stellar parameters that fitted with our model at the
$90$\% confidence level. For clarity, we mark with an `x' cases where our model
produces a single unique solution.}

\label{fig:hardness}
\end{figure*} 

The hardness of the X-ray spectrum determines the depth at which the incident
radiation deposits most of the energy. Soft X-rays are easily absorbed at the
disk surface, while hard X-rays attenuate only at large optical depths. For
incident X-rays with a soft spectrum, a hot surface skin layer is formed at
depths optically thin to optical radiation, and the emission from the skin
layer is at wavelengths shortward of the optical bands. However, for incident
X-rays with a hard spectrum, most of the energy is deposited at depths
optically thick to optical radiation. This heats the internal regions of the
disk plane and hence leads to a more luminous disk at the optical wavelengths.
This was illustrated in figure 6 of \citet{Copp05}, where we see that as the
X-ray hardness is increased, the disk V magnitude similarly increases. We noted
also that the effect of varying the hardness ratio on the stellar luminosity
was very small.

Because we do not know the hardness of the X-ray spectrum incident
on the irradiated surfaces in these systems (as discussed in Section
\ref{sec:mod_irrad}) we have used a fixed hardness ratio of $\xi = 0.1$. We now
discuss the effect of varying the hardness ratio on our results.

In Figure \ref{fig:hardness} we have plotted our determinations of the masses
and radii of the donor stars in the NGC 4559 and NGC 5408 ULXs. We have fixed
the inclination to $\cos(i)=0.5$, the BH mass to $100$\Msun \ and the stellar
age to $10$Myr. If we examine the results for the ULX in NGC 4559, we see that
varying $\xi$ has little to no effect on our determination of the mass of the
donor star. Our determination of the radius however, decreases with increasing
hardness ratio. We see a similar decrease in stellar radius with increasing
$\xi$ in the NGC 5408 case. We see also an appreciable decrease in stellar mass
in this case, but this decrease is still much smaller than the decrease in
stellar radius. 

In a Roche lobe geometry the shapes of the lobes is determined by the mass
ratio of the two components, and the scale of the system is set by their
separation. In our model we determine the scale by setting the volume radius of
the secondary Roche lobe to be equal to the undistorted radius of the donor
star. The results of Figure \ref{fig:hardness} can therefore be understood as
follows. A harder incident X-ray spectrum leads to an accretion disk which is
more luminous at optical wavelengths. If the disk in our model is hotter, then
to keep it consistent with the observation it must be smaller, so the scale of
the Roche lobes must be decreased. Our calculations therefore result in a
smaller fitted stellar radius. Since the mass ratio affects only the shape of
the Roche lobes, the size of the disk is only weakly dependent on stellar mass,
and so varying the hardness ratio will tend to have little effect on this
parameter.

In both ULXs, the changes in stellar parameters are smaller than might be
expected, given the significant variation in disk luminosity with hardness
ratio shown in figure 6 of \citet{Copp05}. This is because that figure did not
include the stellar component. Changing the mass ratio results in a different
determination of the stellar radius, but changing this parameter results in a
change in the luminosity of both the star and the disk. The radius therefore
does not need to be changed by much to have a large effect on the overall
luminosity.

Through examination of the X-ray data, we have found $\xi$ to vary from $\sim
0.1$ to $\sim 1$ in the NGC 4559 ULX. We have assumed this to be a physically
appropriate hardness ratio range for all the ULXs in our sample, and have used
$\xi=0.1$ for all the model fits in this paper. Figure
\ref{fig:hardness} indicates that the general findings in this paper will not
be invalidated if the X-ray spectrum in these systems is harder than we have
assumed. We see that increasing the X-ray hardness to $\xi=1$ causes no change
in the stellar mass in one case and a decrease of $\sim 1$\Msun \ in the other.
We see also a decrease in the stellar radius of $2$ -- $5$\Rsun \ as $\xi$ is
increased from $0.1$ to $1$. We conclude therefore that, if the X-ray spectrum
is harder, our determinations of the spectral type of the donors are still
valid but they may be somewhat smaller and less evolved than our results
suggest.

\subsubsection{Perturbation of the donor as a result of mass transfer} A key
distinguishing feature between the IMBH evolutionary scenarios is the epoch at
which the steady mass transfer began. There are two possibilities: (a) we
observe the system where mass transfer has not yet significantly affected the
state of the star, and so the models we use are a fair description of the
donor; (b) we observe it after the mass transfer has already significantly
altered the state of the star and a binary evolution code is required for
proper modelling.

We note that even in cases when a binary evolution code should be used the
results we present here are not initially invalidated. Our method is to determine the
star which, in the presence of the intense X-ray radiation field produced by
the ULX, will have the appearance of the observed donor, as it exists at the
current epoch. Therefore, assuming our model assumptions (such as the
assumption that the star is in thermal equilibrium) are correct, the masses and
radii will be accurate, since these parameters depend on the current physical
state of the star, not its history. Conversely, the parameters we have
determined such as the stellar age and ZAMS mass will be inaccurate if the star
has lost a significant amount of its mass through Roche lobe overflow.

A binary stellar evolutionary code is necessary when studying a ULX in which the
mass transfer has endured long enough for the appearance of the donor to be
significantly modified. It is unclear as to how long it takes for a star to
deviate in characteristics from the single star evolutionary tracks when it is
undergoing mass loss at the rates we have inferred. The mass loss can be assumed
to have little affect on very massive stars ($\sim 40$\Msun \ and above) since
over the course of their short lives they will transfer only a small percentage
of their mass onto the BH through Roche lobe overflow. On the other hand, it has
been shown that low mass stars ($\simeq 1$\Msun) deviate from single star models
very rapidly when undergoing phases of high mass transfer \citep{Schenker02}.

It is reasonable to expect that a star that has transferred, say, $50$\% of its mass
onto the BH through Roche lobe overflow will be significantly perturbed. When we
examine the accumulated mass loss for star with ZAMS mass $10$ -- $20$\Msun, we
find that to lose $50$\% takes $\sim 2$ -- $3$Myr when an X-ray luminosity of
$10^{40}$\ergss \ is assumed, although this varies depending on the point in the
stellar evolution at which the mass transfer begins. This is a short length of
time, but our model predicts the length of time in which any star can sustain
mass transfer at ULX rates is also short, even when the effect of the mass loss
on the star is not considered. Massive stars can transfer mass at ULX rates from
ZAMS, but have intrinsically short ($< 10$Myr) lifespans. Lower mass donors only begin to
transfer mass at the required rate towards the end of their time on the MS. The
maximum duration of the binary as a ULX, assuming a $10^{40}$\ergss \ X-ray
luminosity and a donor ZAMS mass of $10$ -- $20$\Msun, is around $6$ -- $8$Myr.
Assuming a lower mass transfer rate, as would be expected in ULXs with
X-ray luminosities of $\sim 10^{39}$\ergss, means that the donor takes
significantly longer to be perturbed by the mass transfer. We note again that
these conclusions are also dependent on the accretion efficiency we have assumed
in ULXs being correct.

Note that in this estimation we have assumed the donor is unperturbed at
the beginning of the ultraluminous X-ray emitting phase of the binary. Our
model suggests that low mass stars can transfer mass at sub-ULX rates in the
early part of their lives, assuming they evolve to a semi-detached state
shortly after ZAMS. This would mean that the star would already be appreciably
perturbed before it began its ULX phase. Similarly, if the star has at some
point in its past undergone a phase of extreme mass transfer (such as
thermal-timescale mass transfer, as discussed in Section \ref{sec:ttmt}), then
a binary evolution code would almost certainly be necessary. We note also that
very high rates of mass transfer are possible even when the driver is nuclear
evolution: very massive stars can transfer mass to the BH at a rate of $\sim
10^{-5}$\Msun/yr or more, particularly towards the end of their MS life and
beyond. This would result in rapid and significant deviation from the single
star tracks we have used. Mass transfer at this rate  is in excess of what we
would expect for the ULXs in our sample, but some very X-ray luminous systems
such as the $L_x = \sim 10^{41}$\ergss \ ULX in M82 could have very early type
donors in states of extreme mass loss.

We conclude therefore that a binary evolution code may be necessary in some of
the sources discussed here, but given that the timescales of the ULX active
phase and the timescale for the star to be perturbed by the mass loss are
similar, the use of single star tracks is reasonable for this investigation.


\section{CONCLUSIONS}
\label{sec:conclusions}

We have constructed a model which describes the optical emission from
ultra-luminous X-ray sources (ULXs). Our model assumes an X-ray binary nature
for ULXs, with the donor star overfilling its Roche lobe. We assume radiative
equilibrium and use a radiative transfer formulation to determine the
re-radiated thermal emission \citep{Copp05}. We predict the optical luminosity
of the emission from both the irradiated star and a thin accretion disc. We use
the mass accretion rate as implied by the observed X-ray luminosity as an
additional constraint on the donor star parameters, assuming that the nuclear
evolution of the star is the driver for the mass transfer, and hence
calculating the mass accretion rate from the rate of increase in radius of the
star, compared to the evolution of the Roche lobe.

We have applied our model to optical observations of seven different ULX optical
counterparts, and determined the parameters of the donor stars, by fitting the
multi-band photometric observations to the optical emission predicted by our
model for different sets of stellar parameters. The stellar models were taken
from the Geneva stellar evolution tracks of \citet{Lejeune01}. We
varied the other parameters in our model, such as binary inclination, position
of the star with respect to the observer when the observation was made, BH mass
and stellar metallicity, in order to better determine the nature of the donor
star. 

Previous analysis found that the donor star of ULX are often MS O stars  or
early-type supergiants. Here we have shown that the donor stars are  older and
less massive. Their spectral types are generally consistent  with MS stars or
evolved giants/supergiants of spectral type B or later. They tend to have ages
$\sim 10$ -- $100$Myr. However the accretion rate necessary to fuel an X-ray
luminosity of $10^{40}$\ergss \ will result in a B-star being completely
consumed in $5$ -- $10$Myr. This implies an upper limit on the duration of the
mass transfer. It may be that the capture of the donor by the BH occured within
this period of time, or it could be that binary formed a long time ago but
evolved to a semi-detached state within this time. A further possibility is that
the donor was originally a much earlier-type, more massive star that has been
transferring mass over most of its lifetime, such that it has the appearance of
an older and less massive star in the current epoch. The use of binary stellar
evolution models will be required to fully explore this possibility. 

The ULX in NGC 5204 is an exception to our general findings. Our model cannot
provide a good fit with the observation in this case, owing to the constraint we
apply on the mass accretion rate. When we remove this constraint the best
solution is for an MS O-star in NGC 5204, a more massive donor than has been
suggested for this system by other authors. However, our fits suggest a
much higher mass accretion rate than would be expected from the X-ray
luminosity. A possibility is that this ULX is wind fed, instead of via Roche
lobe overflow as we have assumed. 

There are a number of systematic effects, which we have discussed. It should be
noted that in a number of cases these effects may be greater when we assume a
stellar mass BH rather than an IMBH, due to the physics of super-Eddington
accretion being uncertain. In stellar mass BH systems the disk geometry may be
significantly different from the thin disk approximation we have used, and if
there is strong beaming mass transfer may be occuring on thermal timescales. In
addition, the effect of radiation pressure may significantly perturb the star,
whereas we have shown this effect is small when a more massive BH is used.

We have also discussed the implications for the binary periods and optical
amplitudes, which will be revealed by future observation of these sources. Our
model allows us to determine the relative contribution of the star and the disc
to the emission. For a less massive BH, the optical/IR emission will be
dominated by light from the star. For a more massive BH, the optical/IR
emission will be disc dominated. If we assume an irradiating X-ray luminosity
of $10^{40}$\ergss, the disc will dominate the emission for a BH mass of $>
90$\Msun \ if we assume the donor is a typical B-type MS star, and for a BH
mass of $> 300$\Msun \ if we assume a B supergiant.

In more than half of these systems, we can constrain the mass of the BH based
on the optical observations, although the existence of these constraints are
dependent on an assumed inclination. If we assume an inclination of
$\cos(i)=0.5$ for the ULX X-1 in NGC 1313, we find the BH has a maximum mass of
$\simeq 100$\Msun. If we assume an inclination of $\cos(i)=0.0$ for the ULXs in
NGC 5408 and NGC 4559, we find the BHs in these systems have a maximum mass of
$\simeq 110$\Msun \ and $\simeq 35$\Msun \ respectively. For ULX X-6 in M81, we
found a minimum BH mass of $20$\Msun, or a maximum mass of $33$\Msun, depending
on the inclination. A temporal optical study of this source could determine
with some accuracy the spectral type of the donor, the inclination and the BH
mass, so we conclude that this source in particular is an excellent target for
further observations with the aim of positively discriminating between stellar
mass BH and IMBH scenarios.

\section*{ACKNOWLEDGEMENTS}

We would like to thank the referee for his/her detailed comments which have led
to a number of significant improvements to this paper. 


\bibliography{irrad_paper2}

\end{document}